\documentclass[onecolumn,amsmath,amssymb,nofootinbib,letterpaper,11pt]{article}
\pdfoutput=1
\usepackage{jheppub}
\usepackage{ifpdf}
\usepackage[utf8]{inputenc}
\usepackage[T1]{fontenc}
\usepackage{graphicx}
\input{epsf}
\usepackage{epsfig}
\usepackage{epstopdf}
\usepackage{arcs}
\usepackage{subfigure}
\usepackage{tensor}
\usepackage{braket}
\usepackage{float}
\usepackage[dvipsnames]{xcolor}
\usepackage{amsmath}
\newcommand{\be}{\begin{equation}}
\newcommand{\ee}{\end{equation}}

\usepackage{tikz}
\usetikzlibrary{calc}   
\usetikzlibrary{topaths}

\usepackage{hyperref}
\hypersetup{
	colorlinks =WildStrawberry,
	linkcolor =red,
	pdftitle={Leading corrections to Kerr geometry},
	citecolor=NavyBlue,
	filecolor=WildStrawberry,
	urlcolor=WildStrawberry,
}

\usepackage[toc]{appendix}
\usepackage{color,soul} 
\usepackage{datetime}
\newcommand{\bea}{\begin{eqnarray}}
\newcommand{\eea}{\end{eqnarray}}
\newcommand{\ba}{\begin{eqnarray}}
\newcommand{\ea}{\end{eqnarray}}

\newcommand{\beq}{\begin{equation}}
\newcommand{\eeq}{\end{equation}}
\newcommand{\beqa}{\begin{eqnarray}}
\newcommand{\eeqa}{\end{eqnarray}}
\newcommand{\beqar}{\begin{eqnarray*}}
\newcommand{\eeqar}{\end{eqnarray*}}
\newcommand{\e}{\epsilon}

\newcommand{\X}{\mathcal{X}}
\newcommand{\W}{\mathcal{W}}
\newcommand{\R}{\mathcal{R}}
\newcommand{\Z}{\mathcal{Z}}
\newcommand{\A}{\mathcal{A}}

\newcommand{\C}{\mathcal{C}}
\newcommand{\D}{\mathcal{D}}

\newcommand{\G}{\mathcal{G}}

\renewcommand{\c}{$c$}

\renewcommand{\href}[2]{#2}

\title{Leading higher-derivative corrections to Kerr geometry}
\author[a]{Pablo A. Cano}
\author[a]{and Alejandro Ruip\'erez}

\affiliation[a]{Instituto de F\'isica Te\'orica UAM/CSIC,\\C/ Nicol\'as Cabrera, 13-15, C.U. Cantoblanco, 28049 Madrid, Spain\vspace{0.1cm}}
\vspace{0.1cm}
\emailAdd{pablo.cano@uam.es} 
\emailAdd{alejandro.ruiperez@uam.es}
\abstract{We compute the most general leading-order correction to Kerr solution when the Einstein-Hilbert action is supplemented with higher-derivative terms, including the possibility of dynamical couplings controlled by scalars. The model we present depends on five parameters and it contains, as particular cases,  Einstein-dilaton-Gauss-Bonnet gravity, dynamical Chern-Simons gravity and the effective action coming from Heterotic Superstring theory. By solving the corrected field equations, we find the modified Kerr metric that describes rotating black holes in these theories. We express the solution as a series in the spin parameter $\chi$, and we show that including enough terms in the expansion we are able to describe black holes with large spin. For the computations in the text we use an expansion up to order $\chi^{14}$, which is accurate for $\chi<0.7$, but we provide as well a Mathematica notebook that computes the solution at any given order. We study several properties of the corrected black holes, such as geometry of the horizon, ergosphere, light rings and scalar hair. Some of the corrections violate parity, and we highlight in those cases plots of horizons and ergospheres without $\mathbb{Z}_{2}$ symmetry.}

\preprint{
IFT-UAM/CSIC-19-2
}

\begin{document}
\maketitle

\newpage
\section{Introduction}

General Relativity (GR) describes the gravitational interaction as the effect of spacetime curvature. Einstein's field equations, that rule the dynamics of the gravitational field, can be derived from the Einstein-Hilbert (EH) action
\begin{equation}\label{eq:EH}
S=\frac{1}{16\pi G}\int d^{4}x\sqrt{|g|}R\, ,
\end{equation}
which is essentially the \textit{simplest} non-trivial covariant action one can write for the metric tensor. This beautiful theory has passed a large number of experimental tests --- including the recent detection of gravitational waves coming from black hole and neutron star binaries \cite{Abbott:2016blz,Abbott:2016nmj,TheLIGOScientific:2016src,TheLIGOScientific:2017qsa,Abbott:2017vtc,Abbott:2017oio}--- and it is broadly accepted as the correct description of the gravitational interaction.


However, there are good reasons to think that GR should be modified at high energies. One of these reasons is that GR is incompatible with quantum mechanics. Although we still lack a quantum theory of gravity, it is a common prediction of many quantum gravity candidates that the gravitational action (\ref{eq:EH}) will be modified when the curvature is large enough. For instance, String Theory predicts the appearance of an infinite series of higher-derivative terms \cite{Gross:1986mw,Gross:1986iv,Alvarez-Gaume:2015rwa} correcting the Einstein-Hilbert action. The precise terms and the scale at which they appear depend on the scheme and on the compactification chosen.
Nevertheless, whatever the modification of GR is, it should be possible to describe it following the rules of Effective Field Theory (EFT): we add to the action all the possible terms compatible with the symmetries of the theory and we group them following an increasing order of derivatives (or more generally, an increasing energy dimension). In the case of gravity, we would like to preserve diff. invariance and local Lorentz invariance,\footnote{See \textit{e.g.} \cite{Li:2010cg,Horava:2009uw,deRham:2010kj,Olmo:2011uz} for other possible extensions of GR.} and this means that the corrections take the form of a higher-curvature, or higher-derivative gravity \cite{Stelle:1977ry}. A more general possibility --- that we will also consider here --- is to increase the degrees of freedom in the gravitational sector, by adding other fields that are not active at low energies \cite{EdGB}.

Generically, the introduction of higher-derivative interactions means that Ricci-flat metrics no longer solve the gravitational field equations. As a consequence, the Schwarzschild \cite{Schwarzschild:1916uq} and Kerr \cite{Kerr:1963ud} metrics, that describe static and rotating black holes (BHs) in GR, are not solutions of the modified theories. One has to solve the modified field equations in order to determine the corrected black hole solutions, and it is an interesting task to look at the properties of these corrected geometries. 

On general grounds, the higher-derivative corrections modify the gravitational interaction when the curvature is large, and they usually improve the UV behaviour of gravity \cite{Stelle:1976gc}. The effect of the corrections will be drastic precisely in situations where GR fails, such as in the Big-Bang or black hole singularities, and it is expected that higher-derivative terms can resolve these divergencies \cite{Starobinsky:1980te,Arciniega:2018fxj,Cisterna:2018tgx,Arciniega:2018tnn,Biswas:2011ar,Olmo:2015axa,Menchon:2017qed,Cano:2018aod,delaCruz-Dombriz:2018aal}. However, the corrections can also significantly modify the properties of a black hole at the level of the horizon if its mass is small enough. For example, the divergence of Hawking temperature in the limit $M\rightarrow 0$ in Einstein gravity (EG) black holes can be cured by higher-derivative interactions \cite{Myers:1988ze,Cai:2001dz,PabloPablo4}. In this way, one learns about new high-energy phenomena that might be interpreted as the signature of a UV-complete theory of gravity.

Besides its intrinsic interest, there is another reason why studying higher-derivative-corrected black hole geometries is interesting: they can be used to obtain phenomenological implications of modified gravity. Thanks to the LIGO/VIRGO collaborations \cite{TheLIGOScientific:2014jea,TheVirgo:2014hva} and the Event Horizon Telescope \cite{Falcke:1999pj}, amongst other initiatives \cite{Audley:2017drz}, it will be possible in the next years to test GR with an unprecedented accuracy, and to set bounds on possible modifications of this theory \cite{Giddings:2014ova,Berti:2015itd,Johannsen:2015hib,Cardoso:2016rao,Yunes:2016jcc,Barack:2018yly,Berti:2018cxi}. But in order to do so, we first need to derive observational signatures of modified gravity. In order to measure deviations from GR on astrophysical black holes, the corrections should appear at a scale of the order of few kilometers, which is roughly the radius of the horizon for those BHs.  Although this seems to be an enormous scale for short-distance modifications of gravity, we should only discard it if there is some fundamental obstruction that forbids unnaturally large couplings in the effective theory \cite{Camanho:2014apa}. But if that is not the case, the possibility of observing higher-derivative corrections on astrophysical black holes should be considered \cite{Pani:2009wy}. Hence, studying in a systematic way black hole solutions of modified gravity and their observational implications is a mandatory task for the black hole community in the coming years.

Black hole solutions in alternative theories of gravity have been largely explored in the literature, but for obvious reasons we will restrict our attention to four-dimensional solutions that modify in a continuous way the Einstein gravity black holes, and that do not include matter. This excludes, for example, solutions of pure quadratic gravity, without a linear $R$ term \cite{Riegert:1984zz,Klemm:1998kf,Oliva:2010zd,Oliva:2011xu}. In the same way, theories such as $f(R)$ gravity are not interesting for us, since they do not modify EG solutions in the vacuum (see \textit{e.g.} \cite{delaCruzDombriz:2009et}). Some other theories allow for EG solutions, but additionally possess disconnected branches of different solutions, as is the case of black holes in quadratic gravity \cite{Lu:2015cqa,Lu:2015psa}. We will not consider this case here either, since we are interested in continuous deviations from GR. On the contrary, static black holes correcting Schwarzschild's solution have been studied in the context of Einstein-dilaton-Gauss-Bonnet gravity (EdGB) \cite{EdGB,Kanti:1995vq,Torii:1996yi,Alexeev:1996vs}, and in other scalar-Gauss-Bonnet theories, \textit{e.g.} \cite{Sotiriou:2014pfa,Doneva:2017bvd,Silva:2017uqg,Antoniou:2017acq}.  Those theories contain a scalar that is activated due to the higher-curvature terms. In the case of pure-metric theories, spherically symmetric black holes have been constructed, non-perturbatively in the coupling, in Einsteinian cubic gravity \cite{PabloPablo,Hennigar:2016gkm,PabloPablo2}. Although the profile of the solution has to be determined numerically, this theory has the remarkable property that black hole thermodynamics can be determined analytically. These results have recently been generalized to higher-order versions of the theory \cite{Hennigar:2017ego,Ahmed:2017jod,PabloPablo4}. 

The case of rotating black holes, which is more interesting from an astrophysical perspective, is also more challenging. Obtaining rotating black hole solutions of higher-derivative gravity theories is a very complicated task, and for that reason only approximate solutions or numerical ones are known.  One of the most studied theories in this context is EdGB gravity, where rotating black holes have been constructed perturbatively in the spin and in the coupling \cite{Pani:2009wy,Ayzenberg:2014aka,Maselli:2015tta}, and numerically \cite{Kleihaus:2011tg,Kleihaus:2015aje}. Rotating black holes in dynamical Chern-Simons (dCS) modified gravity\footnote{This theory does not modify spherically symmetric GR solutions, because the corrections are sourced by the Pontryagin density, that vanishes in the presence of spherical symmetry.  However, it does modify rotating black holes.} \cite{Alexander:2009tp} have also been studied, both perturbatively \cite{Konno:2007ze,Yunes:2009hc,Yagi:2012ya} and numerically \cite{Delsate:2018ome}. On the other hand, Ref.~\cite{Pani:2011gy} considers a generalization of EdGB and dCS theories. 
Finally, for pure-metric theories, the recent work \cite{Cardoso:2018ptl} studies rotating black holes in the eight-derivative effective theory introduced in \cite{Endlich:2017tqa}.

A usual approximation, that is used by many of the papers above, consists in obtaining the solution perturbatively in the higher-order couplings. For some purposes it is also interesting to obtain non-perturbative solutions --- for which one usually needs numerical methods--- but, from the perspective of EFT, it does not make any sense to go beyond perturbative level, since the theory will include further corrections at that order. Additionally, the solution is often expanded in a power series of the spin parameter $\chi=a/M$. In most of the literature, only few terms in this expansion are included, so the solutions are only valid for slowly-rotating black holes. However, astrophysical black holes --- and in particular those created after the merging of a black hole binary \cite{Fishbach:2017dwv} --- can have relatively high spin. Moreover, some effects of rotation --- such as the deformation of the black hole shadow \cite{Falcke:1999pj,Amarilla:2010zq,Cunha:2015yba,Younsi:2016azx} --- are barely observable when the spin is low, and other phenomena only happen for rapidly spinning black holes \cite{Hod:2012bw,Yang:2012pj,Hod:2013zza}. Although numerical solutions are not in principle limited to small values of the spin, analytic solutions are most useful for evident reasons. Hence, it would be interesting to provide analytic solutions valid for high-enough angular momentum.  Finally, instead of having a large catalogue of alternative theories of gravity and their black hole solutions, it would be desirable to describe a minimal model that captures all the possible modifications of GR at a given order --- probably, up to field redefinitions --- and to characterize the black holes of that theory. 

The preceding discussion motivates the three main objectives of the present work. First, to establish a general effective theory that can be used to study the leading-order higher-derivative corrections to Einstein gravity vacuum solutions. Second, to obtain the corrections to Kerr black hole in these theories, providing an analytic solution that is accurate for high enough values of the spin. And third, to study in detail some of the properties of these rotating black holes, such as the shape of horizon or the surface gravity, that have often been disregarded in the literature. 

The paper is organized as follows. In Section~\ref{section:effectivetheory} we describe the models with higher-derivative terms and non-minimally coupled scalars that we will use throughout the text, and we argue that they capture the most general corrections of this kind.
In Section~\ref{sec:corrKerr} we describe the ansatz used in order to find the rotating black holes in the previous theories and we solve the equations performing a series expansion in the spin. For the computations in the text we use a series up to order $\mathcal{O}(\chi^{14})$, that is accurate up to $\chi\sim0.7$. We provide as well a Mathematica notebook that computes the solution at any given order (it is contained as an ancillary file in the arXiv source). In Section~\ref{sec:prop} we study some properties of the modified Kerr black holes: horizon, ergosphere, photon rings and scalar hair, and we highlight the interesting geometry of black holes in parity-violating theories, that do not possess $\mathbb{Z}_{2}$ symmetry. We summarize our findings in Section~\ref{conclusions}, commenting on possible extensions and applications of the present work. Finally, there are several appendices with supplemental information.

\section{Leading order effective theory}\label{section:effectivetheory}
The most general diffeomorphism-invariant and locally Lorentz-invariant metric theory of gravity is given by an action of the form
\begin{equation}
S=\int d^4x\sqrt{|g|}\mathcal{L}\left(g^{\mu\nu},R_{\mu\nu\rho\sigma},\nabla_{\alpha}R_{\mu\nu\rho\sigma},\nabla_{\alpha}\nabla_{\beta}R_{\mu\nu\rho\sigma},\ldots\right)\, .
\end{equation}
This is, the most general Lagrangian for such theory will be an invariant formed from contractions and products of the metric, the Riemann tensor, and its derivatives. However, the theory above can be generalized by slightly relaxing some of the postulates. We may construct the Lagrangian using as well the dual Riemann tensor:
\begin{equation}
\tensor{\tilde R}{_{\mu\nu\alpha\beta}}=\frac{1}{2}\epsilon_{\mu\nu\rho\sigma}\tensor{R}{^{\rho\sigma}_{\alpha\beta}}\, .
\end{equation}
These terms generically lead to violation of parity, hence the theory is not (locally) invariant under the full Lorentz group, but only under one of its connected components. However, we know that parity is not a symmetry of nature, so in principle there is no reason to discard terms constructed with $\tilde R_{\mu\nu\alpha\beta}$.
In general, one expands this Lagrangian in terms containing increasing numbers of derivatives, being the first one the Einstein-Hilbert term $R$, with two derivatives. The rest of the terms can symbolically be written as 
\begin{equation}
\nabla^p\mathcal{R}^n\, .
\end{equation}
Since this term contains $2n+p$ derivatives, it should be multiplied by a constant of dimensions of \emph{length}$^{2n+p-2}$ with respect to the Einstein-Hilbert term. This is the length scale $\ell$ at which the higher-derivative terms modify the law of gravitation. When the curvature is much smaller than this length scale ($||R_{\mu\nu\rho\sigma}||<<\ell^{-2}$), the effect of the higher-derivative terms can be treated as a perturbative correction, and terms with increasing number of derivatives become more and more irrelevant. Thus, it is an interesting exercise to obtain the most general theory that includes all the possible leading-order corrections. Here we summarize how we construct this theory, but we refer to the Appendix~\ref{appendix:effectivetheory} for the details. The first terms one may introduce in the action are quadratic in the curvature and hence they contain four derivatives. These terms would induce corrections in the metric tensor at order $\ell^2$, but in four dimensions it turns out that all of these terms either are topological or do not introduce corrections at all. Thus, the first corrections in a metric theory appear at order $\ell^4$ and they are associated to six-derivative terms. As we show in Appendix \ref{appendix:effectivetheory}, it turns out that, up to field redefinitions, there are only two inequivalent six-derivative curvature invariants, one of them parity-even and the other one parity-odd.
However, one could consider a more general theory, allowing the coefficients of the higher-derivative terms to be dynamical \textit{i.e.}, controlled by scalars. This is actually a very natural possibility that is predicted, for instance, by String Theory \cite{EdGB}.  In that case, some of the four-derivative terms do contribute to the equations and they also correct the metric at order $\ell^4$. For simplicity, we will restrict ourselves to massless scalars, but we will allow, in principle, to have an undetermined number of them.  Within this large family of theories, it is possible to show that the most general leading correction to Einstein's theory is captured by the action
\begin{equation}\label{Action}
\begin{aligned}
S=&\frac{1}{16\pi G}\int d^4x\sqrt{|g|}\bigg\{R+\alpha_{1} \phi_{1} \ell^2\mathcal{X}_{4}+\alpha_{2}\left(\phi_2 \cos\theta_{m}+\phi_{1}\sin\theta_{m}\right) \ell^2 R_{\mu\nu\rho\sigma} {\tilde R}^{\mu\nu\rho\sigma}\\\
&+\lambda_{\rm ev}\ell^4\tensor{R}{_{\mu\nu }^{\rho\sigma}}\tensor{R}{_{\rho\sigma }^{\delta\gamma }}\tensor{R}{_{\delta\gamma }^{\mu\nu }}+\lambda_{\rm odd}\ell^4\tensor{R}{_{\mu\nu }^{\rho\sigma}}\tensor{R}{_{\rho\sigma }^{\delta\gamma }} \tensor{\tilde R}{_{\delta\gamma }^{\mu\nu }}-\frac{1}{2}(\partial\phi_{1})^2-\frac{1}{2}(\partial\phi_{2})^2\bigg\}\, ,
\end{aligned}
\end{equation}
where 
\begin{equation}
\mathcal{X}_{4}=R_{\mu\nu\rho\sigma} R^{\mu\nu\rho\sigma}-4R_{\mu\nu}R^{\mu\nu}+R^2
\end{equation}
is the Gauss-Bonnet density and $\phi_{1}$, $\phi_{2}$ are scalar fields.
Besides the overall length scale $\ell$, there are only five parameters: $\alpha_1$, $\alpha_2$, $\lambda_{\rm ev}$, $\lambda_{\rm odd}$ and $\theta_m$. The parameter $\lambda_{\rm odd}$ violates parity, while the ``mixing angle'' $\theta_m$ represents as well a sort of parity breaking phase. For $\theta_m=0,\pi$ (no mixing between scalars), $\phi_2$ is actually a pseudoscalar and the quadratic sector is parity-invariant. For any other value ($\theta_m\neq 0,\pi$), parity is also violated by this sector. 

The theory (\ref{Action}) contains, as particular cases, some well-known models that have been frequently used in the literature. 
The case $\lambda_{\rm ev}=\lambda_{\rm odd}=\theta_m=0$, $\alpha_{2}=-\alpha_{1}=1/8$ corresponds to the prediction of String Theory, where the length scale of the corrections in that case is the string length $\ell^2=\ell_{s}^2\equiv\alpha'$. As we show in the appendix \ref{ap:het}, the corresponding action can be obtained from direct compactification and truncation of the Heterotic superstring effective action at order $\alpha'$. In that case, $\phi_1$ is identified with the dilaton, while $\phi_2$ is the axion, which appears after dualization of the Kalb-Ramond 2-form. 
Another well-known possibility (which is also claimed to proceed from the low-energy limit of String Theory) is $\lambda_{\rm odd}=\lambda_{\rm ev}=\alpha_2=0$, which corresponds to the Einstein-dilaton-Gauss-Bonnet theory.  Rotating black holes in EdGB gravity have been studied both numerically \cite{Kleihaus:2011tg,Kleihaus:2015aje} and in the slowly-rotating limit \cite{Pani:2009wy,Ayzenberg:2014aka,Maselli:2015tta}. The case $\theta_{m}=\pi/2$, which represents an extension of EdGB gravity, has also been considered \cite{Pani:2011gy} (note that this case only contains one dynamical scalar and violates parity). On the other hand, the case $\alpha_{2}\neq 0$ with the rest of couplings set to zero corresponds to dynamical Chern-Simons gravity, whose rotating black holes were studied in Refs.~\cite{Konno:2007ze,Yunes:2009hc,Yagi:2012ya} in the slowly-rotating approximation, while Ref.~\cite{Delsate:2018ome} performs a non-perturbative numerical study. As for the cubic theories, the parity-even term (controlled by $\lambda_{\rm ev}$) can be mapped (modulo field redefinitions) to the Einsteinian cubic gravity (ECG) term \cite{PabloPablo}, for which static black hole solutions have been constructed non-perturbatively in the coupling \cite{Hennigar:2016gkm,PabloPablo2}. 
Phenomenological signatures of static black holes in ECG have also been recently studied in \cite{Hennigar:2018hza,Poshteh:2018wqy}, where a first bound on the coupling was provided, and the possibility to detect deviations from GR in gravitational lensing observations was discussed. Rotating black holes in ECG have not been studied so far. 
Lastly, to the best of our knowledge, the parity odd cubic term has never been used in the context of black hole solutions.

The theory (\ref{Action}) has been constructed following the sole requirement of diff. invariance, but there are some other constraints that could be imposed on physical grounds. For instance, if one wants to preserve parity, then one should set $\theta_{m}=\lambda_{\rm odd}=0$. Nevertheless, we know that nature is not parity-invariant, so keeping these terms is not unreasonable. If one does not wish to include additional light degrees of freedom the scalars should be removed, which amounts to setting $\alpha_{1}=\alpha_{2}=0$ (in that case the scalars are just not activated). On the other hand it is known that higher-derivative terms may break unitarity by introducing ghost modes --- non normalizable states. In the case, for instance, of the Gauss-Bonnet term in (\ref{Action}), this problem does not exist since it produces second-order equations.. The equations of the cubic terms do contain higher-order derivatives ---namely of fourth order---, but the mass scale at which we expect the new modes to appear is

\begin{equation}
m^2\sim \frac{1}{\ell^4 ||R_{\mu\nu\rho\sigma}||}\, .
\end{equation}
This is simply telling us that Effective Field Theory works up to the scale $||R_{\mu\nu\rho\sigma}||\sim \ell^{-2}$, which is something we already knew. Finally, it is also possible to study causality constraints \cite{Gruzinov:2006ie}. In relation to this, the results in \cite{Camanho:2014apa} impose a severe bound on the coupling constants $\lambda_{\rm ev}\ell^4$, $\lambda_{\rm odd}\ell^4$ of the cubic terms. If one wants to observe any effects of higher-derivative corrections on astrophysical black holes, necessarily the corrections should appear at a scale $\ell$ of the order of few kilometers (otherwise the effect would be too small to be detected). Such large couplings are very unnatural, since the natural scale of (quantum) gravity should be Planck length. According to \cite{Camanho:2014apa}, these large couplings lead to violation of causality, that could only be restored by adding an infinite tower of higher-spin particles of mass $\sim\ell^{-1}$. Since, obviously, this is not observed, it was concluded that the couplings associated to the cubic terms should be of the order of Planck scale, hence those corrections would not be viable phenomenologically.\footnote{Let us note that, according to \cite{Endlich:2017tqa}, the conclusion might be different if one considers different UV completions of the effective theory from the one assumed in \cite{Camanho:2014apa}.}

In any case, nothing prevents us from studying the effect of the cubic curvature terms on black holes, no matter the scale at which they appear. These corrections give us valuable information about the effects of modified gravity at high energies, and this is intrinsically interesting, even if those corrections are not viable on an observational basis.

If, for some reason, all the theories in the model (\ref{Action}) were discarded, then the leading correction to GR would be given by the quartic-curvature terms introduced in \cite{Endlich:2017tqa}. These terms modify the metric at order $\mathcal{O}(\ell^6)$ hence they are subleading when the couplings in (\ref{Action}) are non-vanishing. Rotating black holes in those theories were recently studied in \cite{Cardoso:2018ptl} up to order $\chi^4$ in the spin. The methods that we present in this work could be applied to the quartic theories as well and could be used in order to extend some of the results in  \cite{Cardoso:2018ptl}. For instance, one might compute the solution for higher values of the spin or obtain the form of the horizon, as we do in Sec.~\ref{sec:horizon}.

\subsection{Equations of motion}

Our goal is to compute the leading corrections to vacuum solutions of Einstein's theory. Thus, our starting point is a metric $g^{(0)}_{\mu\nu}$ that satisfies vacuum Einstein's equations
\begin{equation}
R^{(0)}_{\mu\nu}=0\ ,
\end{equation}
while the scalars $\phi_1^{(0)}$, $\phi_2^{(0)}$ take a constant value that can be chosen to be zero without loss of generality.\footnote{The action \ref{Action} is invariant (up to a surface term) under constant shifts of the scalars.}
But this field configuration is not a solution when we take into account the higher-derivative terms. First we note that the coupling between scalars and the curvature densities in the action (\ref{Action}) induce source terms in the scalar equations of motion, so that they will not be constant anymore. More precisely the first correction is of order $\ell^2$,
\begin{equation}
\phi_1=\ell^2\phi_1^{(2)}\, ,\quad \phi_2=\ell^2\phi_2^{(2)}\, ,
\end{equation}
and it satisfies 

\begin{eqnarray}
\label{scalarEOM1}
 \nabla^2 \phi_1^{(2)}&=&-\alpha_1 R_{\mu\nu\rho\sigma}R^{\mu\nu\rho\sigma}\Big|_{g=g^{(0)}} -\alpha_2 \sin\theta_m R_{\mu\nu\rho\sigma}\tilde R^{\mu\nu\rho\sigma}\Big|_{g=g^{(0)}}\ , \\
\label{scalarEOM2}
\nabla^2 \phi_2^{(2)}&=&-\alpha_2   \cos \theta_m R_{\mu\nu\rho\sigma}\tilde R^{\mu\nu\rho\sigma}\Big|_{g=g^{(0)}} \ .
\end{eqnarray}
On the other hand, the modified Einstein equations, derived from the action (\ref{Action}), can be written as

\begin{equation}
G_{\mu\nu}=T^{\text{scalars}}_{\mu\nu}+T^{\text{cubic}}_{\mu\nu}\ ,
\end{equation}
where we have passed all the corrections to the right-hand-side in the form of some energy-momentum tensors, that read

\begin{equation}
\begin{aligned}
T^{\text{scalars}}_{\mu\nu}=&-\alpha_1 \ell^2g_{\nu\lambda}\delta^{\lambda \sigma \alpha\beta}_{\mu\rho\gamma\delta}  R^{\gamma\delta}{}_{\alpha\beta} \nabla^\rho\nabla_\sigma \phi_1+4\alpha_2 \ell^2  \nabla^\rho\nabla^\sigma\left[\tilde R_{\rho(\mu\nu)\sigma}\,\left(\cos \theta_m \phi_2+\sin \theta_m \phi_1\right)\right] \\
&+\frac{1}{2}\left[\partial_\mu \phi_1 \partial_\nu \phi_1-\frac{1}{2}g_{\mu\nu} \left(\partial \phi_1\right)^2\right]+\frac{1}{2}\left[\partial_\mu \phi_2 \partial_\nu \phi_2-\frac{1}{2}g_{\mu\nu} \left(\partial \phi_2\right)^2\right] \ ,
\end{aligned}
\end{equation}
and 

\begin{equation}
\begin{aligned}
T^{\text{cubic}}_{\mu\nu}=&\lambda_{\text{ev}}\ell^4\left[3 R_\mu{}^{\sigma \alpha \beta} R_{\alpha\beta}{}^{\rho\lambda} R_{\rho\lambda \sigma \nu}+\frac{1}{2}g_{\mu\nu}\tensor{R}{_{\alpha\beta }^{\rho\sigma}}\tensor{R}{_{\rho\sigma }^{\delta\gamma }}\tensor{R}{_{\delta\gamma }^{\alpha\beta }}-6 \nabla^\alpha\nabla^\beta\left(R_{\mu\alpha\rho \lambda}R_{\nu\beta}{}^{\rho\lambda}\right)\right]\\
&+\lambda_{\text{odd}}\ell^4\bigg[-\frac{3}{2}\tensor{R}{_{\mu}^{\rho\alpha\beta}}\tensor{R}{_{\alpha\beta\sigma\lambda}}\tensor{\tilde R}{_{\nu\rho}^{\sigma\lambda}}-\frac{3}{2}\tensor{R}{_{\mu}^{\rho\alpha\beta}}\tensor{R}{_{\nu\rho\sigma\lambda}}\tensor{\tilde R}{_{\alpha\beta}^{\sigma\lambda}}+\frac{1}{2}g_{\mu\nu}\tensor{R}{_{\mu\nu }^{\rho\sigma}}\tensor{R}{_{\rho\sigma }^{\delta\gamma }} \tensor{\tilde R}{_{\delta\gamma }^{\mu\nu }}\\
&+3\nabla^{\alpha}\nabla^{\beta}\left(\tensor{R}{_{\mu\alpha\sigma\lambda}}\tensor{\tilde R}{_{\nu\beta}^{\sigma\lambda}}+\tensor{R}{_{\nu\beta\sigma\lambda}}\tensor{\tilde R}{_{\mu\alpha}^{\sigma\lambda}}\right)\bigg] \,
\end{aligned}
\end{equation}

\noindent
Since the scalars are of order $\mathcal{O}(\ell^2)$, we can see that the leading correction to the metric associated to the scalar sector is of order $\mathcal{O}(\ell^4)$, the same order at which cubic curvature terms come into play. Thus, we expand the metric as
\begin{equation}\label{mexp}
g_{\mu\nu}=g^{(0)}_{\mu\nu}+ \ell^4 g^{(4)}_{\mu\nu}\ ,
\end{equation}
where $g^{(4)}_{\mu\nu}$ is a perturbative correction. Now, taking into account that $g_{\mu\nu}^{(0)}$ solves Einstein's equations, we get the value of the Einstein tensor to linear order in $g_{\mu\nu}^{(4)}$:

\begin{equation}\label{eq:linearizedEE}
G_{\mu\nu}=\ell^4\left[-\frac{1}{2}\nabla^2\hat{g}^{(4)}_{\mu\nu}-\frac{1}{2}g^{(0)}_{\mu\nu}\nabla^\alpha\nabla^\beta \hat{g}^{(4)}_{\alpha\beta}+\nabla^\alpha\nabla_{(\mu}\hat{g}^{(4)}_{\nu)\alpha}\right]+\mathcal{O}(\ell^6)\, .
\end{equation}
where $\nabla$ is the covariant derivative associated with the zeroth order metric, and $\hat{g}^{(4)}_{\mu\nu}$ is the trace-reversed metric perturbation 
\begin{equation}
\hat{g}^{(4)}_{\mu\nu}=g^{(4)}_{\mu\nu}-\frac{1}{2}g^{(0)}_{\mu\nu} g^{(4)}_{\alpha\beta}g^{(0)\,\alpha\beta}\, .
\end{equation}

\noindent
Then, $\hat{g}_{\mu\nu}^{(4)}$ satisfies the equation
\begin{equation}\label{g4eqn}
-\frac{1}{2}\nabla^2\hat{g}^{(4)}_{\mu\nu}-\frac{1}{2}g^{(0)}_{\mu\nu}\nabla^\alpha\nabla^\beta \hat{g}^{(4)}_{\alpha\beta}+\nabla^\alpha\nabla_{(\mu}\hat{g}^{(4)}_{\nu)\alpha}=\ell^{-4}\left[T^{\text{scalars}}_{\mu\nu}+T^{\text{cubic}}_{\mu\nu}\right]\Big|_{g=g^{(0)}\, ,\, \phi_{i}=\ell^2\phi_{i}^{(2)}}
\end{equation}

\section{The corrected Kerr metric}\label{sec:corrKerr}
After introducing the theory (\ref{Action}), here we present the rotating black hole ansatz that we will use in the rest of the text, and in Sec.~\ref{sec:solving} we sketch how to solve the equations of motion. From now on we set $G=1$. Let us first consider Kerr's metric expressed in Boyer-Lindquist coordinates:
\begin{equation}
\begin{aligned}
ds^2=&-\left(1-\frac{2 M r}{\Sigma}\right)dt^2-\frac{4 Mar\sin^2\theta}{\Sigma}dtd\phi+\Sigma\left(\frac{dr^2}{\Delta}+d\theta^2\right)
\\
&+\left(r^2+a^2+\frac{2 M r a^2\sin^2\theta}{\Sigma}\right)\sin^2\theta d\phi^2\, ,
\end{aligned}
\end{equation}
where 

\begin{equation}
\Sigma=r^2+a^2\cos^2\theta\, ,\quad \Delta=r^2-2Mr+a^2\, .
\end{equation}
Let us very briefly recall some of the properties of this metric. 
\begin{itemize}
\item Being a solution of vacuum Einstein's equations, it is Ricci flat: $R_{\mu\nu}=0$.
\item It is stationary and axisymmetric, with related Killing vectors $\partial_t$ and $\partial_{\phi}$ respectively.
\item It represents an asymptotically flat spacetime with total mass $M$ and total angular momentum $J=a M$.
\item When $M>|a|$ the solution represents a black hole, whose (outer) horizon is placed at the largest radius $r_{+}$ where $\Delta$ vanishes:
\begin{equation}
r_{+}=M+\sqrt{M^2-a^2}\, .
\end{equation}
\end{itemize}

Since Ricci flat metrics do not solve the modified Einstein's equations, the rotating black holes of the theory (\ref{Action}) will not be described by Kerr metric. The search for an appropriate metric ansatz that can be used to parametrize deviations from Kerr metric is a far from trivial problem that has been studied in the literature \cite{Cardoso:2014rha,Konoplya:2016jvv}. However, as long as the mass is much larger than the scale at which the higher-derivative terms appear, $M>>\ell$, the deviation with respect General Relativity will be small --- at least outside the horizon. In that case, we can build the rotating black hole solution of (\ref{Action}) as a perturbative correction over Kerr metric. Since we want to describe an stationary and axisymmetric solution, the corrected metric has to conserve the Killing vectors $\partial_{t}$ and $\partial_{\phi}$. On the other hand, we do not expect to ``activate'' additional components of the metric, so that the corrections appear in the already non-vanishing components. Taking into account these observations, we can write a general ansatz for the corrected Kerr metric

\begin{equation}\label{rotatingmetric0}
\begin{aligned}
ds^2=&-\left(1-\frac{2 M \rho}{\Sigma}-H_1\right)dt^2-\left(1+H_2\right)\frac{4 M a \rho (1-x^2)}{\Sigma}dtd\phi+\left(1+H_3\right)\frac{\Sigma}{\Delta}d\rho^2\\
&+\left(1+ H_5\right)\frac{\Sigma dx^2}{1-x^2}+\left(1+H_4\right)\left(\rho^2+a^2+\frac{2 M  \rho a^2(1-x^2)}{\Sigma}\right)(1-x^2)d\phi^2\, ,
\end{aligned}
\end{equation}

\noindent
where $H_{1,2,3,4,5}$ are functions of $x=\cos\theta$ and $\rho$ only, and they are assumed to be small $|H_{i}|<<1$. Note that we have introduced the coordinate $\rho$ in order to distinguish it from the coordinate $r$ in Kerr metric. We have also introduced the functions
\begin{equation}
\Sigma=\rho^2+a^2x^2\, ,\quad \Delta=\rho^2-2M\rho+a^2\, .
\end{equation} 

However, the ansatz (\ref{rotatingmetric0}) is far too general, and it turns out that we can fix some of the functions $H_i$ by performing a change of coordinates. In particular, it can be shown that there exists a (infinitesimal) change of coordinates $(\rho,x)\rightarrow (\rho', x')$ that preserves the form of the metric and for which $H_{5}'=H_{3}'$. Thus, we are free to choose $H_{3}=H_{5}$, and in that case, the metric reads

\begin{equation}\label{rotatingmetric}
\begin{aligned}
ds^2=&-\left(1-\frac{2 M \rho}{\Sigma}-H_1\right)dt^2-\left(1+H_2\right)\frac{4 M a \rho (1-x^2)}{\Sigma}dtd\phi+\left(1+H_3\right)\Sigma\left(\frac{d\rho^2}{\Delta}+\frac{dx^2}{1-x^2}\right)\\
&+\left(1+H_4\right)\left(\rho^2+a^2+\frac{2 M  \rho a^2(1-x^2)}{\Sigma}\right)(1-x^2)d\phi^2\, .
\end{aligned}
\end{equation}

Note that we are choosing the coordinates $x$ and $\rho$ such that the form of the $(\rho,x)$-metric is respected --- up to a conformal factor --- when the corrections are included. It is easy to see that this choice of coordinates has a crucial advantage: the horizon of the metric (\ref{rotatingmetric}) will still be placed at the (first) point where $\Delta$ vanishes: $\rho_{+}=M+\sqrt{M^2-a^2}$. If we were not careful enough choosing the coordinates, the description of the horizon could be very messy, and this is perhaps the reason why in previous studies the horizon of the corrected solutions has not been studied in depth.
 
We note that, whenever we consider the corrections, the coordinate $\rho$ does not coincide asymptotically with the usual radial coordinate $r$. Advancing the results in next subsection, we get that the functions $H_i$ behave asymptotically  as

\begin{equation}
H_i=h_i^{(0)}+\frac{h_i^{(1)}}{\rho}+\mathcal{O}\left(\frac{1}{\rho^2}\right)\, , \,\, i=1,2,3,4\, ,
\end{equation}
where $h_i^{(k)}$ are constant coefficients. Then, we can see that the usual radial coordinate $r$ that asymptotically measures the area of 2-spheres is related to $\rho$ 
according to

\begin{equation}
\rho=r\left(1-\frac{h_3^{(0)}}{2}\right)-\frac{h_3^{(1)}}{2}+\mathcal{O}\left(\frac{1}{r}\right)\, .
\end{equation}
Using this coordinate, the asymptotic expansion of the metric (\ref{rotatingmetric}) reads

\begin{equation}
\begin{aligned}
ds^2(r\rightarrow\infty)=&-\left(1-h_1^{(0)}-\frac{2 M+Mh_3^{(0)}+h_1^{(1)} }{r}\right)dt^2-\left(1+h_2^{(0)}+h_3^{(0)}/2\right)\frac{4 M a \sin^2\theta}{r}dtd\phi\\
&+dr^2\left(1+\frac{2M+M h_3^{(0)}+h_3^{(1)}}{r}\right)+r^2d\theta^2
+\left(1+h_4^{(0)}-h_3^{(0)}\right)r^2\sin^2\theta d\phi^2\, .
\end{aligned}
\end{equation}

When we solve the equations, we see that we are free to fix the asymptotic values of the coefficients $h_i^{(0)}$. On the other hand, the metric must be asymptotically flat (with the correct normalization at infinity), and we want the parameters $M$ and $a$ to still represent the mass and the angular momentum per mass of the solution, so the asymptotic expansion should read

\begin{equation}
\begin{aligned}
ds^2(r\rightarrow\infty)=&-\left(1-\frac{2 M}{r}\right)dt^2-\frac{4 M a \sin^2\theta}{r}dtd\phi+dr^2\left(1+\frac{2M}{r}\right)+r^2d\theta^2
+r^2\sin^2\theta d\phi^2\, .
\end{aligned}
\end{equation}
From this, we derive the asymptotic conditions that we have to impose on our solution:

\begin{equation}\label{bcond}
h_1^{(0)}=0\, ,\quad h_3^{(0)}=h_4^{(0)}=-\frac{h_3^{(1)}}{M}\, ,\quad h_2^{(0)}=-\frac{h_3^{(0)}}{2}\, .
\end{equation}
Apparently, the condition $Mh_3^{(0)}+h_1^{(1)}=0$ is also required, but this is actually imposed by the field equations.

\subsection{Solving the equations}\label{sec:solving}
Once we have found an appropriate ansatz for our metric, Eq.~(\ref{rotatingmetric}), we have to solve the equations of the theory (\ref{Action}). The first step is to solve the equations for the scalars (\ref{scalarEOM1}, \ref{scalarEOM2}), from where we obtain $\phi_1$ and $\phi_2$ at order $\mathcal{O}(\ell^2)$. Using this result we determine the right-hand-side of (\ref{g4eqn}), while in the left-hand-side we introduce the metric correction $g^{(4)}_{\mu\nu}$, 

\begin{equation}
\begin{aligned}
\ell^4g^{(4)}_{\mu\nu}dx^{\mu}dx^{\nu}=&H_1dt^2-H_2\frac{4 Ma\rho(1-x^2)}{\Sigma}dtd\phi+H_3\Sigma\left(\frac{d\rho^2}{\Delta}+\frac{dx^2}{1-x^2}\right)\\
&+H_4\left(\rho^2+a^2+\frac{2 M \rho a^2(1-x^2)}{\Sigma}\right)(1-x^2)d\phi^2\, ,
\end{aligned}
\end{equation}
which can be read from (\ref{rotatingmetric}). In this way, we get a (complicated) system of equations for the functions $H_i$, that we have to solve. Unfortunately, these equations (including the ones for the scalars) are very intricate and we are not able to obtain an exact solution. However, a possible strategy is to expand the solution in powers of the angular momentum $a$, assuming that it is a small parameter. In previous works \cite{Pani:2009wy,Ayzenberg:2014aka,Maselli:2015tta,Pani:2011gy,Konno:2007ze,Yunes:2009hc}, this method has been employed in order to obtain a few terms in the expansion, which yields an approximate solution for slowly rotating black holes. Nonetheless, if one includes enough terms in the expansion, the result should give a good approximation to the solution also for high values of the spin. One of the goals of this paper is precisely to provide a method that allows for the construction of the solution at arbitrarily high-orders in the spin.

For simplicity, let us first introduce the dimensionless parameter 
\begin{equation}
\chi=\frac{a}{M}\, ,
\end{equation}
that ranges from $0$ to $1$ in Kerr's solution, $\chi=0$ corresponding to static black holes and $\chi=1$ to extremal ones.\footnote{When the corrections are included, we expect that the extremality condition is modified, $\chi_{\rm ext}\neq 1$, but this is not important for our discussion, since we will not deal with extremal or near-extremal geometries here.}  Then, we expand our unknown functions in a power series in $\chi$
\begin{equation}\label{chiexp}
\phi_{1}=\sum_{n=0}^{\infty}\phi_{1}^{(n)}\chi^n\, ,\quad \phi_{2}=\sum_{n=0}^{\infty}\phi_{2}^{(n)}\chi^n\, ,\quad H_{i}=\sum_{n=0}^{\infty}H_{i}^{(n)}\chi^n\, ,\quad i=1,2,3,4\, ,
\end{equation}
where we recall that all the functions depend on $\rho$ and $x$. Then, the idea is to plug these expansions in (\ref{scalarEOM1}, \ref{scalarEOM2}, \ref{g4eqn}), expand the equations in powers of $\chi$, and solve them order by order. The equations satisfied by the $n$-th components are much simpler than the full equations, and we are indeed able to solve them analytically. These are second-order, linear, inhomogeneous, partial differential equations, so that the general solution can be expressed as the sum of a particular solution plus all the solutions of the homogeneous equation. In general, the ``homogeneous part'' of the solution represents infinitesimal changes of coordinates, and the physics is contained in the inhomogeneous part, which is the one sourced by the higher-derivative terms. So, we have to find the solution that captures the corrections but does not introduce unnecessary changes of coordinates.
We observe that the appropriate solution can always be expressed as a polynomial in $x$ and in $1/\rho$. More precisely, we get\footnote{Equivalently, one may expand these functions using Legendre polynomials $P_{p}(x)$.}
\begin{equation}
\phi_{1}^{(n)}=\sum_{p=0}^{n}\sum_{k=0}^{k_{\rm max}}\phi_{1}^{(n,p,k)}x^p\rho^{-k}\, ,\quad \phi_{2}^{(n)}=\sum_{p=0}^{n}\sum_{k=0}^{k_{\rm max}}\phi_{2}^{(n,p,k)}x^p\rho^{-k}\, ,\quad H_{i}^{(n)}=\sum_{p=0}^{n}\sum_{k=0}^{k_{\rm max}}H_{i}^{(n,p,k)}x^p\rho^{-k}\, ,
\end{equation}
where $\phi_{1,2}^{(n,p,k)}$, $H_{i}^{(n,p,k)}$ are constant coefficients and in each case the value of $k_{\rm max}$ depends on $n$ and $p$. When we solve the equations we also observe that all the terms in these series are determined except the constant ones: those with $p=k=0$. However, those coefficients are fixed by the boundary conditions. In the case of the scalars, their value at infinity is arbitrary, so we can set it to zero for simplicity (this does not affect the rest of the solution)
\begin{equation}
\hskip2cm \phi_{1}^{(n,0,0)}=\phi_{2}^{(n,0,0)}=0\, ,\quad n=0,1,2,\ldots\, .
\end{equation}
On the other hand, for the $H_i$ functions we take into account the relations (\ref{bcond}) that we derived previously, which imply that

\begin{equation}\label{bcond2}
H_1^{(n,0,0)}=0\, ,\quad H_3^{(n,0,0)}=H_4^{(n,0,0)}=-\frac{H_3^{(n, 0,1)}}{M}\, ,\quad H_2^{(n,0,0)}=-\frac{H_3^{(n,0,0)}}{2}\, .
\end{equation}
In this way, the solution is completely determined. Since this process is systematic, we can easily program an algorithm that computes the series 
(\ref{chiexp}) at any (finite) order $n$. We provide with the arXiv submission of this paper an ancillary Mathematica notebook that does the job. Using this code, we have computed the solution up to order $\chi^{14}$. As we show in Appendix~\ref{ap:conv}, this expansion provides a minimum accuracy of about $1\%$ everywhere outside the horizon for $\chi=0.7$, and much higher for smaller $\chi$. Thus, we have an analytic solution that works for relatively high values of $\chi$, and we will exploit this fact in next section. Due to the length of the expressions, in Appendix~\ref{ap:sol} we show the solution explicitly up to order $\chi^3$, but the full series up to order $n=14$ is available in the Mathematica notebook. 

Before closing this section, we would like to clarify the following point. In the preceding scheme the corrections are expressed as a powers series in the spin, but we are taking the zeroth-order solution to be the exact Kerr's metric, which is non-perturbative in the spin. Thus, for consistency sake, one should imagine that we also expand the zeroth-order solution in the spin up to the same order at which the corrections were computed. However, for evident reasons we do not do this explicitly. Thus, in the next section, we will write the formulas for several quantities as the result for Kerr's metric, exact in the spin, plus linear corrections, perturbative in the spin, but one should bear in mind that the zeroth-order result should also be expanded.


\section{Properties of the corrected black hole}\label{sec:prop}


In this section we analyze some of the most relevant physical properties of the rotating black hole solutions we have found. We study the geometry of the horizon and of the ergosphere, light rings on the equatorial plane, and scalar hair.

\subsection{Horizon}\label{sec:horizon}

In order for the metric (\ref{rotatingmetric}) to represent a black hole, we have to show that it contains an event horizon. We have argued that, with the choice of coordinates we have made, the horizon is defined by the equation $\Delta=0$, whose roots are $\rho=\rho_{\pm}$, where
\begin{equation}
\rho_{\pm}=M\left(1\pm\sqrt{1-\chi^2}\right)\, .
\end{equation}
The largest root $\rho_{+}$ corresponds to the event horizon, while $\rho=\rho_{-}$ is in principle an inner horizon.\footnote{When the corrections are included, most likely the inner horizon of Kerr's black hole becomes singular. For instance, one expects that the scalars diverge there.} In this work we will only deal with the exterior solution $\rho\ge\rho_{+}$. 
 
Then, let us show that $\rho=\rho_{+}$ is indeed an event horizon. More precisely, we will show that it is a Killing horizon, \textit{i.e.} a null hypersurface where the norm of a Killing vector vanishes. 
Let us first check that the hypersurface defined by $\rho=\rho_{+}$ is null. In order to do so, we consider the induced metric at some constant $\rho$, which is given by 

\begin{equation}
\begin{aligned}
ds^2|_{\rho=\text{const}}&=-\left(1-\frac{2 M \rho}{\Sigma}-H_1\right)dt^2-\left(1+H_2\right)\frac{4 M a \rho (1-x^2)}{\Sigma}dtd\phi+\left(1+H_3\right)\frac{\Sigma dx^2}{1-x^2}\\
&+\left(1+H_4\right)\left(\rho^2+a^2+\frac{2 M  \rho a^2(1-x^2)}{\Sigma}\right)(1-x^2)d\phi^2\, .
\end{aligned}
\end{equation}

Then, we can see that when we evaluate at $\rho=\rho_{+}$, the previous metric is singular, namely it has rank 2. 
Evaluating the determinant of the $(t,\phi)$-metric at $\rho_{+}$, we get 

\begin{equation}\label{eq:vanishingdet0}
\left(g_{tt}g_{\phi\phi}-g^2_{t\phi}\right)\Big|_{\rho=\rho_+}=\frac{4M^2 \rho_+^2(1-x^2)}{\rho_+^2+a^2x^2}\left[H_1-\frac{a^2\left(1-x^2\right)}{\rho_+^2+a^2x^2}\left(2H_2-H_4\right)\right]\Bigg|_{\rho=\rho_+}\, ,
\end{equation}

\noindent
where, for consistency with the perturbative approach, we have expanded linearly in the $H_{i}$ functions.  When we expand the combination between brackets in powers of $\chi$ using the solution we found, we see that all the terms vanish. Thus, the determinant vanishes, 

\begin{equation}\label{eq:vanishingdet1}
\left(g_{tt}g_{\phi\phi}-g^2_{t\phi}\right)\Big|_{\rho=\rho_+}=0\, ,
\end{equation}

\noindent
which proves that this hypersurface is null. The next step is to show that there exists a Killing vector whose norm vanishes at $\rho=\rho_+$. Such vector is a linear combination of the two Killing vectors $\partial_t$ and $\partial_\phi$:

\begin{equation}\label{eq:killingvector}
\xi=\partial_t+\Omega_H \partial_\phi\ ,
\end{equation}
for some constant $\Omega_H$. One can check that the only possible choice of $\Omega_H$ for which $\xi$ is null at $\rho_{+}$ is 

\begin{equation}\label{eq:Omega0}
 \Omega_H=\frac{|g_{t\phi}|}{g_{\phi\phi}}\bigg|_{\rho=\rho_+}=\frac{a}{2M\rho_{+}}\left(1+H_2-H_4
\right)\big|_{\rho=\rho_+}\ ,
\end{equation}
which represents the angular velocity at the horizon. It is then clear that the norm of the vector $\xi$ vanishes at $\rho=\rho_+$, since 

\begin{equation}
\xi^2\Big|_{\rho=\rho_+}=\left(g_{tt}+2 g_{t\phi}\Omega_H +\Omega_H^2 g_{\phi\phi}\right)\Big|_{\rho=\rho_+}=\left(g_{tt}- \frac{g_{t\phi}^2}{g_{\phi\phi}}\right)\Bigg|_{\rho=\rho_+}=0\ ,
\end{equation}
where in the last step we have used (\ref{eq:vanishingdet1}). However, the crucial point here is whether $\Omega_{H}$, given by (\ref{eq:Omega0}), is constant. A priori, this quantity could well depend on $x$, in whose case $\xi$ would not be a Killing vector, and therefore $\rho=\rho_{+}$ would not be a Killing horizon. Nevertheless, expanding this quantity in powers of $\chi$ we do find that it is constant (see (\ref{eq:Omega1}) below), a fact that provides a very strong check on the validity of our results. Thus, we have shown that $\rho=\rho_{+}$ is a Killing horizon, and hence it should correspond to the event horizon of the black hole. 

We can now evaluate the angular velocity in order to study deviations with respect to Kerr's solution. A useful way to express it is the following,

 \begin{equation}\label{eq:Omega1}
\begin{aligned}
 \Omega_H=\frac{\chi}{2 M \left(1+\sqrt{1-\chi^2}\right)}+\frac{\ell^4}{M^5} \left[\alpha_1^2\,\Delta \Omega_H^{(1)}+\alpha_2^2\,\Delta \Omega_H^{(2)}+\lambda_{\text{ev}}\,\Delta \Omega_H^{(\text{ev})}\right]\ ,
\end{aligned}
 \end{equation}
where the first term is the value in Kerr black hole and we made explicit the linear corrections related to the different terms in the action. It turns out that the parity breaking terms do not contribute to this quantity --- nor to many others, as we will see. The dimensionless coefficients $\Delta \Omega_H^{(i)}$ depend on the spin, and the first terms in the $\chi$-expansion read\footnote{The first term in each of the formulas (\ref{eq:OmegaGB}) and (\ref{eq:OmegaCS}) reproduces previous results in the cases of EdGB gravity \cite{Ayzenberg:2014aka} and dCS gravity \cite{Yagi:2012ya}, respectively. The horizon area we obtain (see Eqs.~(\ref{eq:areaGB}) and (\ref{eq:areaCS}) below) also agrees with the results in those works, that computed the area at quadratic order in the spin.}

\begin{figure}[h!]
\begin{center}
\includegraphics[scale=0.6]{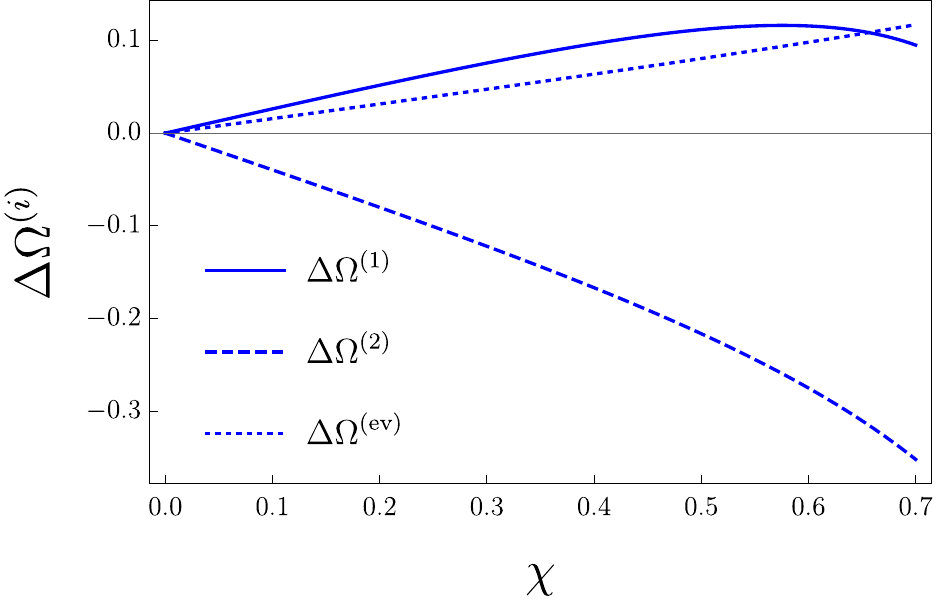}
\caption{Correction to the angular velocity of the black hole associated to every interaction.}
 \label{fig:Omega}
\end{center}
\end{figure}

\begin{eqnarray}
\label{eq:OmegaGB}
\Delta \Omega_H^{(1)}&=&\frac{21\chi}{80} -\frac{21103}{201600}\chi^3-\frac{1356809}{8870400}\chi^5-\frac{78288521}{461260800}\chi^7-\frac{25394183}{143503360}\chi^9+\mathcal O(\chi^{11})\ ,\hskip1cm \\
\label{eq:OmegaCS}
\Delta \Omega_H^{(2)}&=&-\frac{709\chi}{1792} -\frac{169}{1536}\chi^3-\frac{254929}{2365440}\chi^5-\frac{613099}{5271552}\chi^7-\frac{1684631453}{13776322560}\chi^9+\mathcal O(\chi^{11})\ ,\\
\Delta \Omega_H^{(\text{ev})}&=&\frac{5\chi}{32} +\frac{1}{64}\chi^3+\frac{3}{448}\chi^5+\frac{11}{1792}\chi^7+\frac{377}{57344}\chi^9+\mathcal O(\chi^{11})\ .
\end{eqnarray}
The profile of these coefficients is shown in Fig~\ref{fig:Omega}. This plot was done using the expansion up to order $\chi^{15}$, which provides an accurate result up to $\chi=0.7$. Interestingly, we observe that the correction related to $\alpha_{1}$ increases the angular velocity, while the one related to $\alpha_{2}$ decreases it. The one associated to $\lambda_{\rm ev}$ can have either effect, since the sign of $\lambda_{\rm ev}$ is in principle arbitrary. 
We observe that the effect of these terms is larger for smaller masses: the quantity that controls how relevant the corrections are is $\ell^4/M^4$ times the corresponding coupling. They become of order $1$ when $M\sim\ell$, which marks the limit of validity of the perturbative approach. 
 
 \subsubsection*{Surface gravity}

\begin{figure}[ht!]
\begin{center}
\includegraphics[scale=0.6]{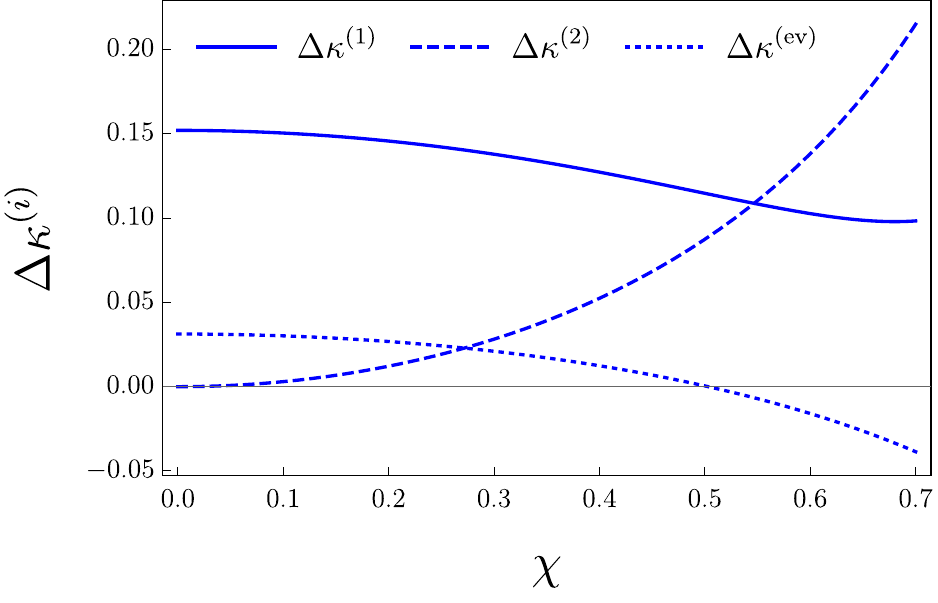}
\caption{Variation of the surface gravity $\Delta \kappa^{(i)}$ due to every correction, as a function of $\chi$. We can observe that contributions coming from curvature-squared terms always increase the temperature, since $\Delta \kappa^{(1)}$ and $\Delta \kappa^{(2)}$ are positive and also the coefficients multiplying them. The contribution from $\lambda_{\rm ev}$ has different sign depending on $\chi$.}
\label{fig:kappa}
\end{center}
\end{figure}

At this stage, the natural step is to compute the surface gravity $\kappa$, defined by the relation

\begin{equation}\label{eq:defsurfacegravity0}
\xi^{\nu}\nabla_{\nu}\xi^{\mu}=\kappa \xi^{\mu}\, ,
\end{equation}

\noindent
that the Killing vector (\ref{eq:killingvector}) must satisfy on the horizon. The computation of $\kappa$ is not straightforward because the coordinates we are using are singular at the horizon. A possibility in order in order to circumvent this problem consists in working in Eddington-Finkelstein coordinates, that cover the horizon. However, there exist a number of alternative methods that can be used in order to obtain the surface gravity even if the coordinates are not well-behaved. Here we will follow a trick proposed in \cite{Poisson:2009pwt}. First, let us rewrite (\ref{eq:defsurfacegravity0}) as 

\begin{equation}\label{eq:defsurfacegravity}
-\partial_\mu \xi^2=2\kappa \xi_{\mu} \ ,
\end{equation}
where we made use of the Killing property $\nabla_{(\mu}\xi_{\nu)}=0$.
Then, let us focus on the left hand side of the equation. 
The norm $\xi^2$ is a function of $x$ and $\rho$, so that $\partial_\mu \xi^2$ only has non-vanishing $\mu=x,\rho$ components. However, one can explicitly check that $\lim_{\rho\rightarrow\rho_{+}}\partial_{x}\xi^2=0$, hence the only non-vanishing component is $\mu=\rho$, and it is given by 

\begin{equation}\label{derivativekillingnorm}
\begin{aligned}
-\partial_\rho \, \xi^2|_{\rho=\rho_+}=&\frac{\left(\rho_+-M\right)}{2M^2\rho^2_+}\left(\rho_+^2+a^2x^2\right)\left[1+2H_2-H_4\right.\\
&\left.+4M^2\rho_+^2\frac{\partial_\rho\left(-H_1\Sigma+a^2(1-x^2)(2H_2-H_4)\right) +2\left(\rho_+-M\right)\left(H_4-2H_2\right)}{2\left(\rho_+-M\right)\left(\rho_+^2+a^2x^2\right)^2}\right]\Bigg|_{\rho=\rho_+}\ ,
\end{aligned}
\end{equation}
where, as usual, we are expanding linearly in the $H_{i}$ functions. On the other hand, since $\xi $ is normal to the horizon, we must have $\xi_\mu= C\,\delta_\mu{}^\rho\ $ for some constant $C$. Of course, this is not true in general: one should imagine that the previous formula holds only on the horizon, where the coordinate $\rho$ is singular. The exact factor $C$ is computed by taking the norm $\xi^2=C^2 g^{\rho\rho}$ and evaluating at the horizon, so that we get

\begin{equation}\label{propfactor}
\begin{aligned}
C=&\lim_{\rho\rightarrow\rho_{+}}\sqrt{\frac{\xi^2}{g^{\rho\rho}}}=\frac{\rho_+^2+a^2x^2}{2M \rho_+}\left[1+H_2+\frac{H_3}{2}-\frac{H_4}{2}\right.\\
&\left.+4M^2 \rho_+^2\frac{\partial_\rho\left(-H_1\Sigma+a^2(1-x^2)(2H_2-H_4)\right) +2\left(\rho_+-M\right)\left(H_4-2H_2\right)}{4\left(\rho_+-M\right)\left(\rho_+^2+a^2x^2\right)^2}\right]\Bigg|_{\rho=\rho_+}\ .
\end{aligned}
\end{equation}
Then, we can plug (\ref{derivativekillingnorm}) and (\ref{propfactor}) into (\ref{eq:defsurfacegravity}) to find 

\begin{equation}
\begin{aligned}
\kappa=&-\frac{\,\partial_\rho\,\xi^2|_{\rho=\rho_+}}{C}=\frac{\left(\rho_+-M\right)}{2M\rho_+}\left[1+H_2-\frac{H_3}{2}-\frac{H_4}{2}\right.\\
&\left.+M^2\rho_+^2\frac{\partial_\rho\left(-H_1\Sigma+a^2(1-x^2)(2H_2-H_4)\right) +2\left(\rho_+-M\right)\left(H_4-2H_2\right)}{\left(\rho_+-M\right)\left(\rho_+^2+a^2x^2\right)^2}\right]\Bigg|_{\rho=\rho_+}\ .
\end{aligned}
\end{equation}
Finally, evaluating this expression on the solution and expanding order by order in $\chi$ we find

\begin{equation}
\kappa=\frac{\sqrt{1-\chi^2}}{2M\left(1+\sqrt{1-\chi^2}\right)}+\frac{\ell^4}{M^5}\left[\alpha_1^2\Delta \kappa^{(1)}+\alpha_2^2 \Delta \kappa^{(2)}+\lambda_{\text{ev}} \Delta \kappa^{(\text{ev})}\right]\ ,
\end{equation}
where the coefficients $\Delta \kappa^{(i)}$ read 

\begin{eqnarray}
\Delta \kappa^{(1)}&=&\frac{73}{480}-\frac{61}{384}\chi^2+\frac{3001}{322560}\chi^4+\frac{5376451}{70963200}\chi^6+\frac{67632847}{615014400}\chi^8+\mathcal O\left(\chi^{10}\right)\ , \\
\Delta \kappa^{(2)}&=&\frac{2127}{7168}\chi^2+\frac{14423}{86016}\chi^4+\frac{429437}{3153920}\chi^6+\frac{125018653}{984023040}\chi^8+\mathcal O\left(\chi^{10}\right)\ , \\
\Delta \kappa^{(\text{ev})}&=&\frac{1}{32}-\frac{7}{64}\chi^2-\frac{3}{64}\chi^4-\frac{7}{256}\chi^6-\frac{157}{8192}\chi^8+\mathcal O\left(\chi^{10}\right)\ .
\end{eqnarray}
Again, we observe that parity-breaking terms do not modify this quantity. In addition, the fact that we obtain a constant surface gravity is another strong check of our solution, since this is a general property that any Killing horizon must satisfy. The profile of these coefficients as functions of $\chi$ is shown in Fig.~\ref{fig:kappa}, using an expansion up to order $\chi^{14}$. We see that both quadratic curvature terms controlled by $\alpha_{1}$ and $\alpha_{2}$ increase the surface gravity, with the difference that the $\alpha_{2}$ correction vanishes for static black holes. On the other hand, the contribution from $\lambda_{\rm ev}$ has a different sign depending on $\chi$. For $\chi<0.5$ the surface gravity is greater than in Kerr black hole, while for $\chi>0.5$ it is lower, or viceversa, depending on the sign of $\lambda_{\rm ev}$.

Another aspect that we can mention is that these contributions do not seem to be vanishing when $\chi\rightarrow 1$. This would imply that $\kappa\big|_{\chi=1}\neq 0$, hence the solutions with $\chi=1$ would not be extremal. In fact, there is a priori no reason to expect that the relation between mass and angular momentum of extremal black holes is preserved when higher-derivative corrections are taken into account.\footnote{In the case of charged black holes, it is known that higher-curvature corrections modify the relation between mass and charges, see \textit{e.g.} \cite{Cano:2018qev,Cano:2018brq}.} Our results in Fig.~\ref{fig:kappa} suggest precisely this; extremality will be reached for a value of $\chi$ slightly different from $1$. However, let us note that the series expansion in $\chi$ breaks down for $\chi=1$, so the perturbative approach is not reliable in order to analyze the corrections to the extremal Kerr solution. One would need to compute directly the corrections to extremal rotating black holes in order to confirm whether the extremality condition is indeed corrected. 

\subsubsection*{Horizon geometry}
\begin{figure}[ht!]
\begin{center}
\includegraphics[scale=0.6]{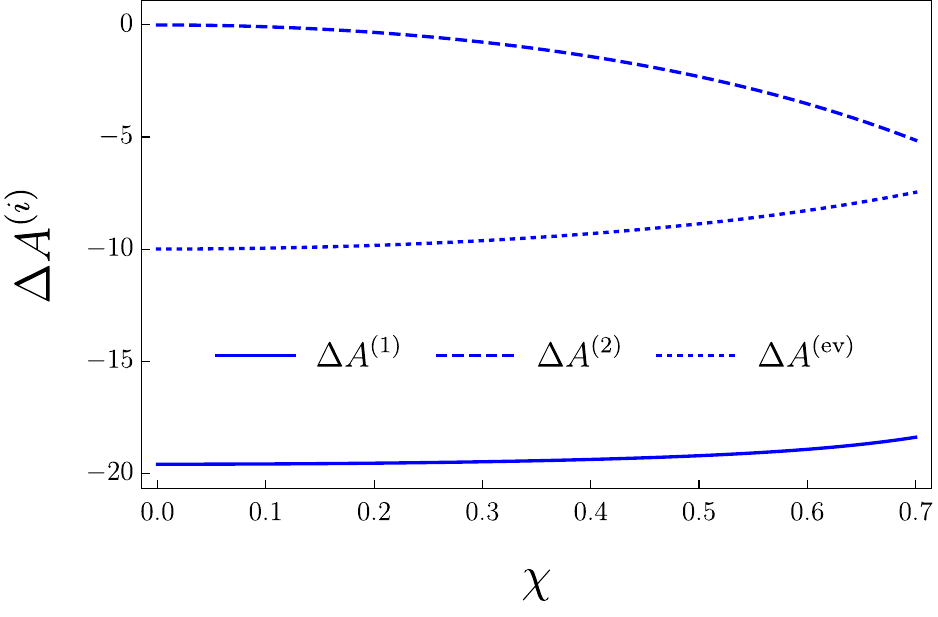}
\caption{Variation of the black hole area $\Delta A^{(i)}$ due to every one of the corrections. The quadratic curvature corrections, controlled by $\alpha_{1}$ and $\alpha_{2}$ always decrease the area with respect to the result in Einstein gravity, while for the even cubic correction the contribution depends on the sign of $\lambda_{\rm ev}$.}
\label{fig:Area}
\end{center}
\end{figure}

Let us finally study the size and shape of the horizon, which will be affected by the corrections. The induced metric at the horizon is 

\begin{equation}\label{horizongeometry}
ds^2_H=\left(1+H_3\right)|_{\rho=\rho_+}\frac{\rho^2_++a^2x^2}{1-x^2}dx^2+\left(1+H_4\right)|_{\rho=\rho_+}\frac{4M^2\rho^2_+\left(1-x^2\right)}{\rho^2_++a^2x^2}d\phi^2\ .
\end{equation}
First, we can find the area, which is given by the integral

\begin{equation}
\begin{aligned}
A_H=&4\pi M \rho_+ \int_{-1}^1dx \left(1+\frac{H_3}{2}+\frac{H_4}{2}\right)\Bigg|_{\rho=\rho_+}\, .\\
 \end{aligned}
\end{equation}
Computing the integral order by order in $\chi$, we can write the area as

\begin{equation}
\begin{aligned}
A_H=8 \pi M^2\left(1+\sqrt{1-\chi^2}\right) +\frac{\pi \ell^4}{M^2}\left(\alpha_{1}^2\, \Delta A^{(1)}+\alpha_{2}^2\, \Delta A^{(2)}+\lambda_{\rm ev} \,\Delta A^{(\rm ev)}\right)\ ,
 \end{aligned}
\end{equation}
where every contribution $\Delta A^{(i)}$ depends on $\chi$, and the first terms read
 
 \begin{eqnarray}
\label{eq:areaGB}
\Delta A^{(1)}&=&-\frac{98}{5}+\frac{11 \chi ^2}{10}+\frac{28267 \chi ^4}{25200}+\frac{11920241 \chi ^6}{7761600}+\frac{2263094063 \chi ^8}{1210809600}+\mathcal O\left(\chi^{10}\right)\, ,\\
\label{eq:areaCS}
\Delta A^{(2)}&=&-\frac{915 \chi ^2}{112}-\frac{25063 \chi ^4}{6720}-\frac{528793 \chi ^6}{295680}-\frac{39114883 \chi ^8}{53813760}+\mathcal O\left(\chi^{10}\right)\, ,\\
 \Delta A^{(\rm ev)}&=&-10+4 \chi ^2+\frac{69 \chi ^4}{40}+\frac{263 \chi ^6}{280}+\frac{183 \chi ^8}{320}+\mathcal O\left(\chi^{10}\right)\, .
 \end{eqnarray}
 In Fig.~\ref{fig:Area} we show the profile of these quantities as functions of $\chi$, using the expansion up to order $\chi^{14}$. We observe that the quadratic corrections always reduce the area (except $\alpha_{2}$ in the static case, that does not contribute). On the other hand, the cubic even correction reduces or increases the area depending on whether $\lambda_{\rm ev}>0$ or $\lambda_{\rm ev}<0$, respectively.

So far, we have not observed the effect of the parity-breaking corrections --- they do not contribute either to the area, the surface gravity or the angular velocity of the black hole. This is expected on general grounds since these corrections contain only odd powers of $x$, and it is easy to see that the contribution, for instance, to the area, must vanish.
Nevertheless, these terms do change the geometry and they will affect the shape of the horizon. Indeed, these parity-breaking corrections  break the $\mathbb{Z}_{2}$ symmetry of the solution, \textit{i.e.} the reflection symmetry on the equatorial plane $x\rightarrow-x$. It is expected that this loss of symmetry is manifest in the form of the horizon. 

In order to visualize the event horizon, we perform an isometric embedding of it in 3-dimensional Euclidean space $\mathbb E^3$. In terms of Cartesian coordinates $(x^{1},x^{2},x^{3})$, we can parametrize the most general axisymmetric surface as

\begin{equation}
x^1=f(x)\sin \phi\ , \quad x^2=f(x)\cos\phi\ , \quad x^3=g(x)\ ,
\end{equation}
where $f(x)$ and $g(x)$ are some functions that must be determined by imposing that the induced metric on the surface, given by 

\begin{equation}
ds^2=\left[\left(f'\right)^2+\left(g'\right)^2\right]\, dx^2+ f^2 \, d\phi^2\ ,
\end{equation}
coincides with (\ref{horizongeometry}). We get immediately that these functions are given by 

\begin{eqnarray}
\label{eq:f}
f(x)&=&2M \rho_+\left(1+\frac{H_4}{2}\right)\Bigg|_{\rho=\rho_+}\left(\frac{1-x^2}{\rho_+^2+a^2x^2}\right)^{1/2}\ ,\\
g(x)&=&\int dx \, \left[\left(1+H_3\right)|_{\rho=\rho_+}\frac{\rho_+^2+a ^2x^2}{1-x^2}-\left(f'\right)^2\right]^{1/2}\ .
\label{eq:g}
\end{eqnarray}
However, it can happen that the solution does not exist if the argument of the square root in the integral becomes negative. In that case, the horizon cannot be embedded completely in $\mathbb{E}^3$. It turns out that this only happens for quite large values of $\chi$ (around $\chi\sim0.9$), and for the values we are considering here, the complete horizon can be embedded. As usual, we expand the expressions (\ref{eq:f}) and (\ref{eq:g}) linearly on $H_{i}$ and at the desired order in $\chi$ and we obtain explicit formulas for $f$ and $g$ that we do not reproduce here for a sake of clarity. \\
\begin{figure}[ht!]
\begin{center}
\includegraphics[scale=0.45]{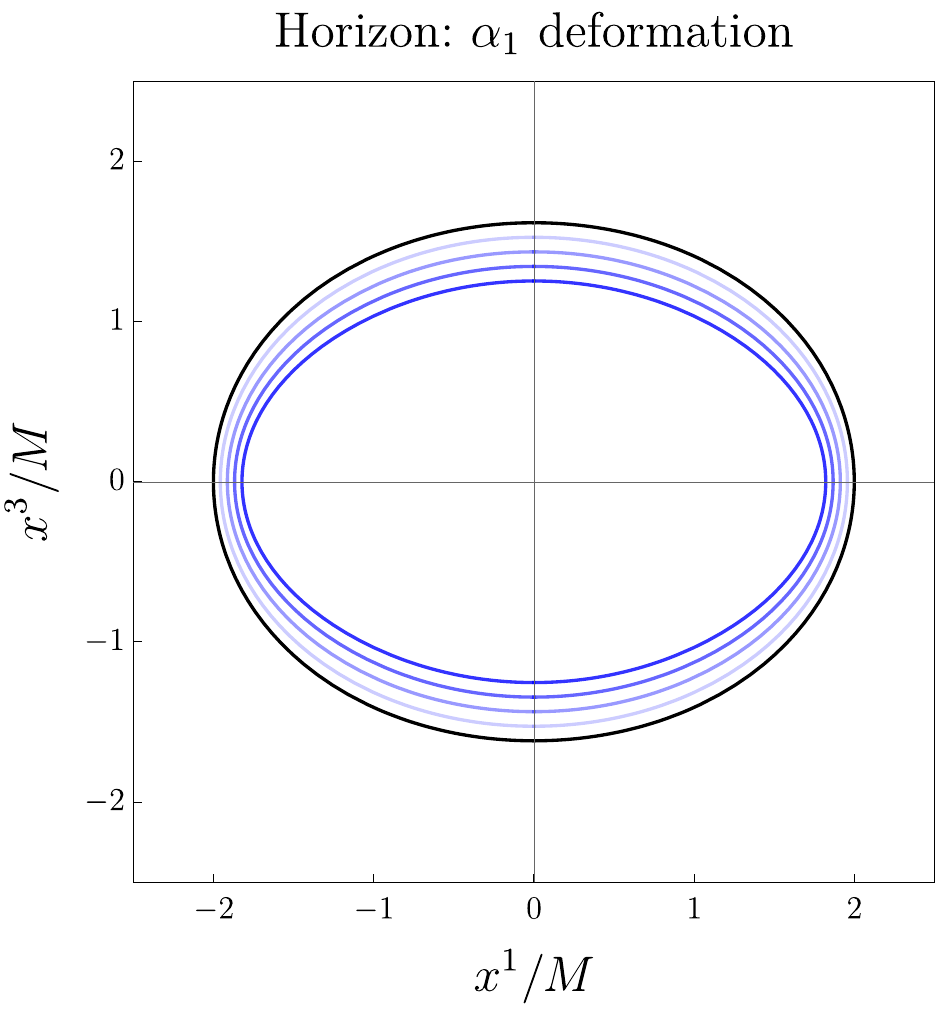}
\includegraphics[scale=0.45]{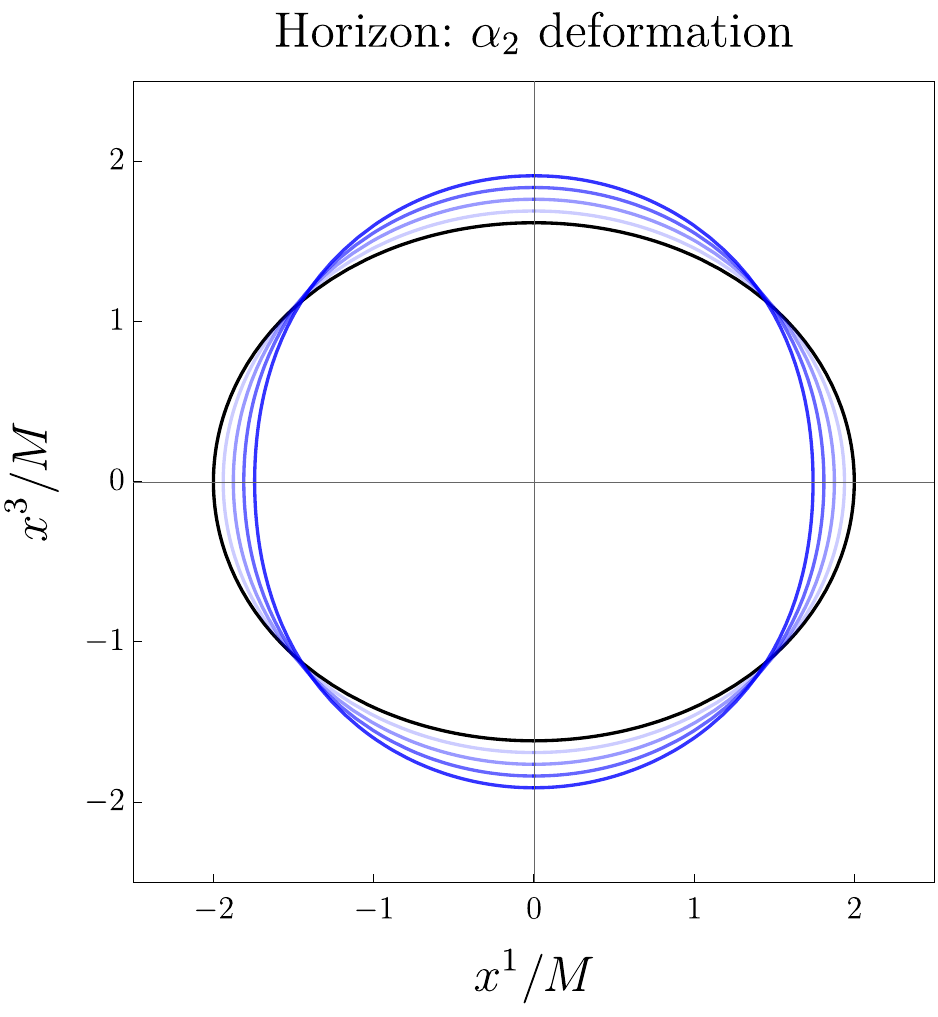}
\includegraphics[scale=0.45]{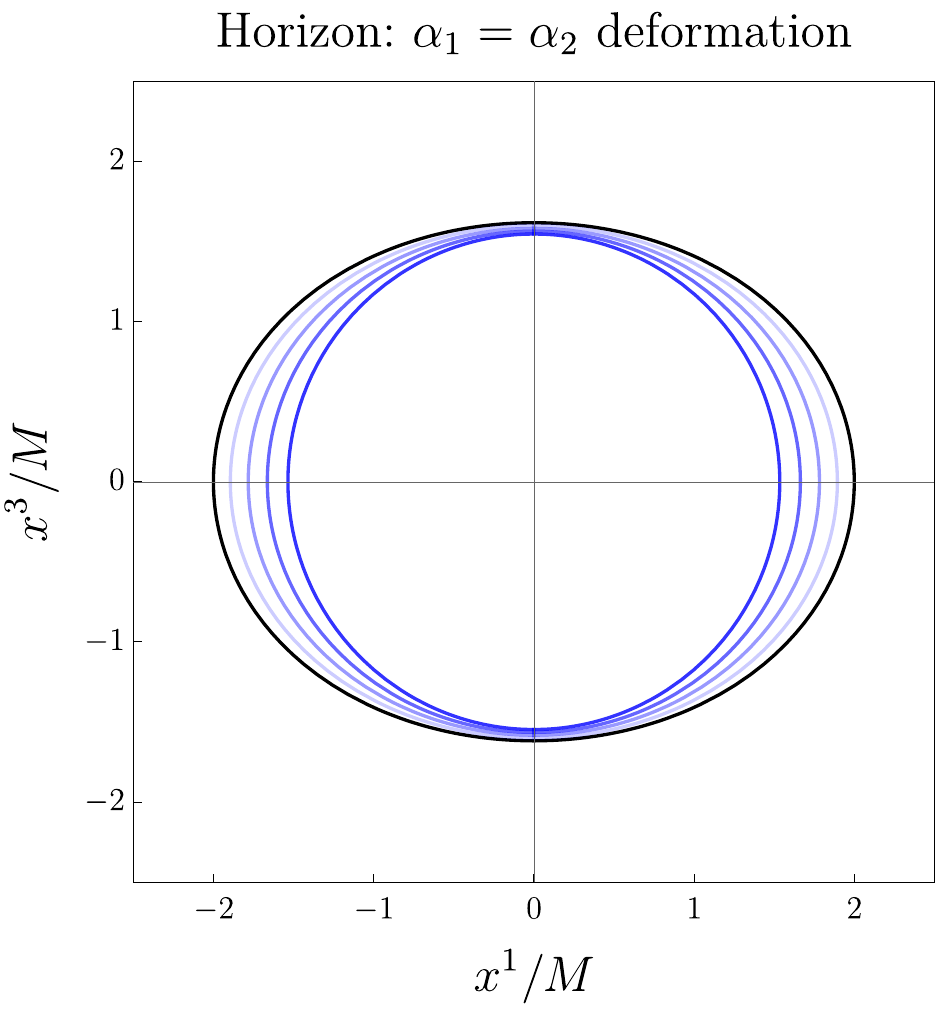}
\includegraphics[scale=0.45]{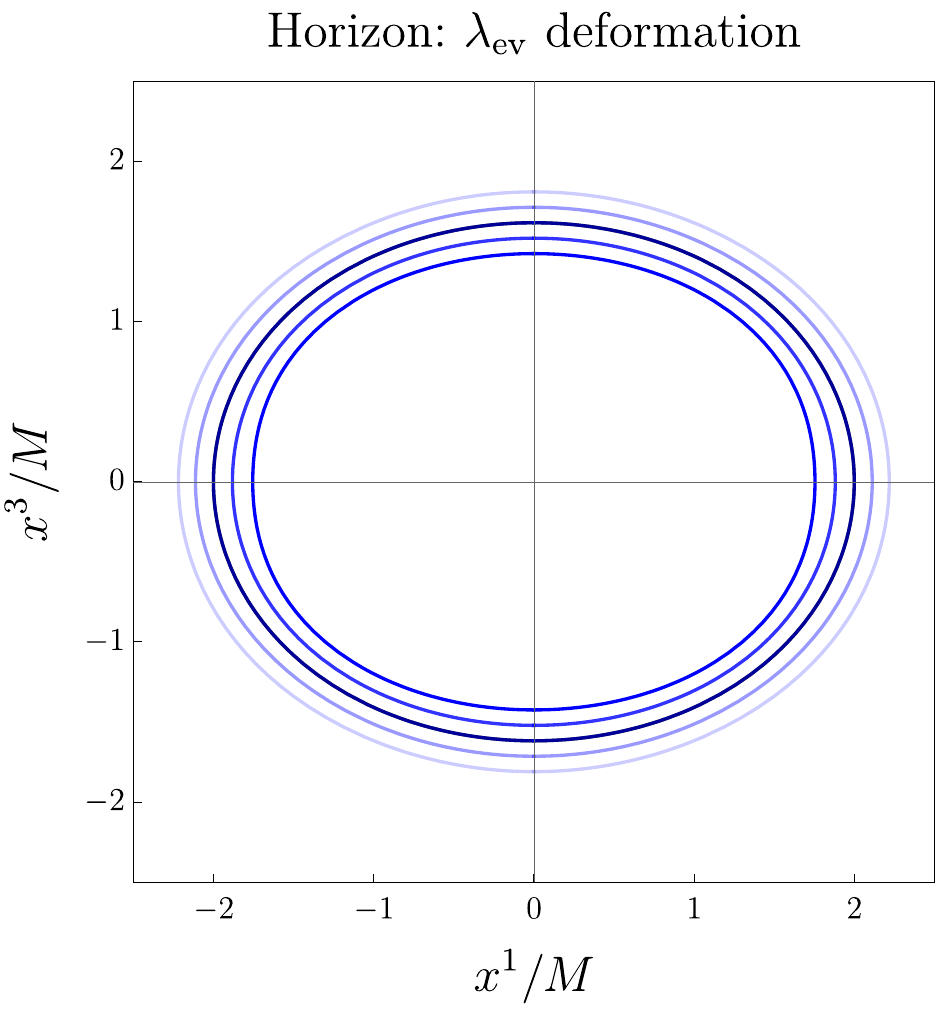}
\caption{Isometric embedding of the horizon in $\mathbb{E}^3$ for different values of the parameters and for $\chi=0.65$. In black we represent the horizon of Kerr black hole and in blue the horizon of the corrected solution for a fixed mass and different values of the couplings. From light to darker blue we increase the value of the corresponding coupling. In each case, only the indicated couplings are non-vanishing. Top left: $\frac{\ell^4}{M^4}\alpha_{1}^2=0.05,0.1,0.15,0.2$,  top right: $\frac{\ell^4}{M^4}\alpha_{2}^2=0.05,0.1,0.15,0.2$, bottom left: $\frac{\ell^4}{M^4}\alpha_{1}^2=\frac{\ell^4}{M^4}\alpha_{2}^2=0.05, 0.1, 0.15, 0.2$, bottom right: $\frac{\ell^4}{M^4}\lambda_{\rm ev}=-0.4, -0.2, 0.2, 0.4$.}
\label{fig:horizonev}
\end{center}
\end{figure}
Now we can use the result to visualize the horizon. In Fig.~\ref{fig:horizonev} we show the horizon for parity-preserving theories. We fix the mass to some constant value and $\chi=0.65$ and we compare the horizon of Kerr black hole with the one in the corrected solutions for different values of the couplings. In this way, we can observe clearly the change in size and in shape of the horizon. As we already noted, both $\alpha_{1}$ and $\alpha_{2}$ reduce the area, but it turns out that they deform the horizon in different ways: $\alpha_{1}$ squashes it while $\alpha_{2}$ squeezes it. We also show the deformation corresponding to the ``stringy'' prediction $\alpha_1=\alpha_2$. In that case we observe that the effect of both terms together is to make to horizon rounder than in Einstein gravity. As for the cubic even correction, it mainly changes the size of the black hole while its shape is almost unaffected. 

\begin{figure}[ht!]
\begin{center}
\flushleft
\includegraphics[scale=0.45]{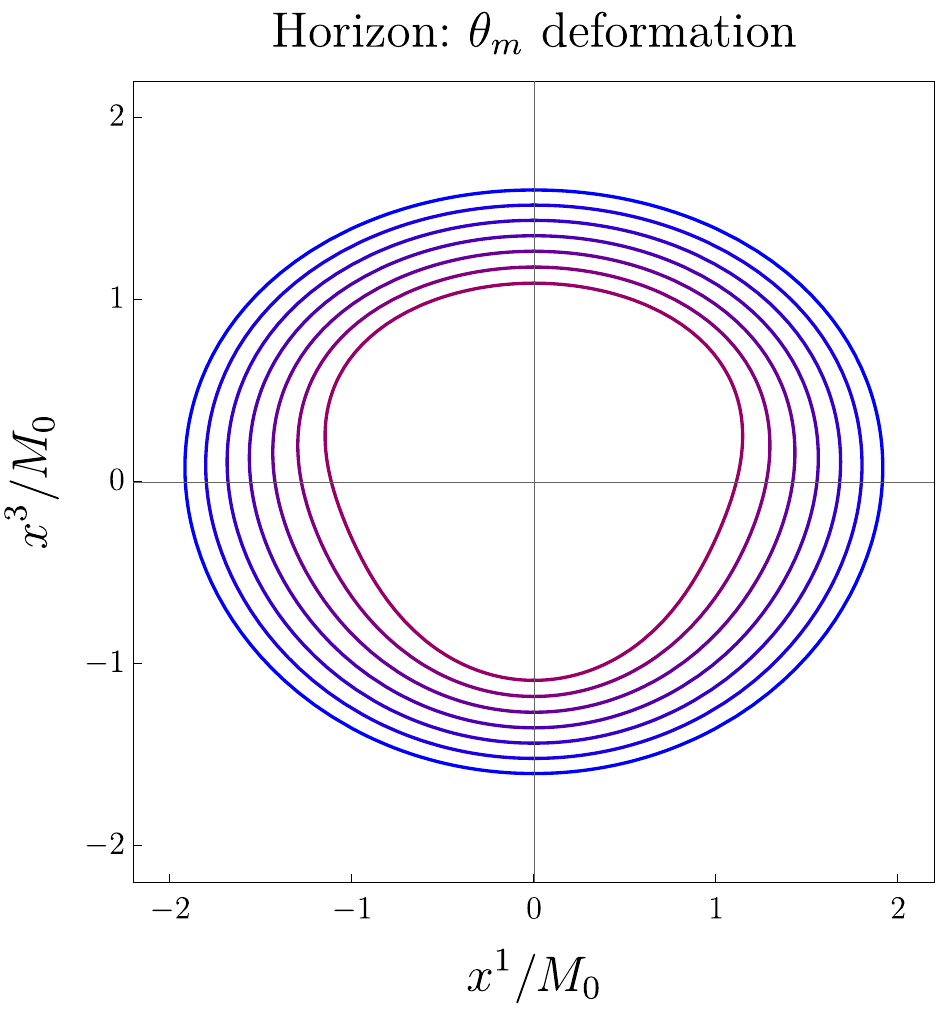}
\hskip0.5cm
\includegraphics[scale=0.45]{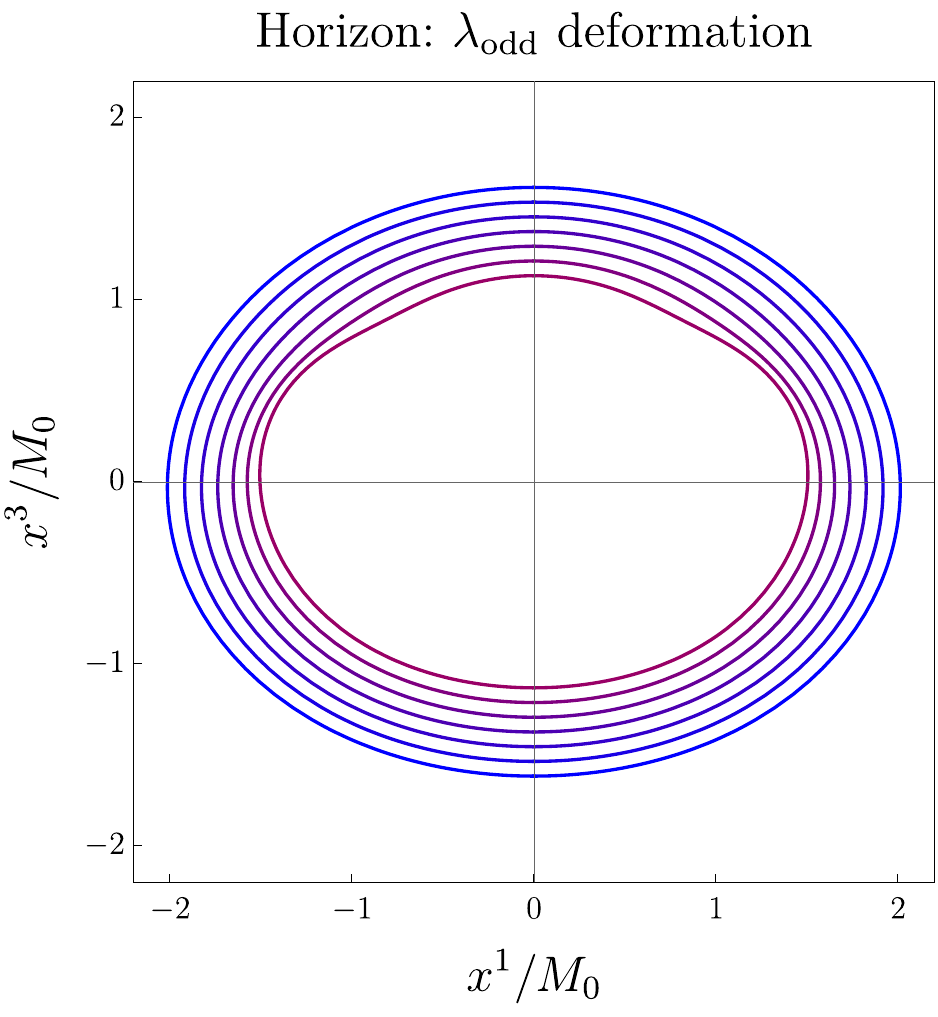}
\includegraphics[scale=0.45]{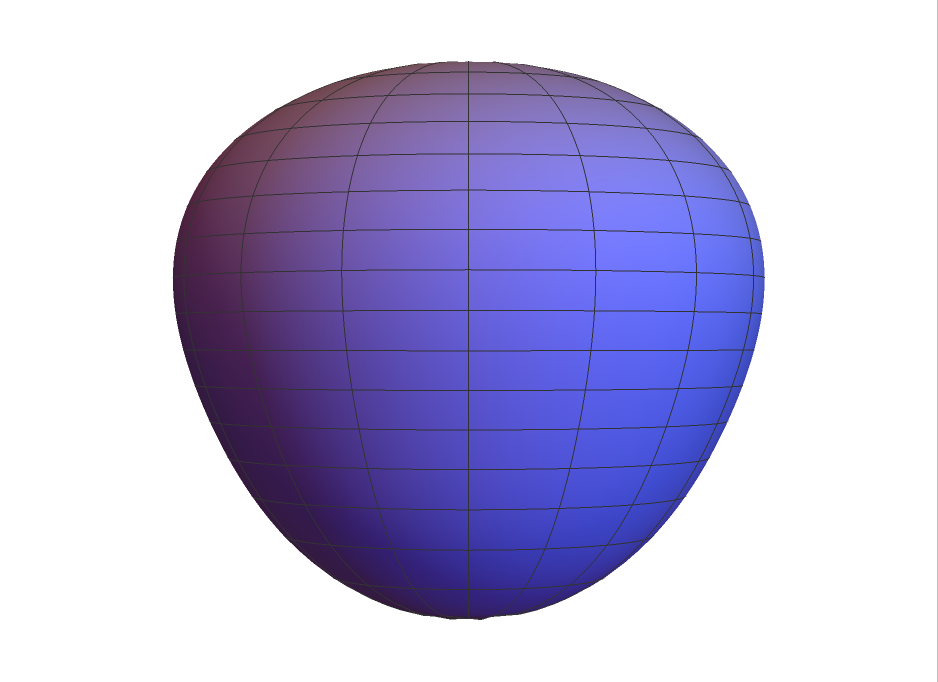}
\hskip1cm
\includegraphics[scale=0.45]{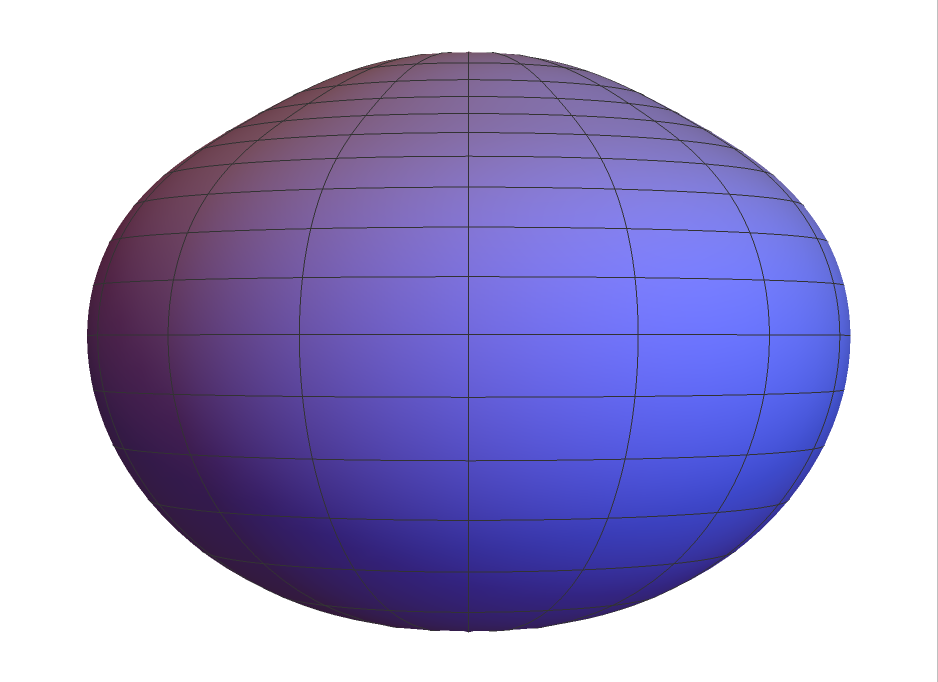}
\caption{Isometric embedding of the horizon in $\mathbb{E}^3$ for parity-breaking theories. For clarity reasons we do not include the comparison with Kerr solution.  In the top row we plot the horizon for different masses ($M_{0}\ge M\ge 0.7 M_{0}$ for some reference mass $M_{0}$) while keeping $\chi=0.65$ and the couplings constant. In each case, only the indicated couplings are non-vanishing. Left:  $\alpha_{1}=\alpha_{2}$, $\theta_{m}=\pi/2$, $M_{0}\approx 2.23 \ell\sqrt{|\alpha_{1}|}$. Right:  $\lambda_{\rm odd}>0$, $M_{0}\approx 1.46 \ell \lambda_{\rm odd}^{1/4}$. Bottom row: 3D embedding of the horizon for $\frac{\ell^4}{M^4}\alpha_{1}^2=\frac{\ell^4}{M^4}\alpha_{2}^2=0.15$, $\theta_m=\pi/2$ (left) and for  $\frac{\ell^4}{M^4}\lambda_{\rm odd}=0.6$ (right). In both cases, the $\mathbb{Z}_2$ symmetry is manifestly broken.}
\label{fig:horizonodd}
\end{center}
\end{figure}

In Fig.~\ref{fig:horizonodd} we present the horizon in the parity-breaking theories (characterized by the two parameters $\theta_{m}$ and $\lambda_{\rm odd}$). In the top row we plot the horizon for a fixed choice of higher-order couplings and for various masses, keeping $\chi=0.65$ constant. The visualization is clearer in this way since these corrections do not change the area. In addition, we can see that for large $M$ the horizon has almost the same form as in EG, but as we decrease the mass the corrections become relevant and it is deformed. We observe in this case that the $\mathbb{Z}_{2}$ symmetry is manifestly broken. Due to exotic form of these horizons we include as well a 3D plot in which we can appreciate them better. Very recently other works have described black hole solutions that do not possess $\mathbb{Z}_{2}$ symmetry \cite{Cunha:2018uzc,Cardoso:2018ptl}. However, to the best of our knowledge, these are the first plots of black hole horizons without $\mathbb{Z}_{2}$ symmetry in purely gravitational theories.

\subsection{Ergosphere}
Another important surface of rotating black holes is the ergosphere, which marks the limit in which an object can remain static outside the black hole. When $g_{tt}<0$, there are no timelike trajectories with constant $(\rho,x,\phi)$, so the ergosphere is identified by the condition $g_{tt}=0$, which for the metric (\ref{rotatingmetric}) can be written as
\begin{equation}\label{eq:ergeq}
1-\frac{2M\rho}{\Sigma}=H_1\, .
\end{equation}
This equation determines the value of the ``ergosphere radius'' $\rho_{\rm erg}$.
Unlike the horizon radius $\rho_{+}$, that does not receive corrections due to the clever choice of coordinates, the ergosphere radius is modified with respect to its value in Kerr metric.  We may express the corrections to $\rho_{\rm erg}$ as
\begin{align}
\rho_{\rm erg}&=M\left(1+\sqrt{1-\chi^2x^2}\right)\\
&+\frac{\ell^4}{M^3}\left[\alpha_{1}^{2}\Delta \rho^{(1)}+\alpha_{2}^{2}\Delta \rho^{(2)}+\alpha_1\alpha_2 \sin\theta_m \Delta \rho^{(m)}+ \lambda_{\rm ev}\Delta \rho^{({\rm ev})}+\lambda_{\rm odd}\Delta \rho^{({\rm odd})} \right]\, ,
\end{align}
where the first term represents the result in Einstein gravity and we have to determine the value of the coefficients $\Delta \rho^{(i)}$. Plugging this into (\ref{eq:ergeq}), we find these coefficients, whose first terms in the $\chi$-expansion are shown in Eq.~(\ref{eq:rhoerg}). In this case, we do get a non-vanishing contribution from the parity-breaking terms, though this is not directly relevant, since $\rho_{\rm erg}$ has no physical meaning by itself.  However, an interesting property that we note by looking at (\ref{eq:rhoerg}) is that all the corrections to $\rho_{\rm erg}$ vanish at $x=\pm1$, corresponding to the north and south poles of the ergosphere. There is a nice interpretation of this fact: the ergosphere and the horizon overlap at the poles. Indeed, the horizon radius $\rho_{+}$ does not have corrections, and the zeroth-order value of the ergosphere radius $\rho^{(0)}_{\rm erg}=M\left(1+\sqrt{1-\chi^2x^2}\right)$ already coincides with $\rho_{+}$ at the poles $\rho^{(0)}_{\rm erg}(x=\pm1)=\rho_{+}$. Hence, the corrections to $\rho^{(0)}_{\rm erg}$ must vanish at $x=\pm 1$ if we want the horizon and the ergosphere to still overlap. 

In order to study the geometry of the ergosphere, we can compute the induced metric for $\rho=\rho_{\rm erg}(x)$ at a constant time $t=t_0$, which reads

\begin{equation}\label{ergometric}
\begin{aligned}
ds^{2}_{\rm erg}=&\left(1+H_3\right)\Sigma\left(\frac{1}{\Delta}\left(\frac{d\rho_{\rm erg}}{dx}\right)^2+\frac{1}{1-x^2}\right)dx^2\\
&+\left(1+H_4\right)\left(\rho^2+a^2+\frac{2 M  \rho a^2(1-x^2)}{\Sigma}\right)(1-x^2)d\phi^2\bigg|_{\rho=\rho_{\rm erg}(x)}\, ,
\end{aligned}
\end{equation}

\begin{figure}[ht!]
\begin{center}
\includegraphics[scale=0.45]{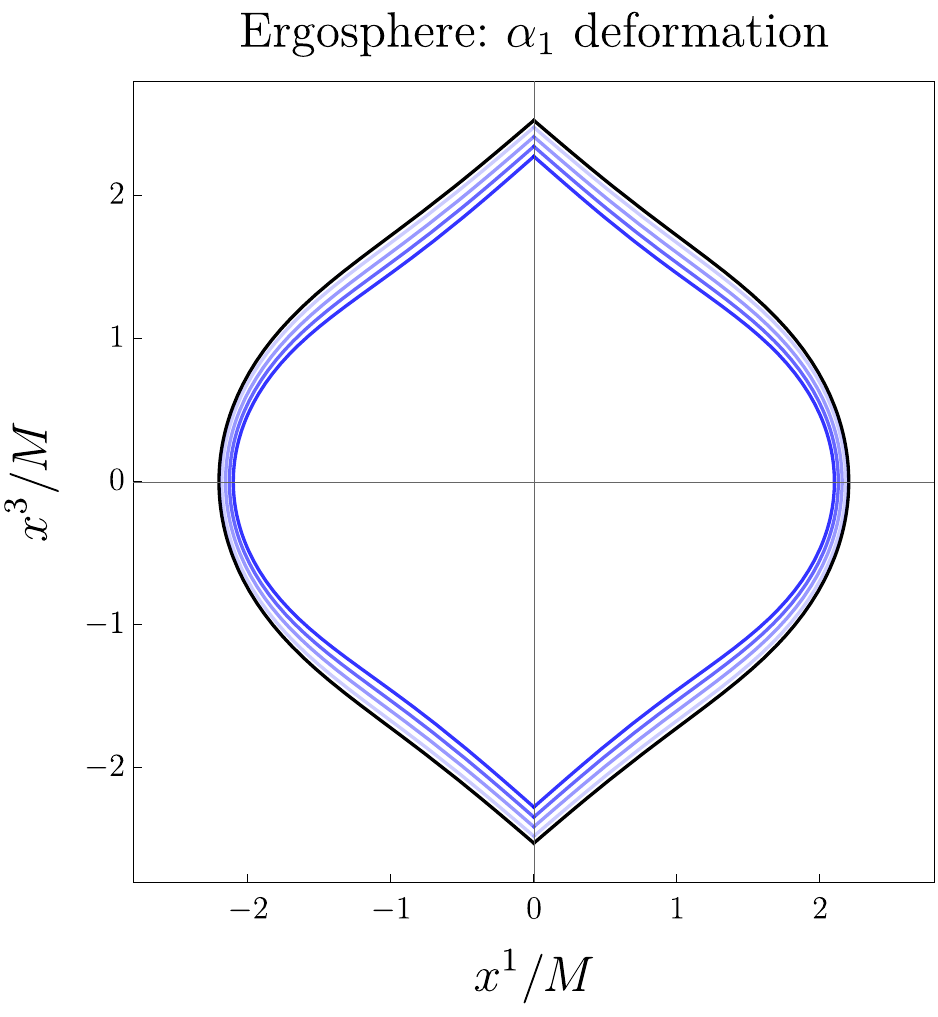}
\includegraphics[scale=0.45]{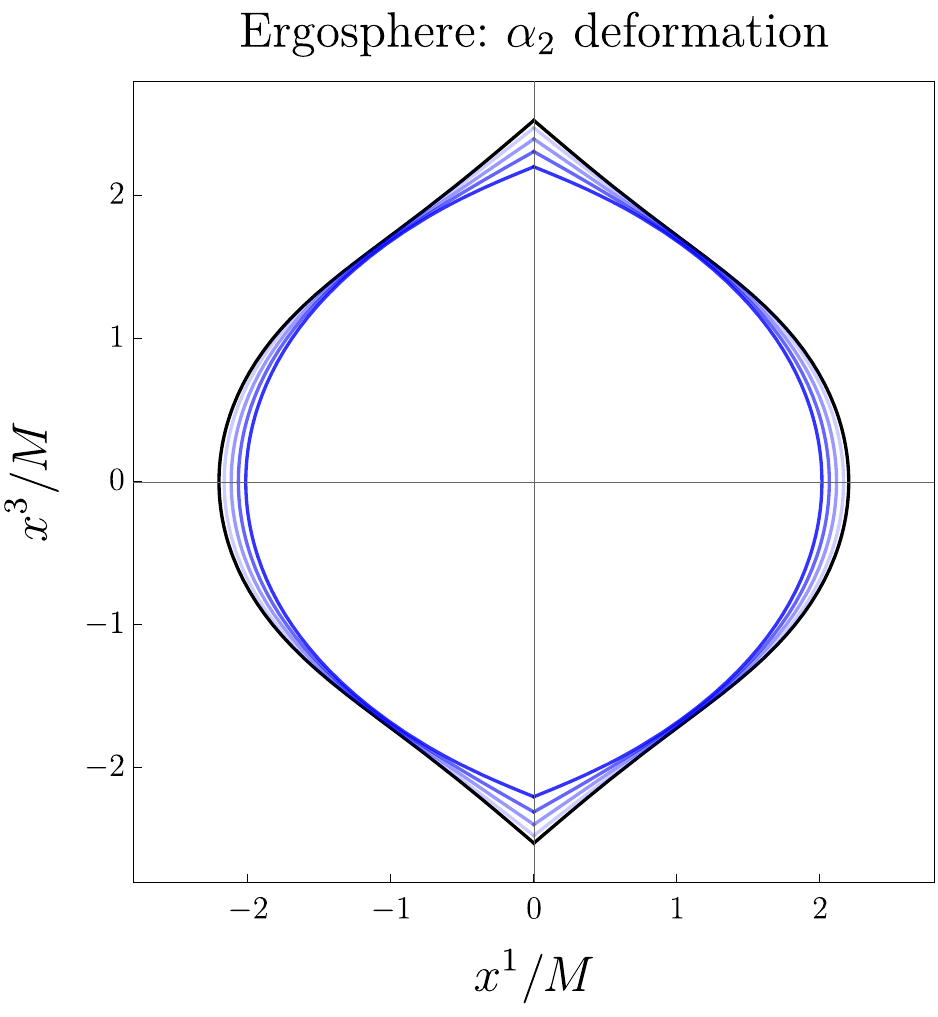}
\includegraphics[scale=0.45]{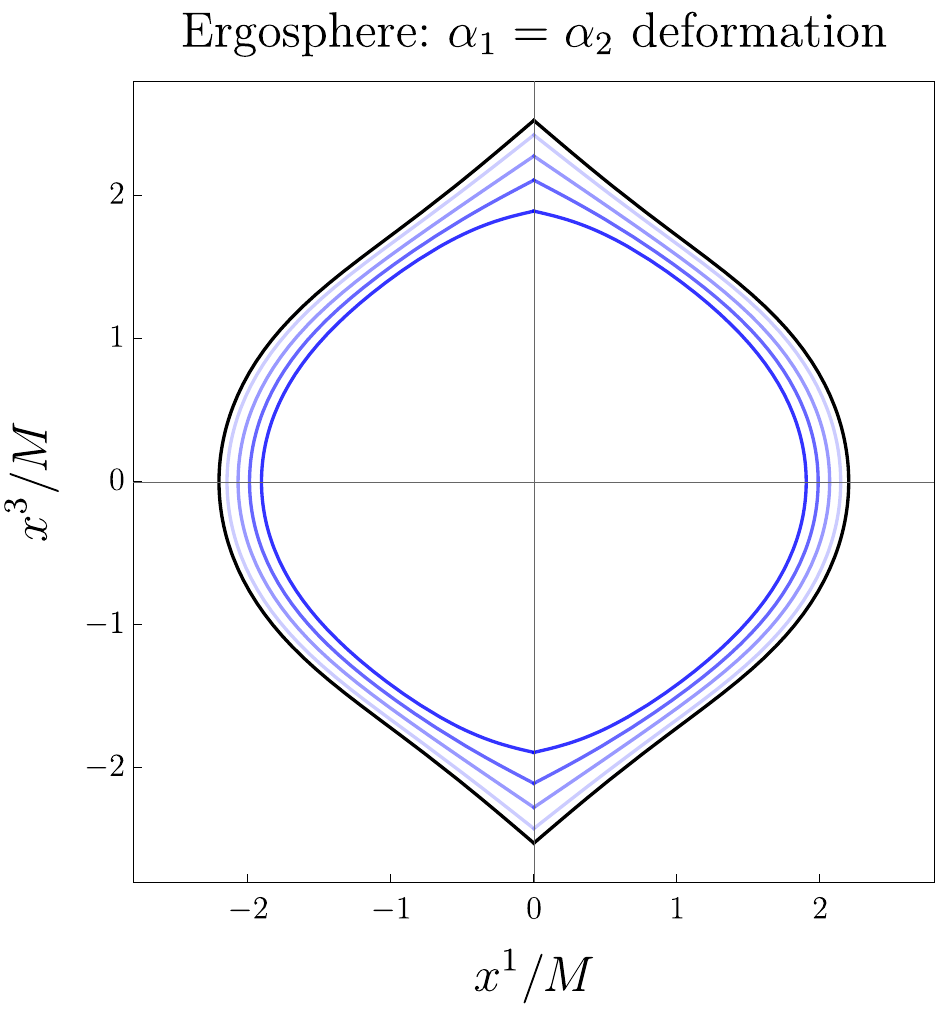}
\includegraphics[scale=0.45]{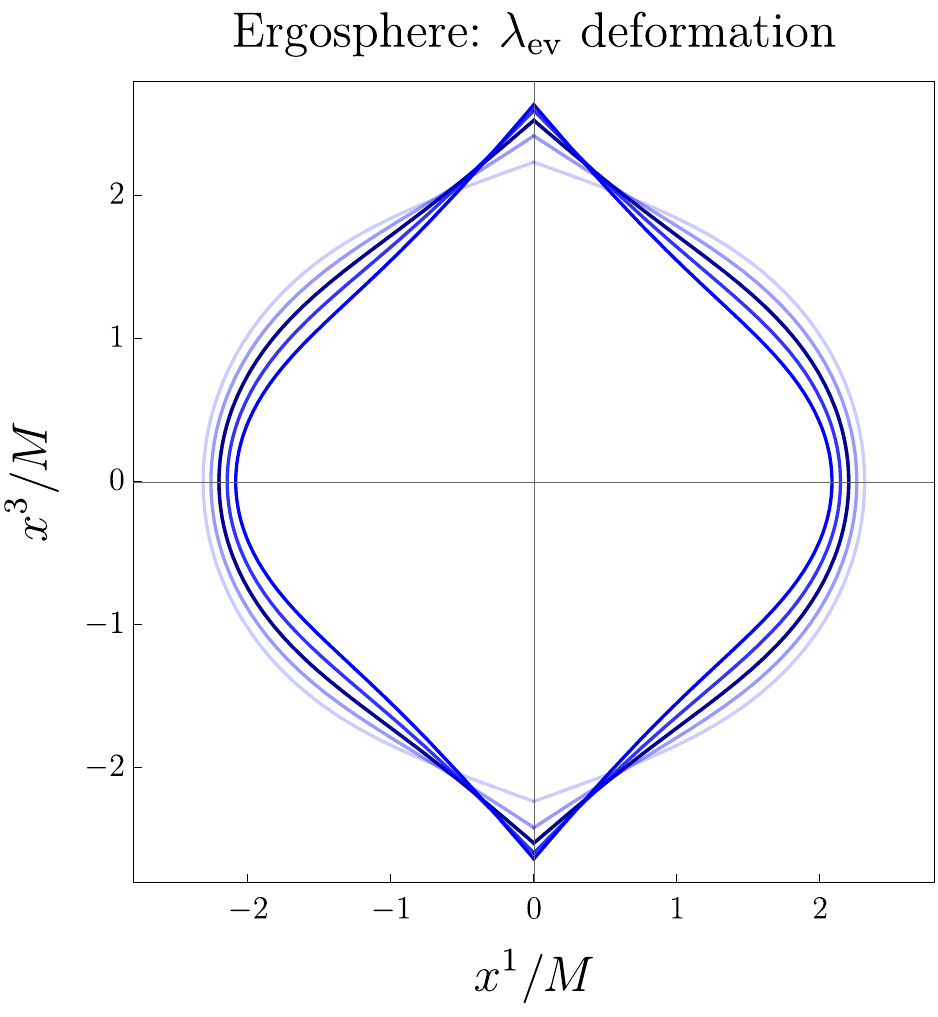}
\caption{Isometric embedding of the ergosphere in $\mathbb{E}^3$ for different values of the parameters and for $\chi=0.65$. In black we represent the ergosphere of Kerr black hole and in blue the ergosphere of the corrected solution, for a fixed mass and different values of the couplings. From light to darker blue we increase the value of the corresponding coupling. In each case, only the indicated couplings are non-vanishing. From left to right and top to bottom: $\frac{\ell^4}{M^4}\alpha_1^2=0.03,0.07,0.11,0.15$,  $\frac{\ell^4}{M^4}\alpha_{2}^2=0.03,0.07,0.11,0.15$, $\frac{\ell^4}{M^4}\alpha_{1}^2=\frac{\ell^4}{M^4}\alpha_{2}^2=0.03,0.07,0.11,0.15$, $\frac{\ell^4}{M^4}\lambda_{\rm ev}=-0.6, -0.3, 0.3, 0.6$.  }
\label{fig:ergev}
\end{center}
\end{figure}

\begin{figure}[ht!]
\begin{center}
\includegraphics[scale=0.45]{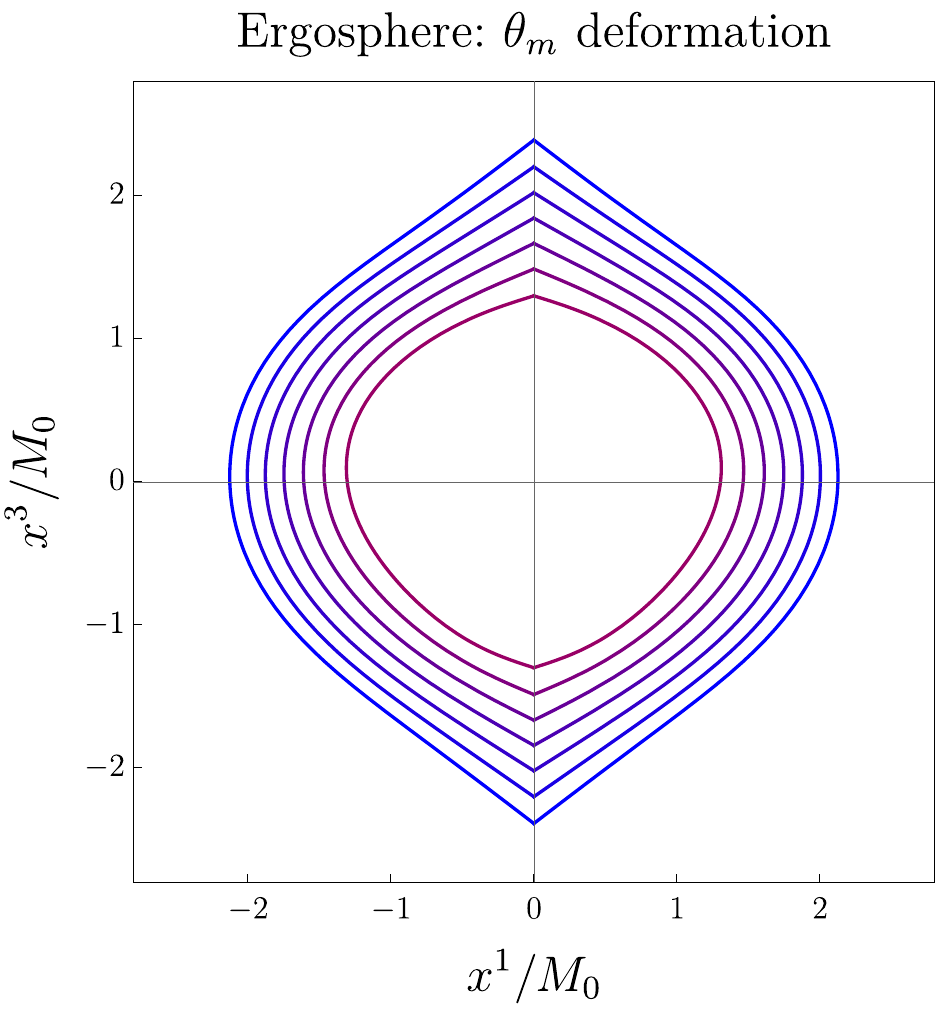}
\includegraphics[scale=0.45]{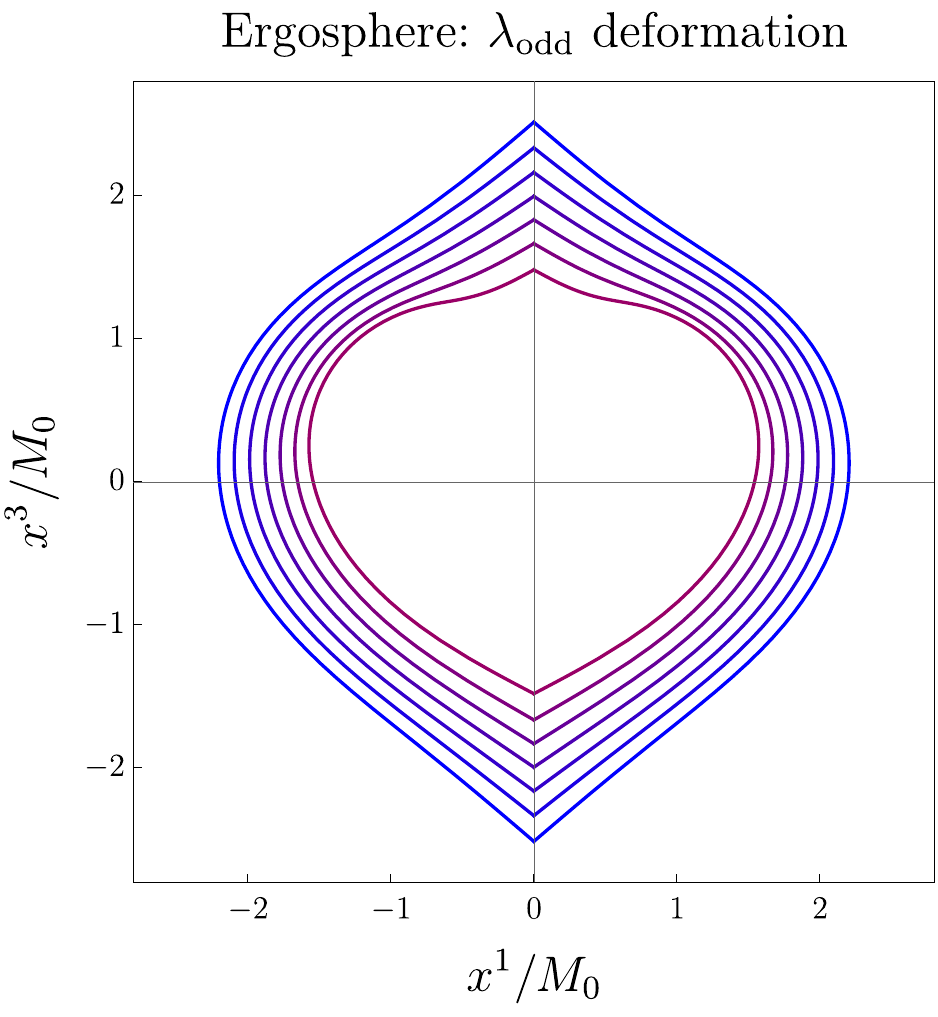}
\includegraphics[scale=0.45]{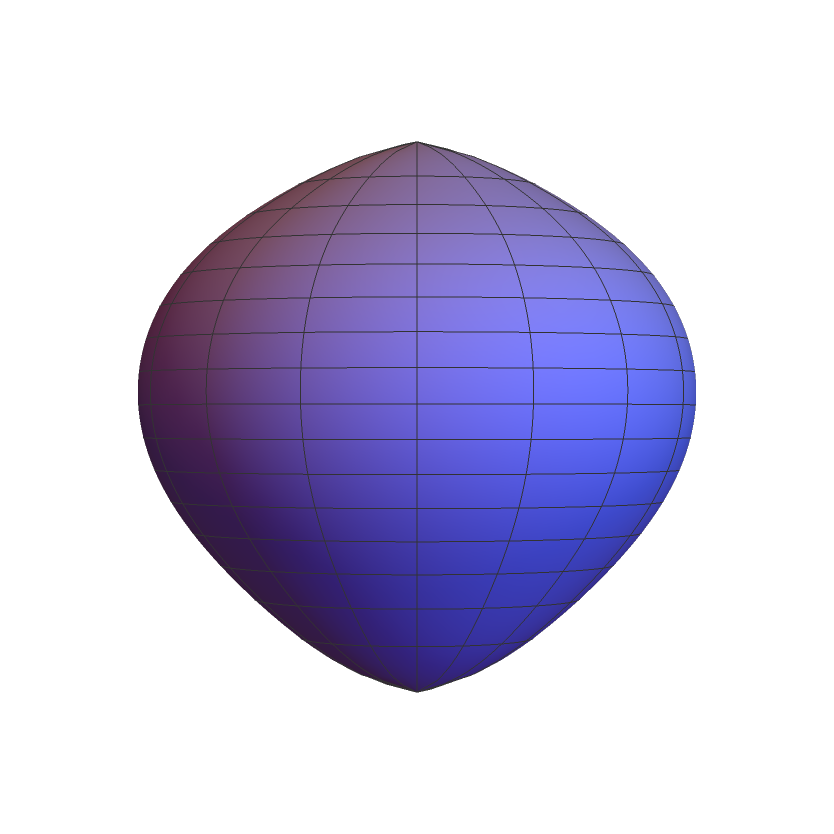}
\hskip1cm
\includegraphics[scale=0.45]{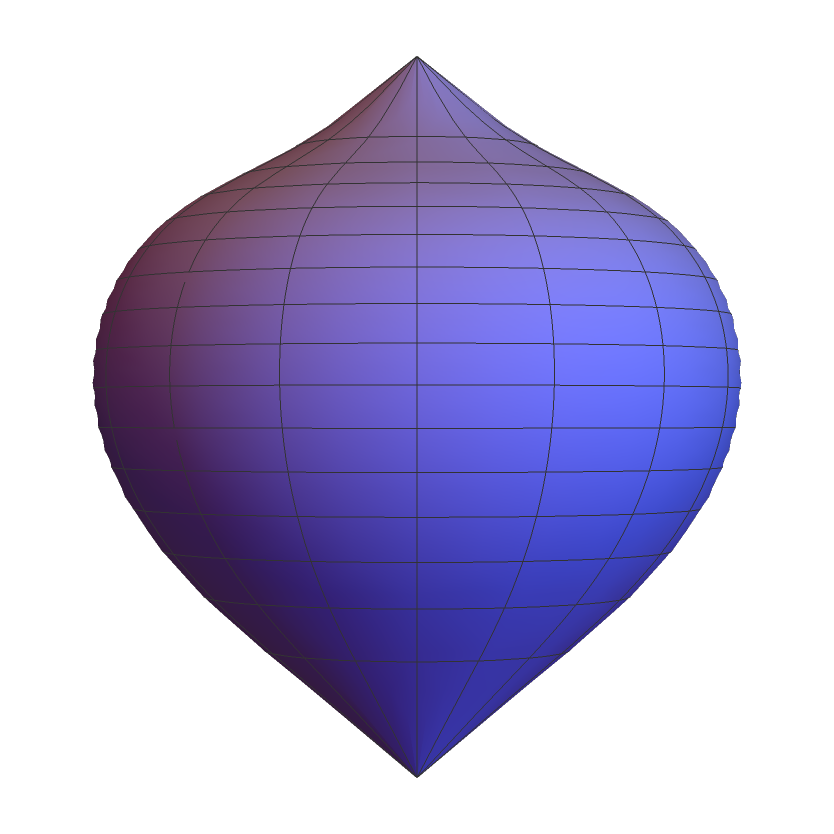}
\caption{Isometric embedding of the ergosphere in $\mathbb{E}^3$ for parity-breaking theories. In the top row we plot the ergosphere for different masses ($M_{0}\ge M\ge 0.7 M_{0}$ for some reference mass $M_{0}$) while keeping $\chi=0.65$ and the couplings constant. In each case, only the indicated couplings are non-vanishing. Left:  $\alpha_{1}=\alpha_{2}$, $\theta_{m}=\pi/2$, $M_{0}\approx 2.23 \ell\sqrt{|\alpha_{1}|}$. Right:  $\lambda_{\rm odd}>0$, $M_{0}\approx 1.35 \ell \lambda_{\rm odd}^{1/4}$. In the bottom row we show a 3D embedding of the ergosphere for $\frac{\ell^4}{M^4}\alpha_{1}^2=\frac{\ell^4}{M^4}\alpha_{2}^2=0.15$, $\theta_m=\pi/2$ (left) and for  $\frac{\ell^4}{M^4}\lambda_{\rm odd}=0.6$ (right). In the latter case we observe clearly that the $\mathbb{Z}_{2}$ symmetry is broken and the ergosphere acquires a characteristic ``trompo'' shape. The effect is more subtle in the left picture, but the $\mathbb{Z}_2$ symmetry is also broken.}
\label{fig:ergodd}
\end{center}
\end{figure}

Using the value of  $\rho_{\rm erg}$ that we have found yields a complicated expression that we omit here for clarity sake. The most useful way to visualize the geometric properties of the ergosphere is to find an isometric embedding of the previous metric in Euclidean space, as we have just done with the horizon. The embedding is shown in Fig. \ref{fig:ergev} for parity-preserving theories, and in Fig.~\ref{fig:ergodd} for parity-breaking ones. In the former case, we plot the ergosphere for a fixed mass and $\chi=0.65$, and for different values of the couplings, including the GR result. We observe that the corrections change the size and shape of the ergosphere. The quadratic terms $\alpha_{1}$ and $\alpha_{2}$ both reduce the area of the ergosphere, while the cubic even term reduces its size for $\lambda_{\rm ev}>0$, and increases it for $\lambda_{\rm ev}<0$. The characteristic conical singularity at the poles of the ergosphere is also considerable affected by some corrections. In particular, we see that $\alpha_{2}$ and $\lambda_{\rm ev}<0$ have the effect of making the cone less sharp. 
In the top row of Fig.~\ref{fig:ergodd} we show instead the embedding of the ergosphere for several values of the mass, while keeping the couplings and $\chi=0.65$ constant. This helps the visualization since parity-breaking interactions do not change the area of the ergosphere. As the mass decreases, the effect of the corrections becomes relevant and we observe, as in the case of the horizon, that the ergosphere does not possess $\mathbb{Z}_{2}$ symmetry. This is more explicit for the cubic odd correction $\lambda_{\rm odd}$ that deforms the ergosphere giving it a characteristic ``trompo'' shape. The effect of $\mathbb{Z}_{2}$ symmetry breaking is less obvious for the $\theta_{m}$ deformation, but nevertheless it can still be observed.  To the best of our knowledge, these are the first examples of ergospheres without $\mathbb{Z}_{2}$ symmetry.

\subsection{Photon rings}
Another aspect of the modified Kerr black holes we would like to explore is their geodesics. The analysis of geodesics is necessary in order to obtain some observable quantities, such as the form of the black hole shadow \cite{Johannsen:2015hib}. However, a detailed analysis of geodesics will require of an independent study due to their intricate character.\footnote{For instance, a preliminary exploration shows that integrability is lost, \textit{i.e.}, there is no Carter constant \cite{Carter:1968rr}.} For that reason, here we consider only a special type of geodesics that are particularly interesting:
circular orbits ($\rho=$ constant) for light rays at the equatorial plane, \textit{i.e}. at $x=0$, known as the photon rings or light rings of the black hole.  However, an appropriate question that we must answer first is whether there are geodesics contained in the equatorial plane at all. In the case of Kerr metric, the reason of their existence is the reflection symmetry $x\rightarrow -x$, but we have seen that in our black holes this symmetry does not exist if we include parity-breaking terms. In fact, in those solutions there is no equatorial plane! Therefore, we should not expect the existence of geodesics contained in the plane $x=0$ if we include those corrections.  In order to understand this better, let us examine the geodesic equations:
\begin{equation}\label{eq:geoeq}
\ddot{x}^{\mu}+\Gamma{_{\alpha\beta}^{\mu}}\dot{x}^{\alpha}\dot{x}^{\beta}=0\, , 
\end{equation}
where $\dot{x}^{\mu}=\tfrac{dx^{\mu}}{d\lambda}$ and $\lambda$ parametrizes the curve $x^{\mu}(\lambda)$. Let us evaluate these equations for a trajectory with $\dot{\rho}=0$ and $x=0$, which represents a circular orbit. We find that the $\mu=x$ component of (\ref{eq:geoeq}) reads 

\begin{equation}\label{eq:xcirculargeodesics}
-\frac{\partial_x H_1 |_{x=0}}{2\rho_\pm^2}\, \dot t^2+\frac{2 M^2 \chi }{\rho_\pm^3}\partial_x H_2 |_{x=0}\,\dot t\,\dot \phi-\frac{\rho_\pm^3+2M^3\chi^2+M^2\chi^2\rho_\pm }{2\rho_\pm^3}\partial_x H_4|_{x=0}\, \dot \phi^2=0\ .
\end{equation}
In order for the truncation $x=0$ to be consistent, the left-hand-side should vanish \emph{independently} of the value of $\dot{t}$ and $\dot{\phi}$. This does not always happens, and the reason is precisely the presence of parity-breaking interactions, controlled by $\lambda_{\rm odd}$ and $\sin\theta_m$. Note that all the terms appearing in (\ref{eq:xcirculargeodesics}) are proportional to $\partial_x H_i |_{x=0}$. When the theory preserves parity, the solution possesses $\mathbb{Z}_{2}$ symmetry and the functions $H_{i}$ only contain even powers of $x$, so that $\partial_x H_i |_{x=0}=0$. On the contrary, the parity-breaking terms introduce odd powers of $x$ in the $H_{i}$ functions --- in particular terms linear in $x$ --- implying that  $\partial_x H_i |_{x=0}\neq0$. Thus, in such theories setting $x=0$ is not consistent and there are no orbits contained in the plane $x=0$ (probably there are no orbits contained in a plane at all, besides the radial geodesics at the axes $x=\pm1$).\footnote{In Ref.~\cite{Cardoso:2018ptl}, rotating black holes were studied in the presence of quartic-curvature corrections, including a parity-violating combination, and it was stated that this interaction does not have effects on equatorial geodesics. Apparently, the analysis of geodesics in that paper missed the fact that those geodesics are not permitted if the parity-violating term is activated. On the other hand, that analysis should be perfectly valid if the problematic term is removed.} For simplicity, from now on we set the parity-violating parameters to zero, $\lambda_{\rm odd}=\theta_m=0$, so that we can study equatorial geodesics. However, we believe that studying the geodesics in those theories is an interesting problem that should be addressed elsewhere.

Let us then focus on the remaining equations. When they are evaluated on $\dot{\rho}=0$ and $x=0$, the $\mu=t$ and $\mu=\phi$ components of the geodesic equations (\ref{eq:geoeq}) tell us that $\dot t=const$ and $\dot\phi=const$ and, consequently, the angular velocity $\omega\equiv d\phi/dt$ is also constant. On the other hand, the component $\mu=\rho$  gives an equation for $\omega$:

\begin{equation}\label{eq:rhocircular}
\Gamma^\rho_{\phi\phi}\omega^2+2\Gamma^\rho_{t\phi}\, \omega+\Gamma^\rho_{tt}=0\ ,
\end{equation}
where the Christoffel symbols are shown in Eq.~(\ref{eq:christ}). Finally, we take into account that for massless particles we have $g_{\mu\nu} \dot{x}^{\mu}\dot{x}^{\nu}=0$, that gives the following equation

\begin{equation}\label{eq:massshellcircular}
\left(1+H_4\right)\left(\rho^3+M^2 \chi^2\rho+2 M^3 \chi^2\right) \omega^2-4M^2\chi \left(1+H_2\right)\, \omega= \rho -2 M-\rho H_1 \ .
\end{equation}

Now, using the equations (\ref{eq:rhocircular}) and (\ref{eq:massshellcircular}) we can solve for $\rho$ and $\omega$. We get two solutions that we can express as the result in Einstein gravity plus corrections:

\begin{align}
\frac{\rho_{{\rm ph}\pm}}{M}&=2\left(1+\cos\left(\frac{2}{3}\arccos \left(\mp \chi\right)\right)\right)+\frac{\ell^4}{M^4}\left[\alpha_1^2 \Delta \rho_{{\rm ph}\pm} ^{(1)}+\alpha_2^2 \Delta \rho_{{\rm ph}\pm} ^{(2)}+\lambda_{\rm ev} \Delta \rho_{{\rm ph}\pm}^{(\rm ev)}\right] \ ,\\
&&\nonumber\\
M\omega_\pm&=\pm\left[\frac{1}{\sqrt{48\cos^4\left(\frac{1}{3}\arccos \left(\mp \chi\right)\right)+\chi^2}}+\frac{\ell^4}{M^4}\left(\alpha_1^2 \Delta \omega^{(1)}_\pm+\alpha_1^2 \Delta \omega^{(1)}_\pm+\lambda_{\rm{ev}} \Delta \omega^{(\rm ev)}_\pm \right) \right]\ ,\hskip1cm
\end{align}
The ``$+$'' solution corresponds to the prograde photon ring (the photons rotate in the same direction as the black hole), while the ``$-$'' solution represents the retrograde photon ring. We reproduce here the values of the coefficients $\Delta \omega^{(i)}_\pm$ expanded up to order $\chi^7$ in the spin

\begin{eqnarray}
\Delta \omega_\pm^{(1)}&=&\frac{4397}{65610 \sqrt{3}}\pm\frac{20596 \chi }{295245}+\frac{1028803 \chi ^2}{14467005 \sqrt{3}}\pm\frac{45262543 \chi ^3}{3906091350}-\frac{3685587061 \chi ^4}{328111673400 \sqrt{3}}\\
&&\mp\frac{110632797883 \chi ^5}{5413842611100}-\frac{910228742414947 \chi ^6}{17151053391964800 \sqrt{3}}\mp\frac{15449837941866829 \chi ^7}{401334649371976320}+\mathcal O\left(\chi^8\right)\ , \nonumber\\
\Delta \omega_\pm^{(2)}&=&\mp\frac{131 \chi }{5103}-\frac{11047 \chi ^2}{381024 \sqrt{3}}\mp\frac{9491513 \chi ^3}{1388832480}-\frac{19022279 \chi ^4}{925888320 \sqrt{3}}\\
&&\mp\frac{353193404087 \chi ^5}{23099061807360}-\frac{2452581602509 \chi ^6}{63522419970240 \sqrt{3}}\mp\frac{5958423964756267 \chi ^7}{222963694095542400}+\mathcal O\left(\chi^8\right)\ ,\nonumber \\
\Delta \omega_\pm^{(\rm ev)}&=&\frac{20}{2187 \sqrt{3}}\pm\frac{320 \chi }{19683}+\frac{26749 \chi ^2}{1928934 \sqrt{3}}\mp\frac{12967 \chi ^3}{104162436}-\frac{4415651 \chi ^4}{1249949232 \sqrt{3}}\\
&&\mp\frac{3101153 \chi ^5}{937461924}-\frac{33998483 \chi ^6}{6629195034 \sqrt{3}}\mp\frac{18127693795 \chi ^7}{6682228594272}+\mathcal O\left(\chi^8\right)\ ,\nonumber
\end{eqnarray}

while the coefficients $\Delta \rho_{{\rm ph}\pm} ^{(i)}$ are shown in Eq.~(\ref{eq:rhoph}) of the Appendix. However, $\rho_{{\rm ph}\pm}$ is a meaningless quantity, since $\rho$ does not have a direct interpretation as a radius. What we should really consider as the radius of the light rings is 
\begin{figure}[ht!]
\begin{center}
\includegraphics[scale=0.45]{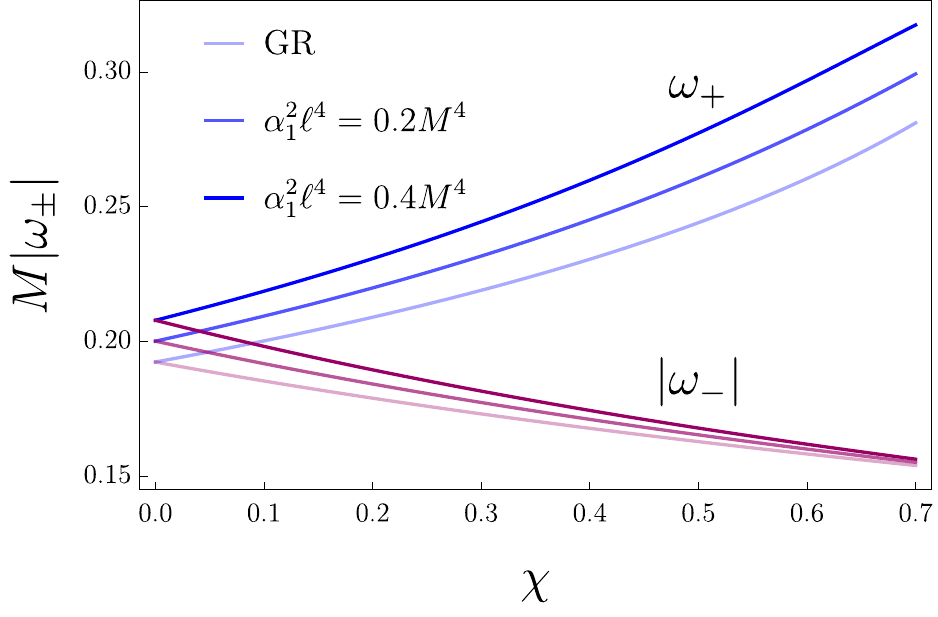} 
\includegraphics[scale=0.45]{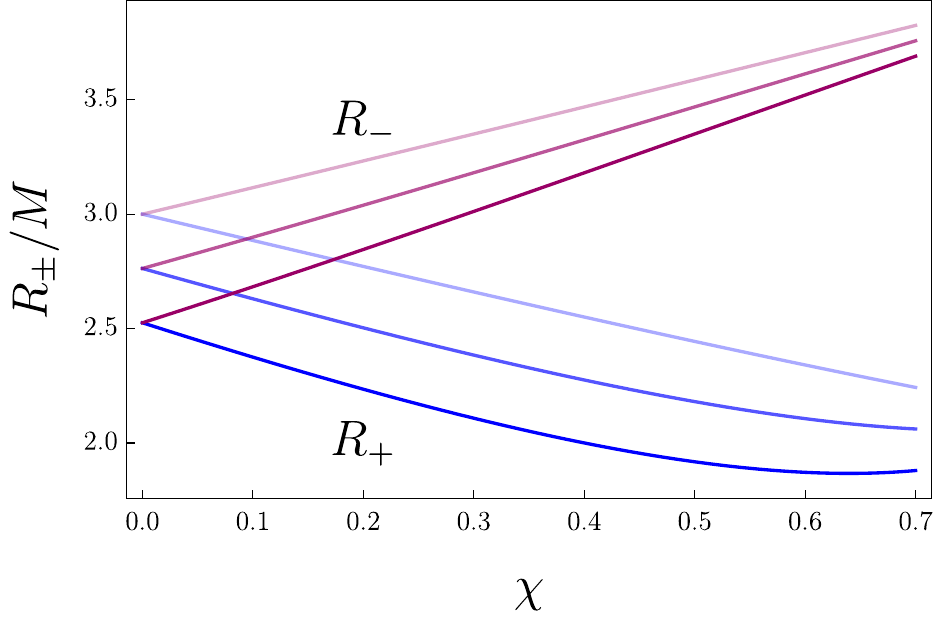}
\includegraphics[scale=0.45]{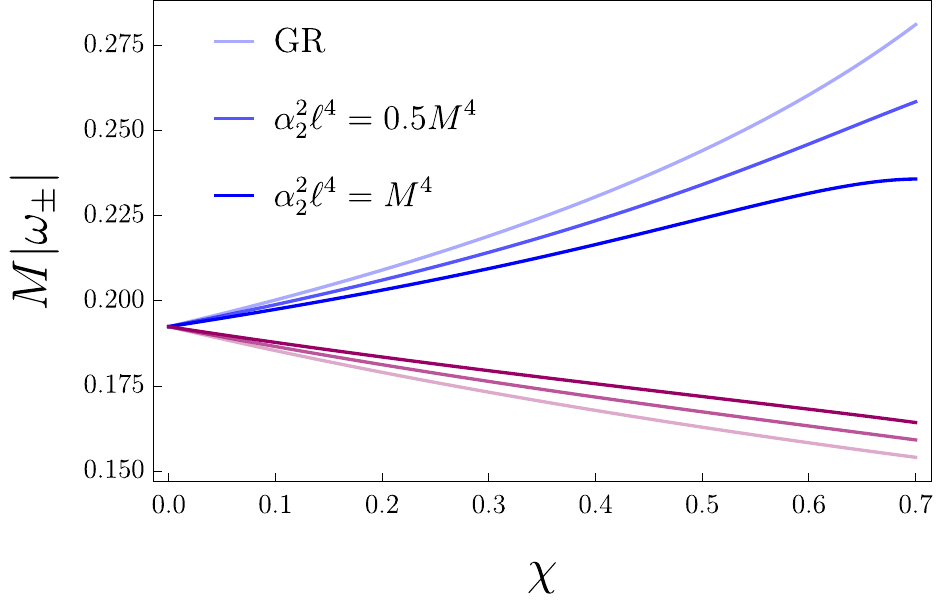}
\includegraphics[scale=0.45]{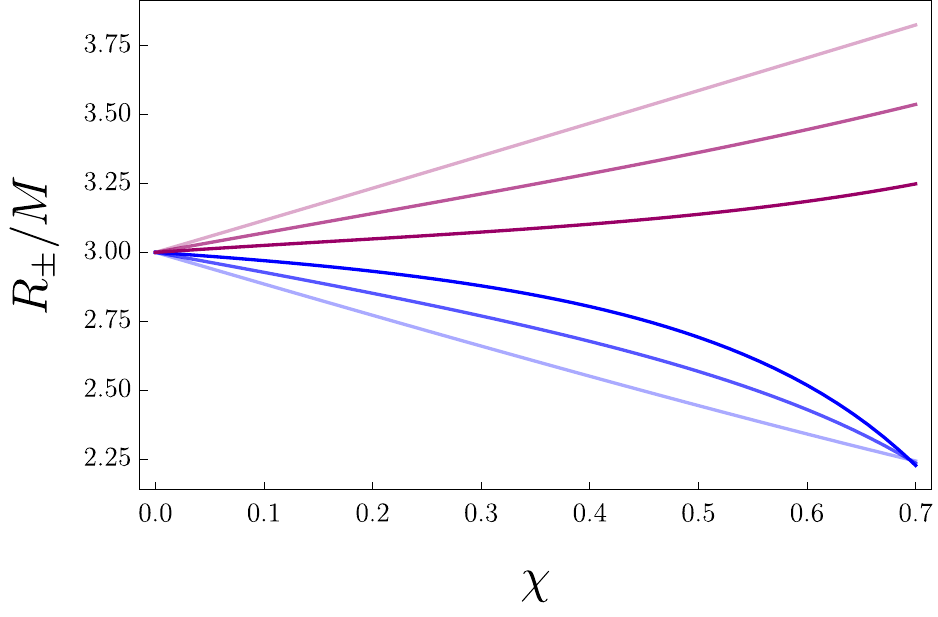}
\includegraphics[scale=0.45]{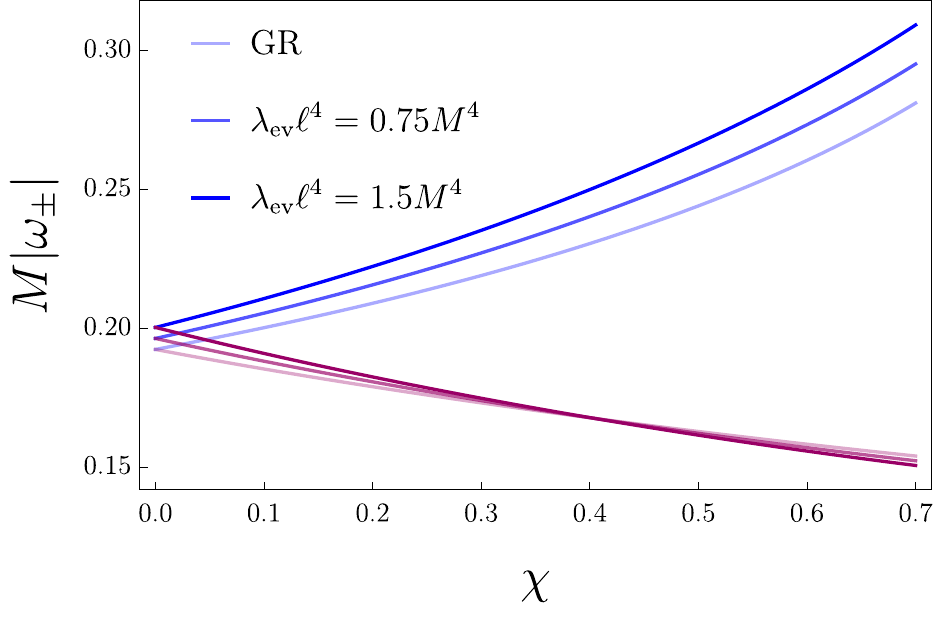}
\includegraphics[scale=0.45]{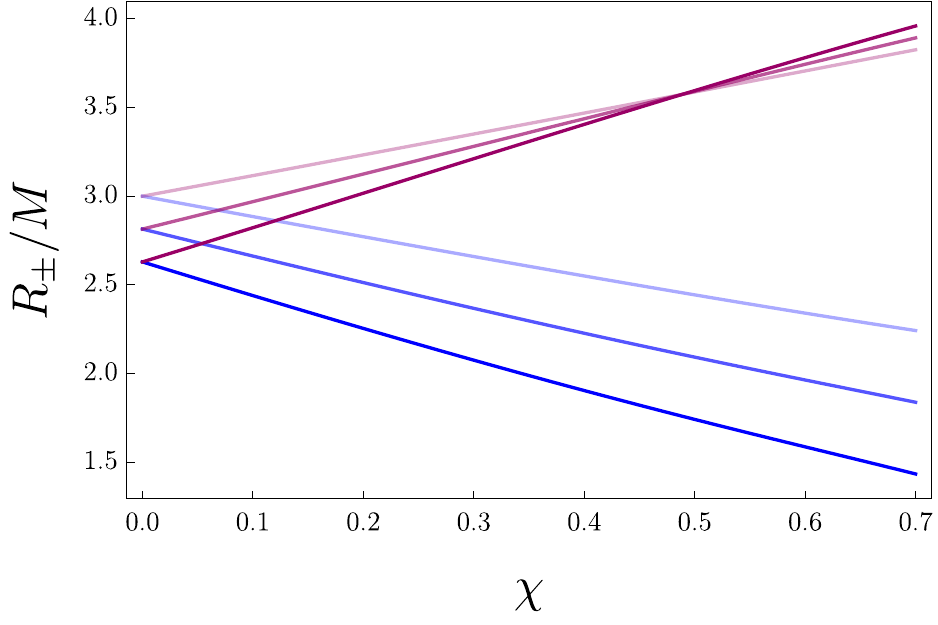}
\caption{Frequencies and radii of the light rings in parity-preserving theories. In blue we plot the quantity corresponding to the prograde orbit and in purple that corresponding to the retrograde one. In the left column we show the frequencies for different values of the couplings and compare them to GR. In the right column we plot the radii $R_{\pm}$ for the same values of the couplings. }
\label{fig:PhR}
\end{center}
\end{figure}

\begin{equation}
R_{\pm}=\sqrt{g_{\phi\phi}}\Big|_{x=0,\,  \rho=\rho_{{\rm ph}\pm}}\, .
\end{equation}
Since the light ring (more precisely, the photon sphere) determines the shape of the black hole shadow, this quantity give us information about the deformation of the shadow (near the equator) due to the corrections. On the other hand, $\omega_{\pm}$ is also an interesting quantity, since it is related to the time-scale of the response of the black hole when it is perturbed. In fact, there is a known quantitative relation between the orbital frequency of the light ring and the quasinormal frequencies of static black holes in the eikonal limit \cite{Cardoso:2008bp,Konoplya:2017wot}. Although the relation probably does not extend to the rotating case, we do expect that $\omega_{\pm}$ captures qualitatively the (real) frequencies of the first quasinormal modes. Hence, we can use $\omega_{\pm}$ in order to perform a first estimation of the effects of the corrections on the black hole quasinormal frequencies.

\begin{figure}[ht!]\label{fig:Ratioomega}
\begin{center}
\includegraphics[scale=0.6]{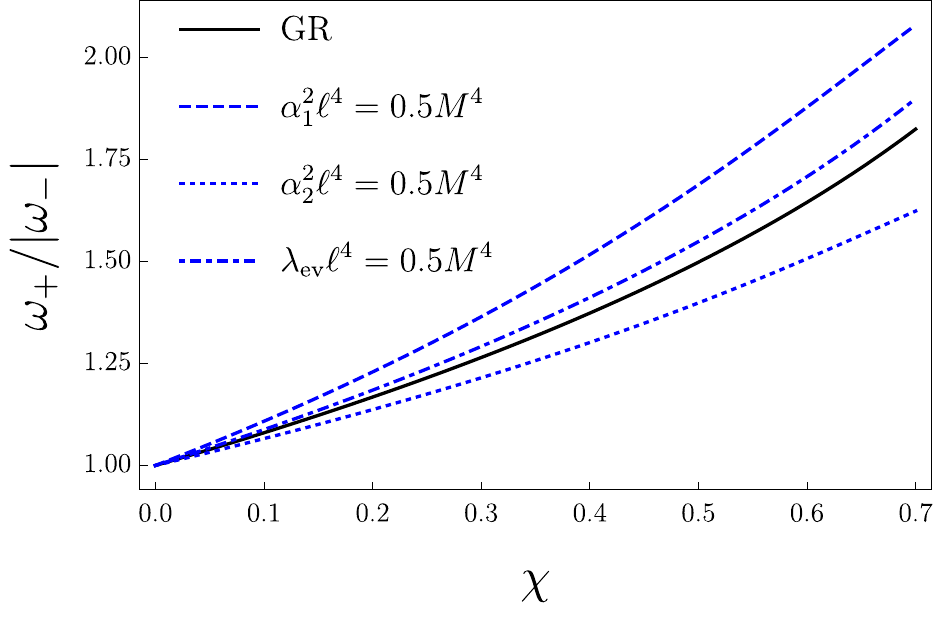} 
\caption{Ratio of light ring frequencies $\omega_{+}/|\omega_{-}|$ in several theories. }
\end{center}
\end{figure}

In Fig.~\ref{fig:PhR} we show the frequencies $\omega_{\pm}$ and the radius $R_{\pm}$ for several values of the higher-order couplings and we compare them to the GR values. These plots were computed using an expansion up to order $\chi^{14}$ of both quantities. We note some characteristic features for each correction. In the case of the quadratic correction controlled by $\alpha_{1}$ we see that both $\omega_{+}$ and $|\omega_{-}|$ increase with respect to the Einstein gravity values. On the other hand, for $\alpha_{2}$ corrections we observe that $\omega_{+}$ decreases while $|\omega_{-}|$ increases so that the difference between the two frequencies is reduced. As for the cubic correction, it increases or decreases $\omega_{+}$ if $\lambda_{\rm ev}>0$ or $\lambda_{\rm ev}<0$ respectively. It has little effect on $\omega_{-}$, but interestingly the sign is different depending on the value of $\chi$. However, in order to characterize deviations from GR it is more useful to look at the ratio of frequencies $\omega_{+}/|\omega_{-}|$, that we show for a few cases in Fig.~\ref{fig:Ratioomega}. In GR, this quantity is completely determined by the spin parameter $\chi$, but in these theories it also depends on the combination $\ell^4/M^4$. Thus, if one is able to determine $\chi$ by other means, the ratio $\omega_{+}/|\omega_{-}|$ can be used to constrain the higher-order couplings.


\subsection{Scalar hair}
So far, we have only focused on the geometry, but one of the most remarkable features of the solutions of (\ref{Action}) is that the scalar fields acquire a non-trivial profile. In fact, the coupling of the scalars to the quadratic curvature invariants prevent these from being constant whenever the invariants are non-vanishing. A slightly less trivial fact ---though also well-known \cite{EdGB,Kanti:1995vq,Torii:1996yi,Alexeev:1996vs,Mignemi:1992pm}--- is that the scalars actually get a charge that can be measured at infinity.  More precisely, the scalar $\phi_1$ gets a charge $Q$ while $\phi_2$ gets dipolar moment $P$, that can be identified by looking at the asymptotic behaviour\footnote{The reason for the negative sign in front of $Q$ is that the charge is conventionally defined as
\begin{equation}
Q=\frac{1}{4\pi}\int d^2\Sigma^{\mu}\partial_{\mu}\phi_1\, ,
\end{equation}
where the integral is taken on spatial infinity.
 }
\begin{equation}
\phi_1\sim -\frac{Q}{\rho}\, , \quad \phi_2\sim P \frac{x}{\rho^2}\, .
\end{equation}
Using the solution in powers of $\chi$ that we have found, we obtain

\begin{eqnarray}\label{Qcharge}
\hskip-0.5cm Q&=&-\frac{\alpha_1\ell^2}{M}\left(2-\frac{\chi ^2}{2}-\frac{\chi ^4}{4}-\frac{5 \chi ^6}{32}-\frac{7 \chi ^8}{64}-\frac{21
   \chi ^{10}}{256}-\frac{33 \chi ^{12}}{512}-\frac{429 \chi ^{14}}{8192}+\ldots \right)\, ,\\
\label{Pcharge}
\hskip-0.5cm P&=&\alpha_2\ell^2\cos\theta_m\left(\frac{5 \chi }{2}-\frac{\chi ^3}{4}-\frac{3 \chi ^5}{32}-\frac{3 \chi ^7}{64}-\frac{7 \chi ^9}{256}-\frac{9 \chi^{11}}{512}-\frac{99 \chi ^{13}}{8192}-\frac{143 \chi^{15}}{16384}+\ldots\right)\, .
\end{eqnarray}

\noindent
Remarkably enough, it is possible to guess the general term of these series and to sum them. We find
\begin{align}
Q&=-\frac{4\alpha_1\ell^2}{M}\frac{\sqrt{1-\chi^2}}{1+\sqrt{1-\chi^2}}\, ,\\
P&=\alpha_2\ell^2\cos\theta_m\frac{2\chi(5-8\chi^2+4\chi^4)}{2-3\chi^2+2\chi^4+2(1-\chi^2)^{3/2}}\, .
\end{align}

\noindent
One can check that the series expansion of these expressions matches those in (\ref{Qcharge}) and (\ref{Pcharge}), so they are most likely correct, and they give the exact value of the charges as functions of the spin. In the case of the charge $Q$, we also check that it agrees with previous results \cite{Yunes:2016jcc,Berti:2018cxi,Prabhu:2018aun}.

Despite having non-vanishing scalar charge, we note however that the solution has no ``hair'', because the charge is completely fixed in terms of the mass and the spin. In other words, the charge cannot be arbitrary.  The reason is that the previous value of the charge is the only one compatible with the requirement of regularity of the solution at the horizon. If we introduce, by hand, any other value of the scalar charge, the resulting solution would develop a singularity at the horizon.

As we mentioned in Sec.~\ref{section:effectivetheory}, in the context of String Theory $\phi_1$ is related to the dilaton, while $\phi_2$ is the axion. In Appendix~\ref{ap:het} we show that the precise identification with the effective action of Heterotic Superstring Theory is $\alpha_1=-\alpha_2=-1/8$ $,\ell^2=\alpha'$, $\varphi=\varphi_{\infty}+\frac{\phi_1}{2}$. Then, the dilaton charge $D$ associated to a rotating black hole reads, at leading order in $\alpha'$,

\begin{equation}
D=\frac{\alpha'}{4M}\frac{\sqrt{1-\chi^2}}{1+\sqrt{1-\chi^2}}\, .
\end{equation}

This can be expressed in a very appealing form as $D=\alpha' \pi T$, where $T=\kappa/(2\pi)$ is the Hawking temperature of the black hole. It turns out that this intriguing connection between asymptotic charge and temperature (or surface gravity) is not a coincidence, but a general phenomenon that happens in EdGB theory with linear coupling \cite{Prabhu:2018aun}.

The field $\phi_{2}$ gets a dipolar moment instead of charge because it is sourced by the parity-violating Pontryagin density --- $\phi_{2}$ is essentially the scalar that appears in dynamical Chern-Simons gravity \cite{Alexander:2009tp}. When the spin vanishes we get $P=0$, and in fact, $\phi_{2}=0$, so that this kind of scalar hair is not present in spherically symmetric solutions \cite{Wagle:2018tyk}.

Besides the asymptotic behaviour, it is also interesting to study the profile of the scalar fields as a function of $x$. The field $\phi_{2}$ is odd under the $\mathbb{Z}_{2}$ transformation $x\rightarrow-x$, while $\phi_{1}$ is even only for $\theta_{m}=n\pi$, $n\in \mathbb{Z}$. For other values of $\theta_{m}$, $\phi_{1}$ does not have a defined parity, which is a manifestation of the breaking of the $\mathbb{Z}_{2}$ symmetry. For instance, when evaluated on the horizon, $\rho=\rho_{+}$, the field $\phi_{1}$ is given by
\begin{equation}
\begin{aligned}
&\phi_{1}\Big|_{\rho_{+}}=\frac{\ell^2}{M^2}\left[\alpha _1 \left(\frac{11}{6}+\left(\frac{5}{16}-\frac{59 x^2}{40}\right) \chi
   ^2+\left(\frac{11}{160}-\frac{117 x^2}{80}+\frac{167 x^4}{224}\right) \chi^4+\ldots\right)\right.\\
  &\left. +\alpha _2 \sin \left(\theta _m\right) \left(\frac{29 x \chi
   }{16}+\left(\frac{187 x}{160}-\frac{13 x^3}{12}\right) \chi ^3+\left(\frac{67
   x}{80}-\frac{629 x^3}{448}+\frac{251 x^5}{512}\right) \chi ^5+\ldots\right)\right]\, .
\end{aligned}
\end{equation}

\begin{figure}[ht!]
\begin{center}
\includegraphics[scale=0.3]{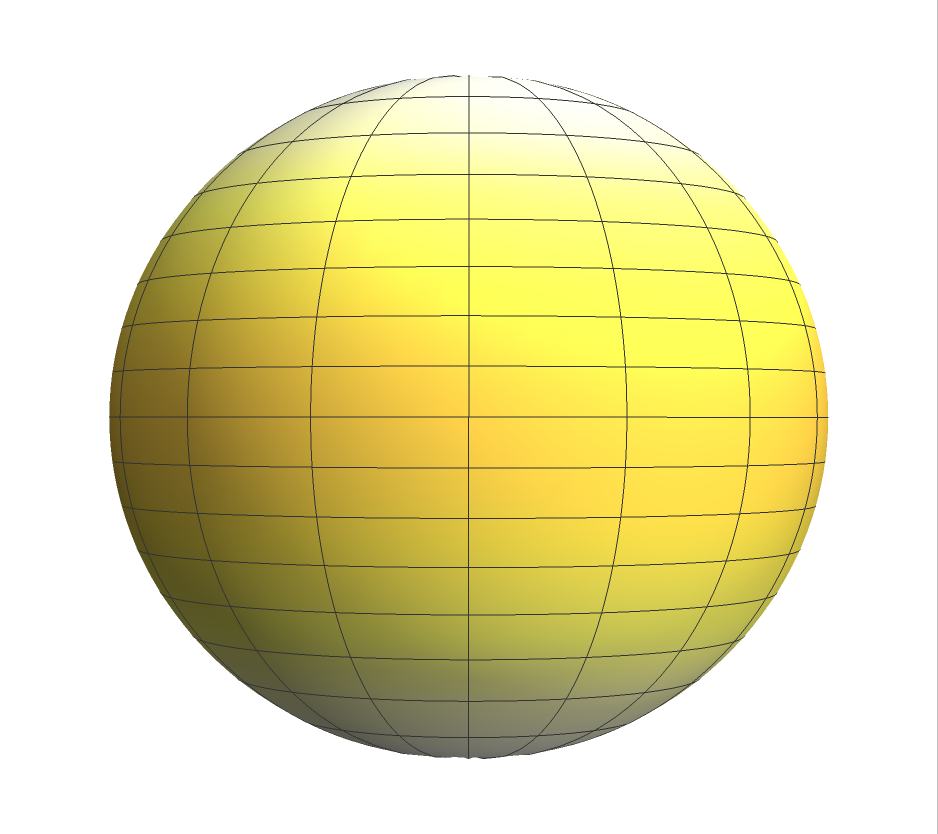} 
\includegraphics[scale=0.3]{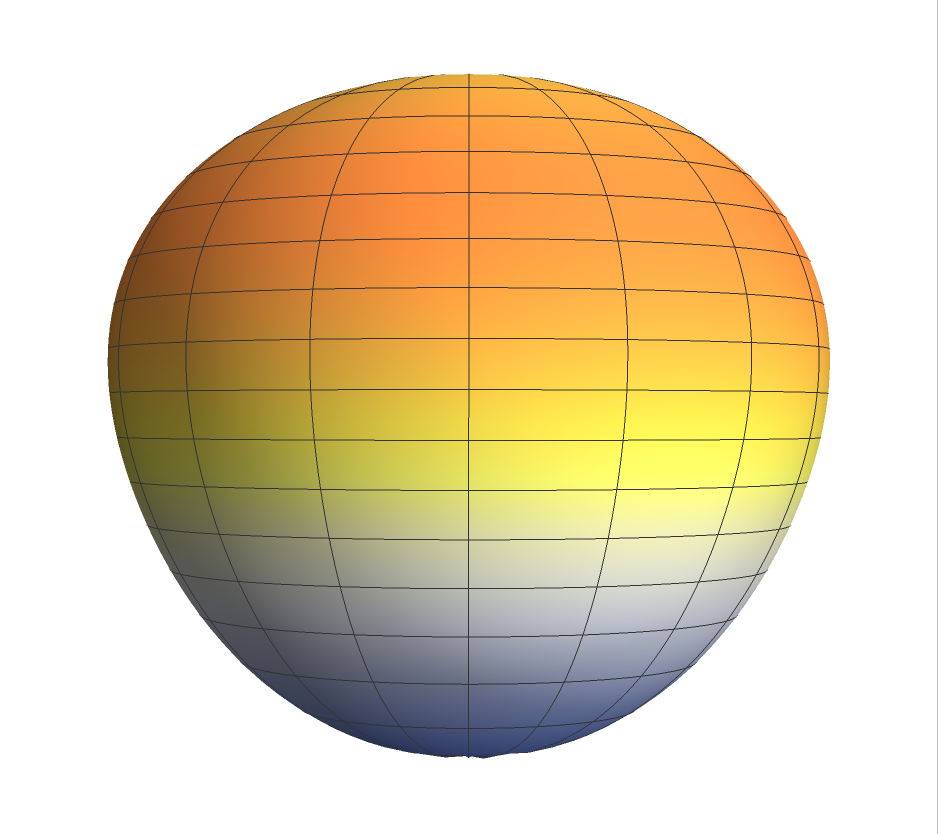} 
\includegraphics[scale=0.34]{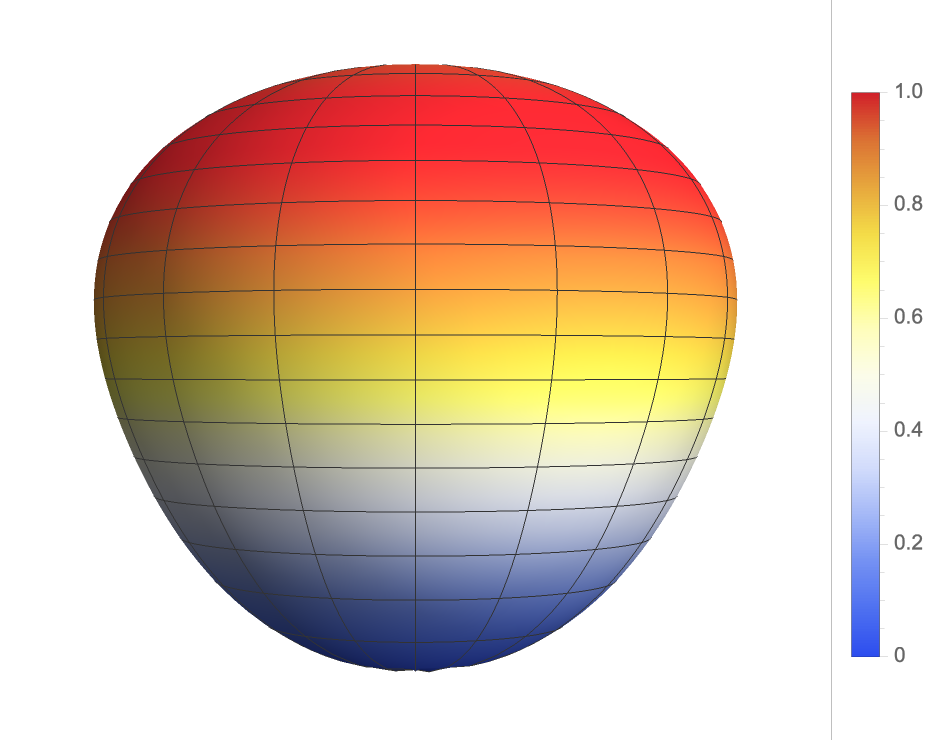} 
\caption{Profile of the scalar field $\phi_{1}$ on the horizon. We show here the case for $\ell^2\alpha_{1}=\ell^2\alpha_{2}=0.4M^2$ and for a parity-breaking phase  $\theta_{m}=0,\pi/4,\pi/2$, from left to right.  }
\label{fig:scalarhorizon}
\end{center}
\end{figure}

We only show here a few terms in the $\chi$-expansion for definiteness, but using the solution up to order $\chi^{14}$ we can determine accurately the profile of $\phi_{1}$ on the horizon for high values of $\chi$. In Fig.~\ref{fig:scalarhorizon} we plot $\phi_{1}$ as a colormap on the horizon for $\chi=0.65$, and $\ell^2\alpha_{1}=\ell^2\alpha_{2}=0.4 M^2$. From left to right, the parity-breaking parameter $\theta_{m}$ takes the values  $\theta_{m}=0, \pi/4, \pi/2$. For $\theta_{m}=0$ the profile is $\mathbb{Z}_2$-symmetric and has a mild variation, taking a maximum value at the equator. When $\theta_{m}\neq 0$, we observe the deformation of the horizon that we reported in Sec.~\ref{sec:horizon}, plus a ``polarization'' of the scalar field, that develops a maximum at the north pole and a minimum at the south one.

Interestingly enough, the scalar profile provides an intuitive picture of the deformation of the horizon. The northern ``hemisphere'' grows due to the $\theta_{m}$ correction, while the southern one has a smaller size, and this coincides with the fact that the scalar field is ``concentrated'' on the northern hemisphere, producing a larger energy density there. Thus, the horizon is enlarged in the region that has a greater scalar energy density.

\section{Conclusions}\label{conclusions}
In this work, we have computed the modified Kerr black hole solution in the effective theory (\ref{Action}), which provides a general framework to study the leading-order deviations from GR associated to higher-derivative corrections.  We expressed the solution as a power series in the spin parameter $\chi$ and we showed that including enough terms we get an accurate result even for large values of $\chi$. In this text we have worked with an expansion up to order $\chi^{14}$, that provides a good approximation for $\chi\le0.7$, but with the software we supply it should be possible to compute the series to higher orders in $\chi$ and to get a solution valid for $\chi\sim 1$. Although the series expansion involves lengthy expressions, it has obvious advantages with respect to numerical solutions, since it allows for many analytic computations, as we have illustrated in Section~\ref{sec:prop}. 

We have studied some of the most remarkable properties of these rotating black holes, with special emphasis on the horizon. We have shown that the corrections modify the shape of the horizon, and in particular, that parity-violating interactions break the $\mathbb{Z}_{2}$ symmetry of Kerr's black hole. We observed the same phenomenon in the case of the ergosphere, and, as far as we know, Fig.~\ref{fig:ergodd} contains the first example of ergospheres without $\mathbb{Z}_{2}$ symmetry.

In addition, we have computed some quantities that were disregarded in previous studies on rotating black holes in modified gravity. In particular, we have obtained the surface gravity of these black holes, from which one obtains the Hawking temperature according to $T=\tfrac{\kappa}{2\pi}$, in natural units. Thus, from the results in Sec.~\ref{sec:horizon} we conclude that the quadratic curvature terms with non-minimally coupled scalars always increase the temperature of black holes, for any value of the spin. On the other hand, the cubic curvature term raises or lowers the temperature depending on the sign of the coupling $\lambda_{\rm ev}$ and on the value of the spin $\chi$. The modification of Hawking temperature may have important consequences for the evaporation process of black holes \cite{PabloPablo4}, and it would be interesting to extend these results by obtaining the value of the temperature non-perturbatively in the coupling and in the spin. 

As a first step in analyzing the geodesics of the modified Kerr black holes, we studied the photon rings, \textit{i.e.} circular light-like geodesics on the equatorial plane. Remarkably, we have found that for parity-breaking theories there are no such orbits: indeed, there are no orbits contained in the equatorial plane because there is no equator at all. Thus, we computed the photon rings for parity-even theories, characterizing the deviations from GR.

Finally, we also noticed the non-trivial scalar fields, and we were able to obtain exact formulas for the monopole and dipole charges. We also computed the profile of the scalar $\phi_{1}$ on the horizon and we observed how the $\mathbb{Z}_{2}$ symmetry is broken when the parity-violating phase $\theta_{m}$ is activated.

Let us now comment on some possible extensions and future directions. As we already mentioned, it would be interesting to obtain the solution for even larger values of the angular momentum, since the effects of rotation are more drastic when the spin is close to the extremal value. It would also be more or less straightforward to extend the results of this paper to other theories that we did not consider here, particularly the quartic ones in \cite{Endlich:2017tqa,Cardoso:2018ptl}. Another possible extension would entail adding a mass term for the scalars in (\ref{Action}), though this would considerably increase the difficulty of finding an analytic solution. 

We have studied some basic properties of the modified Kerr black holes, but the next natural step is to derive observational signatures of these spacetimes. Analyzing the geodesics of these black holes is an interesting task, as one would potentially observe effects coming from the loss of integrability or from the absence of $\mathbb{Z}_{2}$ symmetry in parity-violating theories. Once the geodesics are determined, one could study gravitational lensing or the black hole shadow, similarly as done  \textit{e.g.} in \cite{Amarilla:2010zq,Cunha:2015yba}. However, the most sensitive quantity to the corrections --- and that we expect to measure in the near-future thanks to gravitational wave detectors \cite{Berti:2018vdi} --- is the quasinormal mode spectrum of the black hole. Hence, determining the quasinormal modes and frequencies of the rotating black holes presented here is a very relevant task, for which one needs to perform perturbation theory. Analyzing electromagnetic and scalar perturbations for test fields (or for one of the scalars contained in the model (\ref{Action})) should not involve outstanding complications. On the other hand, the study of gravitational perturbations presents a more challenging problem, since one would need to derive the analogous of the Teukolsky equation \cite{Teukolsky:1973ha} for the modified Kerr black holes. We feel that this problem should be addressed in future work. 

The observation of deviations from General Relativity in astrophysical black holes would represent a tremendous breakthrough that would revolutionize our current understanding of gravity. But even if this is not the case, studying the effects of higher-derivative corrections on black hole geometries provides us with a rich source of new physics, and allows us to learn about new phenomena that could be inherent to an underlying UV-complete theory of gravity.

\acknowledgments
We wish to thank Vitor Cardoso and Leo Stein for useful comments. The work of PAC is funded by Fundaci\'on la Caixa through a ``la Caixa - Severo Ochoa'' International pre-doctoral grant. AR was supported by a ``Centro de Excelencia Internacional UAM/CSIC" FPI pre-doctoral grant and by a grant from the ``Residencia de Estudiantes". Both authors were further supported by the MINECO/FEDER, UE grant FPA2015-66793-P, and from the ``Centro de Excelencia Severo Ochoa" Program grant  SEV-2016-0597.

\appendix

\newpage
\section{Higher-derivative gravity with dynamical couplings}\label{appendix:effectivetheory}

In this appendix, we are going to motivate our choice of effective action (\ref{Action}). Since our goal is to parametrize the leading corrections to vacuum solutions, we will start writing down an action  including all possible curvature invariants containing at most $2n$ derivatives, and then we will discuss which terms are going to induce corrections. By dimensional analysis, a term with $2n$ derivatives will be multiplied by a factor $\ell^{2n-2}$, where $\ell$ is some length scale that we will assume to be small as compared with the size of the black hole, i.e.  $GM>>\ell$. It is clear then that the effective action can be always written as 

\begin{equation}\label{eq:effectiveaction1}
S=\frac{1}{16\pi G}\int d^4x \sqrt{-g} R+ \sum_{n\ge 2}\frac{\ell^{2n-2}}{16\pi G}S^{(2n)}\ ,
\end{equation}
where in $S^{(2n)}$ we will include the terms with $2n$ derivatives.

Up to four-derivative terms, we can add the following terms to the Einstein-Hilbert action 

\begin{equation}\label{eq:effaction4der1}
S^{(4)}=\int d^4x \sqrt{-g} \left[\alpha_1 \mathcal X_4+\alpha_2 R_{\mu\nu\rho\sigma}\tilde R^{\mu\nu\rho\sigma}+\alpha_3 R_{\mu\nu}R^{\mu\nu}+\alpha_4 R^2\right]\ .
\end{equation}
It turns out that, if the coefficients $\alpha_i$ are constants, none of these terms will modify a vacuum solution of GR at $\mathcal O\left(\ell^2\right)$. The reasons are the following: both $\mathcal X_4$ and  $R_{\mu\nu\rho\sigma}\tilde R^{\mu\nu\rho\sigma}$ are topological terms and therefore do not contribute to the equations of motion. The last two terms are quadratic in Ricci curvature, which means that their contributions to the equations of motion will vanish when evaluated on a GR vacuum solution. In other words, Ricci flat metrics are also solutions of EG plus four-derivative terms. 

However, we can think of adding dynamical couplings, i.e. promoting $\alpha_i \rightarrow \alpha_i f_i \left(\phi^1, \dots, \phi^N\right)$, where $\left\{\phi^A\right\}_{A=1, \dots, N}$ is a set of $N$ massless scalars\footnote{A natural extension of this work would be to include a non-vanishing scalar potential.}. To this aim, we have to include also a kinetic term for them in the action (\ref{eq:effectiveaction1}) so that it becomes 

\begin{equation}\label{eq:effectiveaction2}
S=\frac{1}{16\pi G}\int d^4x \sqrt{-g} \left[R-\frac{1}{2}\mathcal M_{AB}(\phi)\partial_\mu\phi^A \partial^\mu \phi^B\right]+ \sum_{n\ge 2}\frac{\ell^{2n-2}}{16\pi G}S^{(2n)}\ ,
\end{equation}
where $\mathcal M_{AB}(\phi)$ is the (symmetric) matrix that characterizes the non-linear $\sigma$-model. However, as we check \textit{a posteriori}, the scalars will be excited by the higher-derivative terms at order $\ell^2$. Then, we only need to include terms that are at most quadratic in the scalars, which contribute to the gravitational equations at order $\ell^4$. Thus, we can expand $\mathcal M_{AB}$ in a Taylor series and only the constant term will contribute at leading order. By means of a redefinition of the scalar fields, this constant term can always be taken to be the identity matrix: $\mathcal M_{AB}|_{\phi^A=0}=\delta_{AB}$. On the other hand, the generalized action for the four-derivative terms, that we denote again by $S^{(4)}$, is

\begin{equation}
S^{(4)}=\int d^4x \sqrt{-g} \left[\alpha_1 f_1\left(\phi\right) \mathcal X_4+\alpha_2f_2\left(\phi\right) R_{\mu\nu\rho\sigma}\tilde R^{\mu\nu\rho\sigma}+\alpha_5 f_5\left(\phi\right)\nabla^2 R\right]\ ,
\end{equation}
where we already neglected the $R_{\mu\nu}R^{\mu\nu}$ and $R^2$ terms, that do not induce corrections at leading order, and we have now added the term $\alpha_5 f_5\left(\phi\right)\nabla^2 R$  that was neglected in (\ref{eq:effaction4der1}) because in the non-dynamical case it is just a total derivative. In the dynamical case, this term can be written (ignoring total derivatives) as $\alpha_5 \nabla^2 f_5\left(\phi\right)\, R$ and, it is possible to prove that it can always be eliminated, at leading order, by a field redefinition of the metric, so that we can set $\alpha_5=0$. 

Indeed it is possible to show that if $K_{\mu\nu}$ is a symmetric tensor and we consider a term  $\ell^4 \, K_{\mu\nu}R^{\mu\nu}$ in the action,  the contribution to the field equations is trivial since it can always be eliminated by a field redefinition: $g_{\mu\nu}\rightarrow g_{\mu\nu}-\ell^4 \hat{K}_{\mu\nu}$, where $\hat{K}_{\mu\nu}=K_{\mu\nu}-\frac{1}{2}g_{\mu\nu} K^{\alpha}_{\,\,\,\, \alpha}\, .$
To show this, let us compute the contribution of this term to the field equations. Passing this contribution to the right-hand-side of the equations. it can be written as an effective energy-momentum tensor 

\begin{equation}
T_{\mu\nu}=\ell^4 \left[\nabla^\rho \nabla_{(\mu}K_{\nu) \rho}-\frac{1}{2}\nabla^2K_{\mu\nu}-\frac{1}{2}g^{(0)}_{\mu\nu}\nabla_\rho\nabla_\sigma K^{\rho\sigma}\right]+\dots\ ,
\end{equation}
where the dots indicate other possible contributions that vanish when evaluated on the zeroth-order Ricci-flat metric. Then by comparison with (\ref{eq:linearizedEE}), it is clear that the corrected Einstein equation is solved by $g_{\mu\nu}=g^{(0)}_{\mu\nu}+\ell^4\hat{K}_{\mu\nu}+\dots$, being $\hat{K}_{\mu\nu}$ trace-reversed with respect to $K_{\mu\nu}$. Since the equation is integrable, it is equivalent to preforming a field redefinition, so this kind of terms do not really contain new physics. We can use this result to demonstrate that other type of terms such as $\phi R$ or $G^{\mu\nu}\nabla_{\mu}\phi\nabla_{\nu}\phi$, that appear for instance in Horndeski theories \cite{Horndeski:1974wa}, can be also removed by a field redefinition.

Let us now analyze the couplings $f_1(\phi)$ and $f_2(\phi)$. The first  we can do is to expand the functions around $\phi^i=0$ and neglect $ \mathcal O\left(\phi^2\right)$ terms, which is equivalent to neglect $\mathcal O\left(\ell^6\right)$ corrections in the metric. Thus, $f_i=a_i+b_{iB}\phi^B+\mathcal O\left(\phi^2\right) $ and, for the same reasons exposed above, the constant coefficients $a_i$ can be neglected. Finally, observe that we still have the freedom to perform a $\mathrm{SO}\left(N\right)$ rotation of the scalars that leaves invariant the kinetic terms.  Using this freedom, up to global factors that can be reabsorbed in a redefinition of $\alpha_1$ and $\alpha_2$, we can always choose

\begin{equation}
f_1=\phi^1\,\ , \qquad f_2=\phi^2\cos\theta_m +\phi^1\sin\theta_m \ .
\end{equation}
This implies that the theory contains at most two active scalars.  In summary, for our purposes the action (\ref{eq:effectiveaction2}) reduces to 

\begin{equation}\label{eq:final2and4der}
\begin{aligned}
S=&\frac{1}{16\pi G}\int d^4x \sqrt{-g} \left\{R-\frac{1}{2}\left(\partial\phi^1\right)^2-\frac{1}{2}\left(\partial\phi^2\right)^2\right.\\
&\left.+\ell^2\left[\alpha_1 \phi^1 \mathcal X_4+\alpha_2\left(\phi^2\cos\theta_m +\phi^1\sin\theta_m \right) R_{\mu\nu\rho\sigma}\tilde R^{\mu\nu\rho\sigma}\right]\right\}+ \sum_{n\ge 3}\frac{\ell^{2n-2}}{16\pi G}S^{(2n)}\ .
\end{aligned}
\end{equation}

Then, corrections to vacuum solutions due to these curvature-squared terms will be parametrized by three parameters: $\alpha_1$, $\alpha_2$ and $\theta_m$. These terms will induce $\mathcal O\left(\ell^4\right)$ corrections in the metric of the solution, since the scalars will be of order $\mathcal O\left(\ell^2\right)$. Therefore, these corrections are equally important to those coming from the six-derivative terms (with constant couplings), which will also induce $\mathcal O\left(\ell^4\right)$ corrections in the metric. Since our goal is to parametrize the leading corrections to vacuum solutions in the most general way possible, we shall also include them.

The most general parity-invariant action formed with curvature invariants with six derivatives is 

\begin{equation}\label{eq:actioncubicterms}
\begin{aligned}
S^{(6)}=\int d^4x\sqrt{|g|}&\left\{\lambda_1R_{\mu\ \nu}^{\ \rho \ \sigma}R_{\rho\ \sigma}^{\ \delta \ \gamma}R_{\delta\ \gamma}^{\ \mu \ \nu}+\lambda_2 R_{\mu\nu }^{\ \ \rho\sigma }R_{\rho\sigma }^{\ \ \delta\gamma }R_{\delta\gamma }^{\ \ \mu\nu }+\lambda_3 R_{\mu\nu\rho\sigma }R^{\mu\nu\rho }_{\ \ \ \delta}R^{\sigma \delta}\right.\\
&\left.+\lambda_4R_{\mu\nu\rho\sigma }R^{\mu\nu\rho\sigma }R+\lambda_5R_{\mu\nu\rho\sigma }R^{\mu\rho}R^{\nu\sigma}+\lambda_6R_{\mu}^{\ \nu}R_{\nu}^{\ \rho}R_{\rho}^{\ \mu}+\lambda_7R_{\mu\nu }R^{\mu\nu }R\right.\\
&\left.+\lambda_8R^3+\lambda_9 \nabla_{\sigma}R_{\mu\nu} \nabla^{\sigma}R^{\mu\nu}+\lambda_{10}\nabla_{\mu}R\nabla^{\mu}R\right\}\, .
\end{aligned}
\end{equation}
There are other six-derivative terms that could be added, such as $\nabla^{\alpha}\nabla^{\beta}R_{\mu\alpha\nu\beta}R^{\mu\nu}$ and $\nabla_{\alpha}R_{\mu\nu\rho\sigma}\nabla^{\alpha}R^{\mu\nu\rho\sigma}$, but these can be reduced to a combination of the terms included in the action. In addition, not all the terms in the previous action are linearly independent. In four dimensions we have two constraints that can be expressed as

\begin{equation}\label{eq:cubic-constraints}
R_{[\mu_1\mu_2 }^{\ \ \ \ \ \mu_1\mu_2 }R_{\mu_3\mu_4 }^{\ \ \ \ \ \mu_3\mu_4 }R_{\mu_5\mu_6] }^{\ \ \ \ \ \mu_5\mu_6 }=0\, ,\quad R_{[\mu_1\mu_2 }^{\ \ \ \ \ \mu_1\mu_2 }R_{\mu_3\mu_4 }^{\ \ \ \ \ \mu_3\mu_4 }R_{\mu_5] }^{\ \ \mu_5}=0\, .
\end{equation}
The first of these constraints actually corresponds to the vanishing of the cubic Lovelock density, $\mathcal{X}_6=0$. These relations allow us to express the terms proportional to $\lambda_1$ and $\lambda_3$ as a combination of the rest of the terms since (\ref{eq:cubic-constraints}) can be rewritten as

\begin{eqnarray}
R_{\mu\ \nu}^{\ \rho \ \sigma}R_{\rho\ \sigma}^{\ \delta \ \gamma}R_{\delta\ \gamma}^{\ \mu \ \nu}&=&\frac{1}{2} R_{\mu\nu }^{\ \ \rho\sigma }R_{\rho\sigma }^{\ \ \delta\gamma }R_{\delta\gamma }^{\ \ \mu\nu } -3R_{\mu\nu\rho\sigma }R^{\mu\nu\rho }_{\ \ \ \delta}R^{\sigma \delta}+\frac{3}{8}R_{\mu\nu\rho\sigma }R^{\mu\nu\rho\sigma }R\\
&&+3R_{\mu\nu\rho\sigma }R^{\mu\rho}R^{\nu\sigma}+2R_{\mu}^{\ \nu}R_{\nu}^{\ \rho}R_{\rho}^{\ \mu}-\frac{3}{2}R_{\mu\nu }R^{\mu\nu }R+\frac{1}{8}R^3\ ,\nonumber\\
R_{\mu\nu\rho\sigma }R^{\mu\nu\rho }_{\ \ \ \delta}R^{\sigma \delta}&=&2\left(R_{\mu\nu\rho\sigma }R^{\mu\rho}R^{\nu\sigma}+R_{\mu}^{\ \nu}R_{\nu}^{\ \rho}R_{\rho}^{\ \mu}-R_{\mu\nu }R^{\mu\nu }R\right)\\
&&+\frac{1}{4}\left( R_{\mu\nu\rho\sigma }R^{\mu\nu\rho\sigma }R+R^3\right)\ .\nonumber
\end{eqnarray}
Hence, we can always set $\lambda_1=\lambda_3=0$. The remaining terms, except those controlled by $\lambda_2$ and $\lambda_4$ are at least quadratic in Ricci curvature and do not induce corrections on Ricci-flat metrics, so we can ignore them: $\lambda_5=\lambda_6=\lambda_7=\lambda_8=\lambda_9=\lambda_{10}=0$. As already discussed, the term proportional to $\lambda_4$ can be eliminated by a field redefinition, since it is proportional to Ricci curvature. Consequently, we will not take it into account from now on, so we set $\lambda_4=0$. Therefore, we are left with only one term out of the initial ten. However, as we did with the four-derivative terms, we can also add parity-breaking densities by using the dual Riemann tensor. One finds again that there is only one independent term, and then the action $S^{(6)}$ reads

\begin{equation}\label{eq:final6der}
\begin{aligned}
S^{(6)}=\int d^4x\sqrt{|g|}&\left\{\lambda_{\rm {ev}}R_{\mu\nu }^{\ \ \rho\sigma }R_{\rho\sigma }^{\ \ \delta\gamma }R_{\delta\gamma }^{\ \ \mu\nu }+\lambda_{\rm {odd}}R_{\mu\nu }^{\ \ \rho\sigma }R_{\rho\sigma }^{\ \ \delta\gamma }\tilde R_{\delta\gamma }^{\ \ \mu\nu }\right\}\, ,
\end{aligned}
\end{equation}
where we have renamed the parameter $\lambda_2$ for evident reasons. Finally, we combine (\ref{eq:final2and4der}) and (\ref{eq:final6der}) to get the action of the effective field theory considered in the main text (\ref{Action}) and that we repeat here for convenience

\begin{equation}
\begin{aligned}
S=&\frac{1}{16\pi G}\int d^4x\sqrt{|g|}\bigg\{R-\frac{1}{2}(\partial\phi_{1})^2-\frac{1}{2}(\partial\phi_{2})^2+\alpha_{2}\left(\phi_2 \cos\theta_{m}+\phi_{1}\sin\theta_{m}\right) \ell^2 R_{\mu\nu\rho\sigma} {\tilde R}^{\mu\nu\rho\sigma}\\\
&+\alpha_{1} \phi_{1} \ell^2\mathcal{X}_{4}+\lambda_{\rm ev}\ell^4\tensor{R}{_{\mu\nu }^{\rho\sigma}}\tensor{R}{_{\rho\sigma }^{\delta\gamma }}\tensor{R}{_{\delta\gamma }^{\mu\nu }}+\lambda_{\rm odd}\ell^4\tensor{R}{_{\mu\nu }^{\rho\sigma}}\tensor{R}{_{\rho\sigma }^{\delta\gamma }} \tensor{\tilde R}{_{\delta\gamma }^{\mu\nu }}\bigg\}+ \sum_{n\ge 4}\frac{\ell^{2n-2}}{16\pi G}S^{(2n)}\, .
\end{aligned}
\end{equation}

\section{Compactification and truncation of the effective action of the Heterotic String}\label{ap:het}

Let us consider the effective action of the Heterotic Superstring, at first-order in the $\alpha'$ expansion, without gauge fields. The ten-dimensional action is given by\footnote{With respect to the conventions of \cite{Cano:2018qev, Chimento:2018kop}, here we are using mostly plus signature $g_{\mu\nu}\rightarrow - g_{\mu\nu}$ and the definition of the Riemann tensor differs by a minus sign, i.e. $R_{\mu\nu\rho}{}^\sigma\rightarrow -R_{\mu\nu\rho}{}^\sigma$.}

\begin{equation}\label{eq:actionheterotic}
S=\frac{g_{s}^2}{16\pi G^{(10)}}\int d^{10}x \sqrt{|g|} \, e^{-2\phi}\left[R+4\left(\partial\phi\right)^2-\frac{1}{2 \cdot 3!} H^2 +\frac{\alpha'}{8}R_{\mu\nu\rho\sigma}R^{\mu\nu\rho\sigma}\right]+\dots ,
\end{equation}
where $\alpha'=\ell_s^2$, being $\ell_s$ the string scale, $G^{(10)}$ is the ten-dimensional Newton's constant, and $g_{s}$ is the string coupling constant. The curvature-squared term\footnote{The curvature-squared term  in the Bergshoeff-de Roo scheme \cite{Bergshoeff:1989de} is $ R_{(-)}{}_{\mu\nu\rho\sigma} R_{(-)}{}^{\mu\nu\rho\sigma}$, where $R_{(-)}{}^a{}_b$ is the curvature of the torsionful spin-connection $\Omega_{(-)}{}^a{}_b=\omega^a{}_b-\frac{1}{2}H_{\mu}{}^a{}_b \, dx^\mu$. For our purposes, however, $R_{(-)}{}_{\mu\nu\rho\sigma} R_{(-)}{}^{\mu\nu\rho\sigma}=R_{\mu\nu\rho\sigma} R^{\mu\nu\rho\sigma}+\dots$, where the dots are terms that can be ignored.} is needed in order to supersymmetrize the action at first order in $\alpha'$, which otherwise would not be supersymmetric due to the presence of the Chern-Simons terms in the definition of the 3-form field strength $H$ (see \cite{Bergshoeff:1989de} for more details). As a consequence, the Bianchi identity is no longer $dH=0$ but it is corrected by 

\begin{equation}\label{eq:Bianchiid}
dH=\frac{\alpha'}{4}R^a{}_b\wedge R^b{}_a+\dots\ ,
\end{equation}
where $R^a{}_b=\frac{1}{2!} R_{\mu\nu}{}^a{}_b \, dx^\mu \wedge dx^\nu $ is the curvature 2-form. 

Now, let us perform the dimensional reduction of (\ref{eq:actionheterotic}) on a six torus, truncating all the Kaluza-Klein degrees of freedom. We get exactly the same action but now with the indices $\mu, \nu$ running from 0 to 4

\begin{equation}\label{eq:actionheterotic4d}
S=\frac{1}{16\pi G}\int d^{4}x \sqrt{|g|} \, e^{-2(\phi-\phi_{\infty})}\left[R+4\left(\partial\phi\right)^2-\frac{1}{2 \cdot 3!} H^2 +\frac{\alpha'}{8}R_{\mu\nu\rho\sigma}R^{\mu\nu\rho\sigma}\right]+\dots\ ,
\end{equation}
where $G$ is the four-dimensional Newton's constant, related to the ten-dimensional one by 

\begin{equation}
G^{(10)}= \left(2\pi \ell_s\right)^6\, G\ ,
\end{equation}
and we have also introduced $e^{\phi_{\infty}}=g_{s}$. Let us show that, ignoring terms whose contribution to the equations of motion is either zero or trivial, this action can be rewritten in a form such that it is manifestly a particular case of (\ref{Action}). First of all, let us rewrite the Bianchi identity (\ref{eq:Bianchiid}) as

\begin{equation}
\frac{1}{3!}\sqrt{|g|}\nabla_{\mu} H_{\nu\rho\sigma}\epsilon^{\mu\nu\rho\sigma}=-\frac{\alpha'}{8}\sqrt{|g|} R_{\mu\nu\rho\sigma} \tilde R^{\mu\nu\rho\sigma}+\dots \ .
\end{equation}

Secondly, we have to dualize the 3-form into a (pseudo)scalar $\varphi$. Following the usual procedure, we introduce a Lagrange multiplier into the action (\ref{eq:actionheterotic4d}), 

\begin{equation}\label{eq:actionheterotic4d1}
\begin{aligned}
S=&\frac{1}{16\pi G}\int d^{4}x\sqrt{|g|}  \left\{\, e^{-2(\phi-\phi_{\infty})}\left[R+4\left(\partial\phi\right)^2-\frac{1}{2 \cdot 3!} H^2 +\frac{\alpha'}{8}R_{\mu\nu\rho\sigma}R^{\mu\nu\rho\sigma}\right]\right.\\
&+\left.\varphi \left(\frac{1}{3!}\nabla_\mu H_{\nu\rho\sigma}\epsilon^{\mu\nu\rho\sigma}+\tfrac{\alpha'}{8}\,R_{\mu\nu\rho\sigma}\tilde R^{\mu\nu\rho\sigma}\right)\right\}+\dots \ .
\end{aligned}
\end{equation}
The relation between the 3-form field strength and the scalar  is found by imposing that the variation of the action with respect to $H$ vanishes,

\begin{equation}
\frac{\delta S}{\delta H}=0\quad  \Rightarrow \quad H^{\mu\nu\rho}=e^{2(\phi-\phi_{\infty})}\epsilon^{\mu\nu\rho\sigma}\partial_\sigma \varphi  \ .
\end{equation}
Now, we rewrite (\ref{eq:actionheterotic4d1}) in terms of $\varphi$, getting

\begin{equation}
\begin{aligned}
S=\frac{1}{16\pi G}\int d^{4}x \sqrt{|g|} \, e^{-2(\phi-\phi_{\infty})}&\left[R+4\left(\partial\phi\right)^2-\frac{e^{4(\phi-\phi_{\infty})}}{2}\left(\partial\varphi\right)^2 +\frac{\alpha'}{8}R_{\mu\nu\rho\sigma}R^{\mu\nu\rho\sigma}\right.\\
&+\left. \frac{\alpha' e^{2(\phi-\phi_{\infty})}}{8} \varphi R_{\mu\nu\rho\sigma}\tilde R^{\mu\nu\rho\sigma}\right]+\dots\ .
\end{aligned}
\end{equation}
Since this action is not written in the Einstein frame, let us rescale the metric $g_{\mu\nu}\rightarrow e^{-2(\phi-\phi_{\infty})}g_{\mu\nu}$ in order to eliminate the conformal factor.  Expanding in $(\phi-\phi_{\infty})$ and keeping only the leading terms, we get  

\begin{equation}
\begin{aligned}
S=&\frac{1}{16\pi G}\int d^{4}x \sqrt{|g|} \left[R-2\left(\partial\phi\right)^2-\frac{1}{2} \left(\partial\varphi\right)^2 +\frac{\alpha'}{8}(1-2\phi+2\phi_{\infty})R_{\mu\nu\rho\sigma}R^{\mu\nu\rho\sigma}\right.\\
&\left.+\frac{\alpha'}{8}\varphi\,R_{\mu\nu\rho\sigma}\tilde R^{\mu\nu\rho\sigma}\right]+\dots
\end{aligned}
\end{equation}
where we have dropped some terms that can be removed with a field redefinition. 
Finally, defining $\phi^1=2\phi-2\phi_{\infty}$ and $\phi^2=\varphi$, and ignoring terms that do not contribute to the equations of motion at leading order, we can write the action in the following final form

\begin{equation}
\begin{aligned}
S=\frac{1}{16\pi G}\int d^{4}x \sqrt{|g|} \left[R-\frac{1}{2}\left(\partial\phi^1\right)^2-\frac{1}{2} \left(\partial\phi^2\right)^2 -\frac{\alpha'}{8}\phi^1\mathcal X_4+\frac{\alpha'}{8}\varphi\,R_{\mu\nu\rho\sigma}\tilde R^{\mu\nu\rho\sigma}\right]\ .
\end{aligned}
\end{equation}
We have upgraded the Riemann squared term to the Gauss-Bonnet density $\mathcal{X}_{4}$ since both give the same contribution at leading order (the Ricci$^2$ and $R^2$ terms do not contribute). This can also be done by means of a field redefinition. 
Then, the choice of parameters that gives us the corrections  predicted by the simplest compactification of the effective action of the Heterotic Superstring is 

\begin{equation}
\alpha_1= -\frac{1}{8}\, ,\quad   \alpha_2= \frac{1}{8}\, ,\quad \theta_m=0\,, \quad \lambda_{\rm ev}=\lambda_{\rm odd}=0\, ,\quad  \ell=\ell_s\ .
\end{equation}

\section{The solution}\label{ap:sol}

In this appendix, we show the metric functions $H_1, H_2, H_3, H_4$ as well as the scalars of the solution, $\phi_1$ and $\phi_2$ up to order $\mathcal O\left(\chi^3\right)$,

\begin{align}
\phi_{1}&=\alpha_{1}\ell^2\left\{\frac{8 M}{3 \rho^3}+\frac{2}{\rho^2}+\frac{2}{M \rho}\right.\\
&\left.+\left[-\frac{M^2}{5 \rho^4}-\frac{2 M}{5
   \rho^3}-\frac{1}{2 \rho^2}-\frac{1}{2 M \rho}+\left(-\frac{96 M^3}{5 \rho^5}-\frac{42 M^2}{5
   \rho^4}-\frac{14 M}{5 \rho^3}\right) x^2\right] \chi ^2\right.\\
  & \left.+\left[-\frac{2 M^3}{35
   \rho^5}-\frac{M^2}{7 \rho^4}-\frac{3 M}{14 \rho^3}-\frac{1}{4 \rho^2}-\frac{1}{4 M
   \rho}+\left(\frac{4 M^4}{7 \rho^6}+\frac{24 M^3}{35 \rho^5}+\frac{3 M^2}{7 \rho^4}+\frac{M}{7
   \rho^3}\right) x^2\right.\right.\\
   &\left.\left.+\left(\frac{360 M^5}{7 \rho^7}+\frac{110 M^4}{7 \rho^6}+\frac{22 M^3}{7
   \rho^5}\right) x^4\right] \chi ^4\right\}\\
   &+\alpha_{2}\ell^2\sin\theta_m \left\{x \left(\frac{9 M^2}{\rho ^4}+\frac{5 M}{\rho ^3}+\frac{5}{2 \rho ^2}\right) \chi
   +\left[x^3 \left(-\frac{100 M^4}{3 \rho ^6}-\frac{12 M^3}{\rho ^5}-\frac{3 M^2}{\rho
   ^4}\right)\right.\right.\\
   &\left.\left.+x \left(-\frac{2 M^3}{5 \rho ^5}-\frac{3 M^2}{5 \rho ^4}-\frac{M}{2 \rho
   ^3}-\frac{1}{4 \rho ^2}\right)\right] \chi ^3\right\}+\mathcal{O}(\chi^5)\,,
\end{align}

\begin{align}
\phi_{2}&=\alpha_{2}\ell^2\cos\theta_m \left\{x \left(\frac{9 M^2}{\rho ^4}+\frac{5 M}{\rho ^3}+\frac{5}{2 \rho ^2}\right) \chi
   +\left[x^3 \left(-\frac{100 M^4}{3 \rho ^6}-\frac{12 M^3}{\rho ^5}-\frac{3 M^2}{\rho
   ^4}\right)\right.\right.\\
   &\left.\left.+x \left(-\frac{2 M^3}{5 \rho ^5}-\frac{3 M^2}{5 \rho ^4}-\frac{M}{2 \rho
   ^3}-\frac{1}{4 \rho ^2}\right)\right] \chi ^3\right\}+\mathcal{O}(\chi^5)\, .
\end{align}

\begin{equation}
\begin{aligned}
H_{1}&=\alpha_{1}^2\ell^4\left\{\frac{416 M^3}{11 \rho ^7}+\frac{112 M^2}{165 \rho ^6}+\frac{428 M}{1155 \rho
   ^5}-\frac{3202}{385 \rho ^4}-\frac{122}{385 M \rho ^3}-\frac{1117}{1155 M^2 \rho
   ^2}+\frac{1117}{1155 M^3 \rho }\right.\\
   &\left.+\chi ^2\left[x^2 \left(-\frac{87008 M^5}{165 \rho
   ^9}+\frac{3377728 M^4}{35035 \rho ^8}+\frac{903092 M^3}{15015 \rho ^7}+\frac{493638556
   M^2}{7882875 \rho ^6}-\frac{22915196 M}{7882875 \rho ^5}\right.\right.\right.\\
   &\left.\left.\left.-\frac{169553}{160875 \rho
   ^4}-\frac{7721321}{1126125 M \rho ^3}\right)-\frac{3635392 M^4}{105105 \rho
   ^8}-\frac{2245064 M^3}{105105 \rho ^7}-\frac{87538336 M^2}{7882875 \rho
   ^6}+\frac{995398 M}{2627625 \rho ^5}\right.\right.\\
   &\left.\left.+\frac{2988737}{1126125 \rho
   ^4}+\frac{736487}{1126125 M \rho ^3}-\frac{787153}{450450 M^2 \rho
   ^2}+\frac{787153}{450450 M^3 \rho }\right] 
   \right\}\\
   &+\alpha_{2}^2\ell^4\chi ^2\left\{x^2 \left(\frac{342 M^5}{\rho ^9}-\frac{9279 M^4}{637 \rho ^8}-\frac{19280
   M^3}{1001 \rho ^7}-\frac{1094689 M^2}{42042 \rho ^6}+\frac{298393 M}{84084 \rho
   ^5}+\frac{80291}{24024 \rho ^4}\right.\right.\\
   &\left.\left.+\frac{80291}{24024 M \rho ^3}\right)-\frac{20268
   M^4}{637 \rho ^8}-\frac{11710 M^3}{637 \rho ^7}-\frac{30707 M^2}{3234 \rho
   ^6}+\frac{1074 M}{7007 \rho ^5}-\frac{271}{12012 \rho ^4}-\frac{271}{12012 M \rho
   ^3}\right.\\
   &\left.+\frac{72185}{48048 M^2 \rho ^2}-\frac{72185}{48048 M^3 \rho }\right\} \\
   &+\alpha_{1}\alpha_{2}\sin(\theta_{m})\ell^4 \left\{ \chi x \left[\frac{21120 M^4}{91 \rho ^8}-\frac{21352 M^3}{1001 \rho ^7}-\frac{43564 M^2}{2145
   \rho ^6}-\frac{551776 M}{15015 \rho ^5}-\frac{5618}{15015 \rho ^4}\right.\right.\\
   &\left.\left.+\frac{89989}{30030
   M^2 \rho ^2}\right] +\chi^3\left[x^3 \left(-\frac{11556352 M^6}{4641 \rho
   ^{10}}+\frac{1402164667 M^5}{3828825 \rho ^9}+\frac{12014583319 M^4}{53603550 \rho
   ^8}\right.\right.\right.\\
   &\left.\left.\left.+\frac{879521737 M^3}{4873050 \rho ^7}-\frac{895892573 M^2}{76576500 \rho
   ^6}-\frac{611550767 M}{153153000 \rho ^5}-\frac{43683743}{12252240 \rho ^4}\right)+x
   \left(-\frac{555211 M^5}{49725 \rho ^9}\right.\right.\right.\\
   &\left.\left.\left.-\frac{417419143 M^4}{53603550 \rho
   ^8}-\frac{278633 M^3}{17867850 \rho ^7}+\frac{2744165393 M^2}{536035500 \rho
   ^6}+\frac{244492811 M}{30630600 \rho ^5}+\frac{157764391}{306306000 \rho
   ^4}\right.\right.\right.\\
   &\left.\left.\left.-\frac{75784931}{61261200 M^2 \rho ^2}\right)\right] \right\}\\
   &+\lambda_{\rm ev}\ell^4 \left\{-\frac{48 M^3}{11 \rho ^7}-\frac{8 M^2}{33 \rho ^6}-\frac{40 M}{231 \rho
   ^5}-\frac{32}{231 \rho ^4}-\frac{32}{231 M \rho ^3}-\frac{64}{231 M^2 \rho
   ^2}+\frac{64}{231 M^3 \rho }\right.\\
   &\left.+\chi^2\left[x^2 \left(\frac{1728 M^5}{11 \rho ^9}+\frac{1752
   M^4}{7007 \rho ^8}-\frac{800 M^3}{1001 \rho ^7}-\frac{8660 M^2}{7007 \rho
   ^6}-\frac{9518 M}{7007 \rho ^5}-\frac{1005}{1001 \rho ^4}-\frac{2669}{1001 M \rho
   ^3}\right)\right.\right.\\
   &\left.\left.-\frac{5952 M^4}{7007 \rho ^8}-\frac{520 M^3}{1617 \rho ^7}-\frac{68
   M^2}{1911 \rho ^6}+\frac{830 M}{7007 \rho ^5}+\frac{587}{3003 \rho ^4}+\frac{587}{3003
   M \rho ^3}-\frac{865}{1001 M^2 \rho ^2}+\frac{865}{1001 M^3 \rho }\right] \right\}\\
   &+\lambda_{\rm odd}\ell^4\left\{\chi x \left[-\frac{3456 M^4}{91 \rho ^8}-\frac{1152 M^3}{1001 \rho ^7}-\frac{96 M^2}{143 \rho
   ^6}-\frac{384 M}{1001 \rho ^5}-\frac{192}{1001 \rho ^4}+\frac{768}{1001 M^2 \rho
   ^2}\right] \right.\\
   &\left. +\chi^3\left[x^3 \left(\frac{745344 M^6}{1547 \rho ^{10}}-\frac{86140
   M^5}{17017 \rho ^9}-\frac{13190 M^4}{2431 \rho ^8}-\frac{515974 M^3}{119119 \rho
   ^7}-\frac{47015 M^2}{17017 \rho ^6}-\frac{274763 M}{238238 \rho
   ^5}\right.\right.\right.\\
   &\left.\left.\left.-\frac{511811}{476476 \rho ^4}\right)+x \left(-\frac{596 M^5}{221 \rho
   ^9}-\frac{8530 M^4}{119119 \rho ^8}+\frac{100326 M^3}{119119 \rho ^7}+\frac{111563
   M^2}{119119 \rho ^6}+\frac{153991 M}{238238 \rho ^5}\right.\right.\right.\\
   &\left.\left.\left.+\frac{111151}{476476 \rho
   ^4}-\frac{13735}{68068 M^2 \rho ^2}\right)\right]\right\}+\mathcal O\left(\chi^4\right)\, ,
\end{aligned}
\end{equation}

\begin{equation}
\begin{aligned}
H_2&=\alpha_1^2 \,\ell^4\left\{\frac{1117}{2310}+\frac{208}{11 \rho^6}-\frac{208}{165 \rho^5}-\frac{142}{231 \rho^4}-\frac{5188}{1155 \rho^3}-\frac{337}{1155 \rho^2}-\frac{1117}{2310 \rho}\right.\\
&\left.+\chi^2\left[\frac{787153}{900900}-\frac{1817696}{105105 \rho^7}-\frac{3258916}{315315 \rho^6}-\frac{42983383}{7882875 \rho^5}+\frac{93497}{5255250 \rho^4}+\frac{36396163}{31531500
   \rho^3}\right.\right.\\
&\left.\left.+\frac{2033089}{7882875 \rho^2}-\frac{787153}{900900 \rho}+x^2\left(-\frac{40384}{165 \rho^8}-\frac{4002832}{105105 \rho^7}-\frac{1675328}{315315 \rho^6}+\frac{190462}{9625
   \rho^5}+\frac{80274479}{15765750 \rho^4}\right.\right.\right.\\
&\left.\left.\left.+\frac{8052437}{6306300 \rho^3}-\frac{988269}{350350 \rho^2}\right) \right]\right\}\\
&+\alpha_1\alpha_2 \sin \theta_m\ell^4  \left\{\chi x\left(\frac{10560}{91 \rho^7}+\frac{2570584}{45045 \rho^6}+\frac{5030294}{225225 \rho^5}-\frac{1352913}{175175 \rho^4}-\frac{2310579}{350350 \rho^3}-\frac{2964341}{750750
   \rho^2}\right.\right.\\
&\left.\left.-\frac{12761}{17875 \rho}\right) +\chi^3\left[x\left(-\frac{555211}{99450 \rho^8}-\frac{33606289}{4123350 \rho^7}-\frac{1807455889}{321621300 \rho^6}-\frac{185905529}{160810650 \rho^5}+\frac{5347197771}{1667666000
   \rho^4}\right.\right.\right.\\
&\left.\left.\left.+\frac{20139019441}{15008994000 \rho^3}+\frac{827344753}{1000599600 \rho^2}-\frac{7696421}{25014990 \rho}\right) +x^3\left(-\frac{5239616}{4641
   \rho^9}-\frac{2886836873}{7657650 \rho^8}\right.\right.\right.\\
&\left.\left.\left.-\frac{3454981169}{53603550 \rho^7}+\frac{18272087263}{321621300 \rho^6}+\frac{3088505012}{134008875 \rho^5}+\frac{88918288507}{15008994000
   \rho^4}+\frac{24278291273}{45026982000 \rho^3}\right)  \right]\right\}\\
&+\alpha_2^2\,\ell^4\left\{-\frac{27}{2 \rho^5}-\frac{60}{7 \rho^4}-\frac{5}{\rho^3}+\chi^2\left[-\frac{72185}{96096}-\frac{10134}{637 \rho^7}-\frac{447949}{57330 \rho^6}-\frac{564161}{194040 \rho^5}+\frac{154675}{96096 \rho^4}\right.\right.\\
&\left.\left.+\frac{1153277}{1345344 \rho^3}+\frac{457841}{3363360
   \rho^2}+\frac{72185}{96096 \rho}+x^2\left(\frac{171}{\rho^8}+\frac{81219}{637 \rho^7}+\frac{4701743}{126126 \rho^6}-\frac{32689}{5544 \rho^5}\right.\right.\right.\\
&\left.\left.\left.-\frac{1852791}{224224 \rho^4}-\frac{3310225}{1345344
   \rho^3}+\frac{462029}{672672 \rho^2}\right) \right]\right\}\\
&+\lambda_{\rm ev}\,\ell^4\left\{\frac{32}{231}-\frac{24}{11 \rho^6}-\frac{4}{33 \rho^5}-\frac{20}{231 \rho^4}-\frac{16}{231 \rho^3}-\frac{16}{231 \rho^2}-\frac{32}{231 \rho}+\chi^2\left[\frac{865}{2002}-\frac{2976}{7007 \rho^7}\right.\right.\\
&\left.\left.-\frac{920}{1617 \rho^6}-\frac{853}{1911 \rho^5}-\frac{7349}{28028 \rho^4}-\frac{15739}{168168 \rho^3}+\frac{1783}{84084 \rho^2}-\frac{865}{2002
   \rho}+x^2\left(\frac{840}{11 \rho^8}+\frac{624704}{21021 \rho^7}\right.\right.\right.\\
&\left.\left.\left.+\frac{328360}{21021 \rho^6}+\frac{1781}{231 \rho^5}+\frac{276011}{84084 \rho^4}+\frac{156647}{168168 \rho^3}-\frac{78439}{84084
   \rho^2}\right) \right]\right\}\\
&+\lambda_{\rm odd}\,\ell^4\left\{\chi x\left(-\frac{1728}{91 \rho^7}-\frac{6560}{1001 \rho^6}-\frac{4040}{1001 \rho^5}-\frac{17676}{7007 \rho^4}-\frac{11112}{7007 \rho^3}-\frac{948}{1001 \rho^2}-\frac{135}{1001 \rho}\right) \right.\\
&\left.+\chi^3\left[x\left(-\frac{298}{221 \rho^8}-\frac{176734}{119119 \rho^7}-\frac{255071}{357357 \rho^6}-\frac{2333}{51051 \rho^5}+\frac{64221}{238238 \rho^4}+\frac{257645}{952952 \rho^3}\right.\right.\right.\\
&\left.\left.\left.+\frac{10575}{68068
   \rho^2}-\frac{1185}{34034 \rho}\right) +x^3\left(\frac{343296}{1547 \rho^9}+\frac{208174}{2431 \rho^8}+\frac{4665742}{119119 \rho^7}+\frac{5774105}{357357 \rho^6}+\frac{657785}{119119
   \rho^5}\right.\right.\right.\\
&\left.\left.\left.+\frac{22837}{17017 \rho^4}+\frac{12379}{408408 \rho^3}\right) \right]\right\}+\mathcal O\left(\chi^4\right)\, ,
\end{aligned}
\end{equation}

\begin{equation}
\begin{aligned}
H_3&=\alpha_1^2 \,\ell^4\left\{-\frac{1117}{1155}-\frac{368}{33 \rho ^6}-\frac{1168}{165 \rho ^5}-\frac{1102}{231 \rho ^4}-\frac{404}{1155 \rho ^3}-\frac{19}{1155 \rho ^2}+\frac{1117}{1155 \rho }\right.\\
&\left.+\chi^2\left[-\frac{787153}{450450}+\frac{210256}{105105 \rho ^7}+\frac{358564}{105105 \rho ^6}+\frac{29284144}{7882875 \rho ^5}+\frac{2871703}{1126125 \rho ^4}+\frac{888572}{1126125 \rho
   ^3}\right.\right.\\
&\left.\left.+\frac{10139}{150150 \rho ^2}+\frac{787153}{450450 \rho }+x^2 \left(\frac{23488}{165 \rho ^8}+\frac{6074176}{105105 \rho ^7}+\frac{1857368}{105105 \rho
   ^6}-\frac{64561864}{7882875 \rho ^5}-\frac{124199}{25025 \rho ^4}\right.\right.\right.\\
&\left.\left.\left.-\frac{329289}{125125 \rho ^3}+\frac{43252}{20475 \rho ^2}\right)\right]\right\}\\
&+\alpha_1\alpha_2\sin \theta_m\,\ell^4  \left\{\chi x\left[ -\frac{6144}{91 \rho ^7}-\frac{34044}{1001 \rho ^6}-\frac{31664}{2145 \rho ^5}+\frac{16854}{5005 \rho ^4}+\frac{16241}{5005 \rho ^3}+\frac{89989}{30030 \rho ^2}\right]\right.\\
&\left.+\chi^3 \left[x\left(\frac{1399373}{232050
   \rho ^8}+\frac{76931759}{8933925 \rho ^7}+\frac{237697639}{35735700 \rho ^6}+\frac{197152489}{76576500 \rho ^5}-\frac{142049293}{102102000 \rho ^4}\right.\right.\right.\\
&\left.\left.\left.-\frac{7454231}{5105100 \rho
   ^3}-\frac{75784931}{61261200 \rho ^2}\right)+x^3\left(\frac{9126080}{13923 \rho ^9}+\frac{1870089271}{7657650 \rho ^8}+\frac{1696574476}{26801775 \rho ^7}\right.\right.\right.\\
&\left.\left.\left.-\frac{320470309}{15315300 \rho
   ^6}-\frac{42765071}{4504500 \rho ^5}-\frac{29731159}{8751600 \rho ^4}+\frac{132059}{2356200 \rho ^3}\right)  \right]\right\}\\
&+\alpha_2^2\,\ell^4\left\{\chi^2\left[\frac{72185}{48048}+\frac{639}{1274 \rho ^7}+\frac{2005}{2548 \rho ^6}+\frac{41549}{42042 \rho ^5}+\frac{8581}{12012 \rho ^4}+\frac{887}{1716 \rho ^3}+\frac{270}{1001 \rho
   ^2}\right.\right.\\
&\left.\left.-\frac{72185}{48048 \rho }+x^2 \left(-\frac{99}{\rho ^8}-\frac{57843}{1274 \rho ^7}-\frac{455055}{28028 \rho ^6}+\frac{5891}{3822 \rho ^5}+\frac{3425}{8008 \rho
   ^4}-\frac{2969}{4004 \rho ^3}-\frac{14015}{6864 \rho ^2}\right)\right]\right\}\\
&+\lambda_{\rm ev}\,\ell^4\left\{-\frac{64}{231}-\frac{392}{11 \rho ^6}+\frac{8}{33 \rho ^5}+\frac{40}{231 \rho ^4}+\frac{32}{231 \rho ^3}+\frac{32}{231 \rho ^2}+\frac{64}{231 \rho }+\chi^2\left[-\frac{865}{1001}\right.\right.\\
&\left.\left.+\frac{3664}{7007 \rho ^7}+\frac{11380}{21021 \rho ^6}+\frac{752}{1911 \rho ^5}+\frac{213}{1001 \rho ^4}+\frac{139}{3003 \rho ^3}-\frac{32}{429 \rho
   ^2}+\frac{865}{1001 \rho }+x^2 \left(\frac{7960}{11 \rho ^8}\right.\right.\right.\\
&\left.\left.\left.-\frac{372584}{21021 \rho ^7}-\frac{199660}{21021 \rho ^6}-\frac{101648}{21021 \rho ^5}-\frac{6455}{3003 \rho
   ^4}-\frac{1921}{3003 \rho ^3}+\frac{3043}{3003 \rho ^2}\right)\right]\right\}\\
&+\lambda_{\rm odd}\,\ell^4\left\{\chi x  \left(-\frac{19008}{91 \rho ^7}+\frac{4320}{1001 \rho ^6}+\frac{384}{143 \rho ^5}+\frac{1728}{1001 \rho ^4}+\frac{1152}{1001 \rho ^3}+\frac{768}{1001 \rho ^2}\right)\right.\\
&\left.+\chi^3\left[x \left(\frac{2802}{1547 \rho ^8}+\frac{167628}{119119 \rho ^7}+\frac{71475}{119119 \rho ^6}-\frac{2087}{119119 \rho
   ^5}-\frac{148143}{476476 \rho ^4}-\frac{768}{2431 \rho ^3}-\frac{13735}{68068 \rho ^2}\right)\right.\right.\\
&\left.\left.+x^3 \left(\frac{2964736}{1547 \rho ^9}-\frac{838886}{17017 \rho ^8}-\frac{2761840}{119119 \rho ^7}-\frac{1189205}{119119 \rho ^6}-\frac{3665}{1001 \rho ^5}-\frac{486569}{476476
   \rho ^4}+\frac{601}{34034 \rho ^3}\right)\right]\right\}\\
&+\mathcal O\left(\chi^4\right)\, ,
\end{aligned}
\end{equation}

\begin{equation}
\begin{aligned}
H_4&=\alpha_1^2 \ell^4\left\{-\frac{1117}{1155}-\frac{368}{33 \rho ^6}-\frac{1168}{165 \rho ^5}-\frac{1102}{231 \rho ^4}-\frac{404}{1155 \rho ^3}-\frac{19}{1155 \rho ^2}+\frac{1117}{1155 \rho }\right.\\
&\left.+\chi^2\left[-\frac{787153}{450450}+\frac{1984}{33 \rho ^9}+\frac{3232}{165 \rho ^8}+\frac{529624}{21021 \rho ^7}+\frac{92864}{9555 \rho ^6}+\frac{91595428}{7882875 \rho
   ^5}+\frac{489939}{125125 \rho ^4}\right.\right.\\
&\left.\left.+\frac{5646292}{1126125 \rho ^3}+\frac{981961}{450450 \rho ^2}+\frac{787153}{450450 \rho }+x^2 \left(-\frac{1984}{33 \rho ^9}+\frac{6752}{55
   \rho ^8}+\frac{1212104}{35035 \rho ^7}+\frac{1194428}{105105 \rho ^6}\right.\right.\right.\\
&\left.\left.\left.-\frac{126873148}{7882875 \rho ^5}-\frac{7126703}{1126125 \rho ^4}-\frac{7721321}{1126125 \rho ^3}\right)\right]\right\}\\
&+\alpha_1\alpha_2\sin \theta_m\,\ell^4  \left\{\chi x \left[-\frac{6144}{91 \rho ^7}-\frac{34044}{1001 \rho ^6}-\frac{31664}{2145 \rho ^5}+\frac{16854}{5005 \rho ^4}+\frac{16241}{5005 \rho
   ^3}+\frac{89989}{30030 \rho ^2}\right]\right.\\
&\left.+\chi^3 \left[x\left(\frac{33408}{91
   \rho ^{10}}+\frac{17348416}{45045 \rho ^9}+\frac{736024381}{2552550 \rho ^8}+\frac{1195689244}{8933925 \rho ^7}+\frac{5169579091}{107207100 \rho ^6}+\frac{11447047}{1392300
   \rho ^5}\right.\right.\right.\\
&\left.\left.\left.-\frac{124881623}{102102000 \rho ^4}-\frac{43008619}{30630600 \rho ^3}-\frac{75784931}{61261200 \rho ^2}\right)+x^3\left(-\frac{33408}{91 \rho ^{10}}+\frac{69003776}{255255 \rho ^9}-\frac{291804563}{7657650 \rho ^8}\right.\right.\right.\\
&\left.\left.\left.-\frac{1659697979}{26801775 \rho
   ^7}-\frac{957111191}{15315300 \rho ^6}-\frac{1159441303}{76576500 \rho ^5}-\frac{43683743}{12252240 \rho ^4}\right) \right]\right\}\\
&+\alpha_2^2\,\ell^4\left\{\chi^2\left[\frac{72185}{48048}-\frac{54}{\rho ^8}-\frac{27897}{637 \rho ^7}-\frac{40995}{1274 \rho ^6}-\frac{18587}{1617 \rho ^5}-\frac{12993}{2002 \rho ^4}-\frac{12241}{3432 \rho
   ^3}-\frac{85145}{48048 \rho ^2}\right.\right.\\
&\left.\left.-\frac{72185}{48048 \rho }+x^2 \left(-\frac{45}{\rho ^8}-\frac{705}{637 \rho ^7}+\frac{234445}{14014 \rho ^6}+\frac{294806}{21021 \rho
   ^5}+\frac{183353}{24024 \rho ^4}+\frac{80291}{24024 \rho ^3}\right)\right]\right\}\\
&+\lambda_{\rm ev}\,\ell^4\left\{-\frac{64}{231}-\frac{392}{11 \rho ^6}+\frac{8}{33 \rho ^5}+\frac{40}{231 \rho ^4}+\frac{32}{231 \rho ^3}+\frac{32}{231 \rho ^2}+\frac{64}{231 \rho }\right.\\
&\left.\chi^2\left[-\frac{865}{1001}+\frac{736}{11 \rho ^9}+\frac{384}{11 \rho ^8}+\frac{393088}{21021 \rho ^7}+\frac{5356}{539 \rho ^6}+\frac{107504}{21021 \rho ^5}+\frac{6109}{3003 \rho
   ^4}+\frac{2075}{1001 \rho ^3}+\frac{2819}{3003 \rho ^2}\right.\right.\\
&\left.\left.+\frac{865}{1001 \rho }+x^2 \left(-\frac{736}{11 \rho ^9}+\frac{7576}{11 \rho ^8}-\frac{251560}{7007 \rho
   ^7}-\frac{132388}{7007 \rho ^6}-\frac{66960}{7007 \rho ^5}-\frac{3975}{1001 \rho ^4}-\frac{2669}{1001 \rho ^3}\right)\right]\right\}\\
&+\lambda_{\rm odd}\,\ell^4\left\{\chi x \left(-\frac{19008}{91 \rho ^7}+\frac{4320}{1001 \rho ^6}+\frac{384}{143 \rho ^5}+\frac{1728}{1001 \rho ^4}+\frac{1152}{1001 \rho ^3}+\frac{768}{1001 \rho ^2}\right)\right.\\
&\left.+\chi^3\left[x \left(\frac{34560}{91 \rho ^{10}}+\frac{175360}{1001 \rho ^9}+\frac{1340034}{17017 \rho
   ^8}+\frac{3834240}{119119 \rho ^7}+\frac{25469}{2431 \rho ^6}+\frac{138241}{119119 \rho ^5}-\frac{122901}{476476 \rho ^4}\right.\right.\right.\\&\left.\left.\left.-\frac{10151}{34034 \rho ^3}-\frac{13735}{68068 \rho
   ^2}\right)+x^3 \left(-\frac{34560}{91 \rho ^{10}}+\frac{29630976}{17017 \rho ^9}-\frac{2148098}{17017 \rho ^8}-\frac{6428452}{119119 \rho ^7}-\frac{2365711}{119119 \rho
   ^6}\right.\right.\right.\\
&\left.\left.\left.-\frac{576463}{119119 \rho ^5}-\frac{511811}{476476 \rho ^4}\right)\right]\right\}+\mathcal O\left(\chi^4\right)\ .
\end{aligned}
\end{equation}

\newpage

\section{Convergence of the $\chi$-expansion}\label{ap:conv}
\begin{figure}[ht!]
\begin{center}
\includegraphics[scale=0.6]{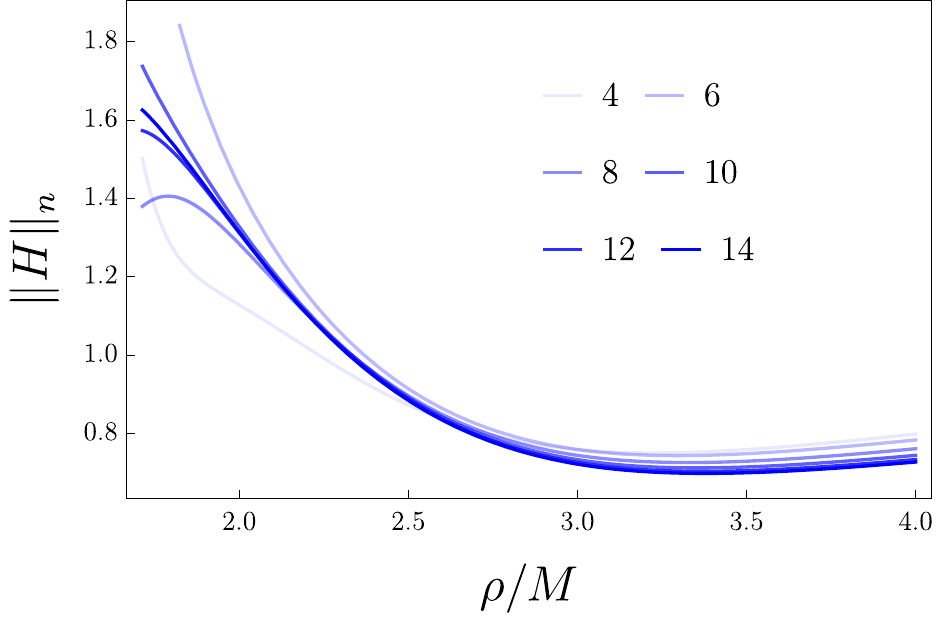} 
\caption{Convergence of the norm of the $H_i$ functions, $\|H\|_{n}$. We show the profile of $\|H\|_{n}$ in the axis $x=1$, $\rho\ge\rho_{+}$ for the values of $n$ indicated, for spin  $\chi=0.7$, and for couplings $\alpha_{1}=0.5$, $\alpha_{2}=0.7$, $\theta_{m}=\pi/4$, $\lambda_{\rm ev}=\lambda_{\rm odd}=1$. The accuracy of the expansion up to $n=14$ at $\rho=\rho_{+}$ is around $1\%$, but the convergence is much better as we move far from the horizon.}
\label{fig:conv}
\end{center}
\end{figure}

In this section we analyze the convergence of the solution presented in Sec.~\ref{sec:solving}, and whose first terms in the $\chi$-expansion are shown in Appendix~\ref{ap:sol}. In order to study the convergence, first we must consider the partial sums
\begin{equation}\label{eq:partialsum}
H_{i,n}=\sum_{k=0}^{n}H_{i}^{(k)}\chi^{k}\, ,\quad i=1,2,3,4\ .
\end{equation}
Then, we have to investigate if the sequence of functions $H_{i,n}$ converges to a function $H_i$, this is, 
\begin{equation}
\lim_{n\rightarrow\infty}H_{i,n}= H_{i}\, ,
\end{equation}
and what is the radius of convergence for a given domain  $(\rho,x)\in \Omega$.
We can study the convergence of the four functions $H_{i}$ at the same time by introducing the ``norm" $\|H\|_{n}$, as 
\begin{equation}
\|H\|_{n}:=\sqrt{H_{1,n}^2+H_{2,n}^2+H_{3,n}^2+H_{4,n}^2}\, .
\end{equation}
Thus, every $H_{i,n}$ converges if and only if $\|H\|_{n}$ converges. Since we are only interested in the exterior region of the black hole, it is sufficient to look at the convergence for $\rho\ge\rho_{+}$ , $-1<x<1$. Using the terms of the solution up to order $n=14$, we observe that the sequence $\|H\|_{n}$ converges in the exterior region if the spin is small enough. We wish to determine the maximum value of $\chi$ for which the expansion up to order $n=14$ --- the one we use thorough the text --- provides an accurate approximation to the full series. This value of course depends on the point of the spacetime. For instance, far from the black hole the few first terms in the expansion already provide a very precise result, even for $\chi\sim1$. On the contrary, the convergence is worse at the horizon $\rho=\rho_{+}$, and, specially, at the axes $x=\pm 1$. Thus, we should look at the convergence at those points. It is useful to define the relative differences, 
\begin{equation}
d_{n}=\frac{\|H\|_{n+1}-\|H\|_{n}}{\|H\|_{n}}.
\end{equation}
For instance, when we evaluate this for $\alpha_{1}=0.5$, $\alpha_{2}=0.7$, $\theta_{m}=\pi/4$, $\lambda_{\rm ev}=\lambda_{\rm odd}=1$ and $\chi=0.7$, at the north pole of the horizon, $\rho=\rho_{+}$, $x=1$, we get the following sequence (starting at $n=0$): 1.3, -0.79, 2.7, -0.74, 2.9, -0.61, -0.52, 0.25, 0.72, -0.27, -0.21, 0.14, 0.076, -0.039, \ldots  We see that the sequence starts converging from $n=8$, and the difference between $\|H\|_{13}$ and $\|H\|_{14}$ is barely a $4\%$. Since the difference with the term $n=15$ will be even smaller, we are confident that the series up to $n=14$ provides a precision of the order of $1\%$, for $\chi=0.7$ when evaluated at the north pole of the horizon. In the rest of points, and for smaller values of $\chi$, the accuracy is significantly greater.  We illustrate this in In Fig.~\ref{fig:conv} , where we show the profile of $\|H\|_{n}$, for several values of $n$, in the line $x=1$, $\rho\ge\rho_{+}$. Also, if for the same values of the couplings we set $\chi=0.65$, we get $d_{13}=-0.81 \%$, so the series up to order $n=14$ is around five times more accurate than for $\chi=0.7$.

Finally, one could try to determine what is the maximum value of $\chi$ for which the series will converge all the way up to the horizon. In order to find the radius of convergence, one can apply for example the root test to the coefficients $H_{i}^{(k)}$ in (\ref{eq:partialsum}):

\begin{equation}
\chi_{\rm max}=\inf\left\{\lim_{k\rightarrow\infty} \left|H_{i}^{(k)}\right|^{-1/k}\bigg|\ \  i=1,2,3,4, \ \ \rho \ge \rho_{+}, \ \ -1< x<1 \right\}\, .
\end{equation}

Using the coefficients up to order $n=14$ it is difficult to provide a definitive answer, but the results seem to be consistent with $\chi_{\rm max}\sim 1$. So, it could be possible to get close to the extremal limit adding enough terms in the series expansion, though the number of terms required to get a good approximation increases quite rapidly as we approach $\chi=1$.

\section{Some formulas}\label{ap:aux}
\subsubsection*{Radius of the ergosphere}
\begin{align}\label{eq:rhoerg}
\Delta \rho^{(1)}&=\chi ^2(1-x^2) \frac{53}{120}+\chi ^4 (1-x^2)\left(\frac{56750791}{84084000}-\frac{41115397 x^2}{63063000}\right)\\
\nonumber
&+\chi ^6 (1-x^2)\left(-\frac{679368329719 x^4}{912546835200}+\frac{10245671873 x^2}{165917606400}+\frac{336187298257}{1825093670400}\right)+\mathcal{O}(\chi^8)\, ,\\
\Delta \rho^{(2)}&=-\chi ^2 (1-x^2)\frac{709 x^2}{448}-\chi ^4(1-x^2) \left(\frac{4433503 x^2}{2690688}+\frac{504467}{1345344}\right)\\
\nonumber
&-\chi ^6(1-x^2) \left(\frac{915791950769 x^4}{625746401280}+\frac{148163587307 x^2}{312873200640}+\frac{8754619243}{18962012160}\right)+\mathcal{O}(\chi^8)\, ,\\
\Delta \rho^{(m)}&=\chi ^5 x(1-x^2)\left(\frac{401316913 x^2}{22870848000}-\frac{401316913 }{22870848000}\right)\\
\nonumber
&+\chi ^7 x(1-x^2)\left(\frac{1222303361 x^5}{85085952000}+\frac{116649901427 x^3}{12167291136000}-\frac{39651603 x}{1655413760}\right)+\mathcal{O}(\chi^9)\, ,\\
\Delta \rho^{({\rm ev})}&=\chi ^2 (1-x^2)\frac{1}{2}+\chi ^4(1-x^2) \left(\frac{1245 x^2}{5096}+\frac{3243}{10192}\right)\\
\nonumber
&+\chi ^6 (1-x^2)\left(\frac{14596973 x^4}{79008384}+\frac{24066599 x^2}{158016768}+\frac{1021961}{4051712}\right)+\mathcal{O}(\chi^8)\, ,\\
\Delta \rho^{({\rm odd})}&=\chi ^5 x(1-x^2)\left(\frac{669 }{106624}-\frac{669 x^2}{106624}\right)\\
\nonumber
&+\chi ^7 x(1-x^2) \left(-\frac{109 x^4}{14896}+\frac{131 x^2}{144704}+\frac{6495 }{1012928}\right)+\mathcal{O}(\chi^9)\, .
\end{align}
\subsubsection*{Some Christoffel symbols}
\begin{eqnarray}\label{eq:christ}
\Gamma^\rho_{tt}&=&\frac{\rho^2_\pm -2M\rho_\pm +M^2 \chi^2}{\rho_\pm^4}\left[M+\frac{\ell^4}{2M^4}\left(2M H_3+\rho^2 \partial_\rho H_1\right)\right]\Bigg|_{\rho=\rho_\pm, x=0}\ ,\\
\Gamma^\rho_{t\phi}&=&-\frac{\left(\rho^2_\pm -2M\rho_\pm +M^2 \chi^2\right) M^2\chi}{\rho_\pm^4}\left[1-\frac{\ell^4}{M^4}\left(H_3-H_2+\rho \partial_\rho H_2\right)\right]\Bigg|_{\rho=\rho_\pm, x=0}\ ,\\
\Gamma^\rho_{\phi\phi}&=&-\frac{\rho^2_\pm -2M\rho_\pm +M^2 \chi^2}{\rho_\pm^4}\left[\left(\rho_\pm^3-M^3\chi^2\right)\left(1+\frac{\ell^4}{ M^4}\left(H_4-H_3\right)\right)\right.\nonumber\\
&+&\left.\frac{\ell^4}{ 2M^4}\left(\rho_\pm^4+2 M^3\chi^2\rho_\pm +M^2\chi^2\rho_\pm^2\right)\partial_\rho H_4\right]\Bigg|_{\rho=\rho_\pm, x=0}\ .
\end{eqnarray}

\subsubsection*{Photon rings}
\begin{eqnarray}\label{eq:rhoph}
\Delta \rho_{{\rm ph}\pm} ^{(1)}&=&-\frac{11833}{280665}\mp\frac{1894454 \chi }{841995 \sqrt{3}}+\frac{7366829759 \chi ^2}{3831077250}\mp\frac{63500581373 \chi ^3}{51719542875 \sqrt{3}}+\frac{4499912684330179 \chi ^4}{5613018549138000}\\
&&\mp\frac{2518625711779631 \chi ^5}{16839055647414000 \sqrt{3}}+\frac{39043683908212961 \chi ^6}{237415044638772000}\pm\frac{88176541508946559687 \chi ^7}{153370118836646712000 \sqrt{3}}+\mathcal O\left(\chi^8\right)\ , \nonumber\\
\Delta \rho_{{\rm ph}\pm} ^{(2)}&=&\pm\frac{124 \chi }{81 \sqrt{3}}-\frac{27237253 \chi ^2}{27243216}\pm\frac{7143579103 \chi ^3}{3677834160 \sqrt{3}}-\frac{4930918052561 \chi ^4}{8018597927340}\\
&&\pm\frac{941808834424915 \chi ^5}{538849780717248 \sqrt{3}}-\frac{105521162301612787 \chi ^6}{151476660579404160}\pm\frac{328617664140943525921 \chi ^7}{184044142603976054400 \sqrt{3}}+\mathcal O\left(\chi^8\right)
\ , \nonumber\\
\Delta \rho_{{\rm ph}\pm} ^{(\rm ev)}&=&
\frac{424}{6237}\mp\frac{656 \chi }{693 \sqrt{3}}+\frac{11087308 \chi ^2}{15324309}\mp\frac{88055819 \chi ^3}{91945854 \sqrt{3}}+\frac{18900112949 \chi ^4}{44547766263}\\
&&\mp\frac{2387981426735 \chi ^5}{3207439170936 \sqrt{3}}+\frac{965001464261 \chi ^6}{2874198737592}\mp\frac{4056120100091 \chi ^7}{6384037580613 \sqrt{3}}+\mathcal O\left(\chi^8\right)\ .\nonumber
\end{eqnarray}

\bibliography{Gravities}
\bibliographystyle{JHEP-2}
\label{biblio}

\end{document}